\documentclass[11pt]{article}
\usepackage{amssymb,amsmath,amsfonts}
\usepackage{graphicx}
\usepackage{graphics}
\usepackage{eepic,epsfig}
\usepackage{dcolumn}

\@addtoreset{equation}{section}

\makeatother

\begin{document}

\title{Finite temperature fermionic charge and current densities \\
in conical space with a circular edge}
\author{A. A. Saharian\thanks{%
E-mail: saharian@ysu.am}, V. F. Manukyan\thanks{%
E-mail: vardan.manukyan@ysu.am}, T. A. Petrosyan\thanks{%
E-mail: tigran.petrosyan@ysu.am} \vspace{0.3cm} \\
\textit{Institute of Physics, Yerevan State University,}\\
\textit{1 Alex Manoogian Street, 0025 Yerevan, Armenia} \vspace{0.3cm} }
\maketitle

\begin{abstract}
We study the finite temperature and edge induced effects on the charge and
current densities for a massive spinor field localized on a 2D conical space
threaded by a magnetic flux. The field operator is constrained on a circular
boundary, concentric with the cone apex, by the bag boundary condition and
by the condition with the opposite sign in front of the term containing the
normal to the edge. In two-dimensional spaces there exist two inequivalent
representations of the Clifford algebra and the analysis is presented for
both the fields realizing those representations. The circular boundary
divides the conical space into two parts, referred as interior (I-) and
exterior (E-) regions. The radial current density vanishes. The edge induced
contributions in the expectation values of the charge and azimuthal current
densities are explicitly separated in the both regions for the general case
of the chemical potential. They are periodic functions of the magnetic flux
and odd functions under the simultaneous change of the signs of magnetic
flux and chemical potential. An important difference from the fermion
condensate, considered previously in the literature, is that the mean charge
and current densities are finite in the limit when the observation point
tends to the boundary. In the E-region all the spinorial modes are regular
and the total charge and current densities are continuous functions of the
magnetic flux. In the I-region the corresponding expectation values are
discontinuous at half-integer values of the ratio of the magnetic flux to
the flux quantum. Those discontinuities come from the contribution of the
irregular mode in the I-region. 2D fermionic models, symmetric under the
parity and time-reversal transformations (in the absence of magnetic fields)
combine two spinor fields realizing the inequivalent representations of the
Clifford algebra. The total charge and current densities in those models are
discussed for different combinations of the boundary conditions for separate
fields. Applications are discussed for electronic subsystem in graphitic
cones described by the 2D Dirac model.
\end{abstract}

\bigskip


\bigskip

\section{Introduction}

The investigation of (2+1)-dimensional field-theoretical models is motivated
by several reasons. Despite the low dimensionality, these models may exhibit
a rich variety of properties that can reveal insights into theories with
higher dimensionality, providing a testing ground for effects which are more
complicated for the analysis in 4-dimensional spacetime. The 2D models often
allow for exact solutions and detailed analytical studies that are not
feasible in higher dimensions. These models appear as high temperature
limits of (3+1)-dimensional field theories and describe the low energy
dynamics of excitations in condensed matter systems \cite{Frad13,Naga99}. An
interesting area of research is the lower dimensional gravity \cite{Jack85}
which provides unique tools for studying gravitational interactions with
reduced complexity. The corresponding models can shed light on the nature of
spacetime, black holes, and quantum gravity, potentially extending our
understanding of higher-dimensional theories of gravity. The 2D theories
foster interdisciplinary research between high-energy physics, condensed
matter physics, and mathematical physics. These models involve rich
mathematical structures, such as modular forms and braid groups. Many
physical phenomena in Dirac materials such as graphene, topological
insulators and Weyl semimetals can be effectively explained by 2D theories.
For those systems, the long-wavelength excitations of electronic subsystem
are described by the Dirac equation with the velocity of light replaced by
the Fermi velocity which is nearly 300 times smaller \cite{Gusy07}-\cite%
{Xiao11}. As a result, an opportunity is opened for the investigation of the
relativistic effects at lower velocities. Moreover, the Coulomb interaction
is strongly renormalized on the graphite sheet. In 3D topological
insulators, the two dimensional massless fermionic excitations appear as
edge states on the surface. The quantum Hall effect and other topological
phases of matter, observed in condensed matter physics, exhibit exotic
properties such as fractional statistics and topologically protected edge
states, which can be effectively described using (2+1)-dimensional field
theories. In (2+1) dimensions, gauge theories with Chern-Simons terms are of
particular interest. These terms lead to novel features like topological
mass generation and have applications in both high-energy physics and
condensed matter systems, such as anyonic excitations in fractional quantum
Hall systems \cite{Muno20}.

In a number of 2D models of quantum field theory, including those describing
condensed matter systems, the physical degrees of freedom are confined to
finite regions by imposing different types of periodicity and boundary
conditions. Additional conditions on the fields are also imposed in problems
with different sorts of topological defects. Understanding how the boundary
conditions affect the behavior of quantum fields in (2+1) dimensions can
lead to more realistic and applicable models. Investigating the edge-induced
effects in (2+1) spacetime dimensions can also provide insights into the
AdS/CFT correspondence, where the (2+1)-dimensional quantum theory is
considered on the boundary of a (3+1)-dimensional AdS spacetime. The latter
can enhance our understanding of the importance of holography related to the
string theory and quantum gravity. In the context of field theories the
boundary conditions modify the spectrum of quantum fluctuations, thereby
giving rise to boundary-induced contributions in the expectation values of
physical characteristics. This type of shift is known as the Casimir effect,
which has been the subject of extensive research due to its significance in
diverse areas of fundamental physics and micromechanical applications.
Investigations have been conducted for a broad spectrum of boundary and
background geometries (for reviews see \cite{Most97}-\cite{Casi11}). A
series of analogous effects emerge in models with compact spatial dimensions
as a consequence of the compactification conditions imposed on the fields
(topological Casimir effect).

Another field of active research is the investigation of the effects caused
by a finite temperature on 2D systems, impacting their physical properties
and behavior. Considering these temperature-dependent effects is critical
for the practical application of 2D materials in various technologies,
including optoelectronics, sensors, and nanotechnology. Finite temperature
induces thermal fluctuations that can affect the structural integrity by
causing sometimes to the formation of ripples or other deformations in the
material. Additionally, the electronic, magnetic and optical properties can
be significantly influenced by finite temperature. By increasing the thermal
energy it is possible to affect the mobility of charge carriers,
conductivity, and band structure. For example, in semiconducting 2D
materials, temperature can impact the band gap and carrier concentration.
For instance, the effects of the Coulomb interaction in multi-walled carbon
nanotubes have been observed by several electrical transport experiments. At
temperatures lower than 1 K, the Coulomb blockade regime appears where the
tunneling transparency of a barrier vanishes in the device due to the
interactions between electrons. When the temperature increases, the
Luttinger liquid regime may be formed. The transition from one regime to
another is theoretically studied in \cite{Bell06}. At very high
temperatures, the thermal agitation can disrupt magnetic ordering in 2D
materials, potentially leading to a transition from ferromagnetic to
paramagnetic states. Changes in temperature can alter the absorption and
emission properties, the band structure and exciton binding energy, as a
result, changing the way how the material interacts with light.

In the present paper we investigate the combined effects of nontrivial
topology and boundary on the finite temperature fermionic charge and current
densities for a massive field in (2+1)-dimensional conical spacetime. The
latter appears as an effective geometry in field-theoretical models
describing 2D condensed matter systems and as a subspace of the geometry
outside straight cosmic strings. The formation of topological defects of
cosmic string type during phase transitions in the early universe may have
significant implications for astrophysics and cosmology \cite{Kibb76,Vile94}%
. An illustrative example of condensed matter systems exhibiting effective
conical geometry is that of graphitic cones \cite{Ge94}-\cite{Matt23}. The
effects of nontrivial topology of a conical space on the local
characteristics of the vacuum state for quantum fields have been extensively
examined in the existing literature (see, e.g., the references in \cite%
{Bell20}). In this study, the expectation values of the field squared,
fermion condensate and energy-momentum tensor were considered as key
characteristics. In conical geometries with magnetic fluxes within the core,
another noteworthy phenomenon is the emergence of azimuthal vacuum currents
\cite{Srir01}-\cite{Site18} (the vacuum currents in more general curved 2D
geometries were studied in \cite{Saha24}). The boundary-induced Casimir
effect for the current density in conical geometries has been investigated
in \cite{Bell20,Beze10}. The influence of the defect core of finite
thickness on the vacuum current was discussed in \cite{Beze15}. The current
density in conical spacetimes with local de Sitter and anti-de Sitter
geometries were studied in \cite{Oliv19}-\cite{Oliv24} (for vacuum currents
in de Sitter and anti-de Sitter spacetimes with a part of dimensions
compactified on a torus see \cite{Saha24b} and references therein).

The organization of the paper is as follows. In the next section we describe
the field, the boundary condition and the background geometry. The complete
set of mode functions is given for the regions outside and inside a circular
boundary. In Section \ref{sec:BF} we summarize the results for the vacuum
expectation values in a conical space with a circular boundary and for the
finite temperature charge and current densities in a boundary-free conical
space. The finite temperature expectation values of the charge and current
densities outside and inside a circular boundary in 2D conical space are
evaluated in Sections \ref{sec:Exter} and \ref{sec:Inter}. The
boundary-induced contributions are explicitly separated in the expectation
values. The asymptotic behavior of the boundary-induced parts in the
limiting regions of the parameters is described and numerical analysis is
presented in Section \ref{sec:Num}. The main results of the paper are
summarized in Section \ref{sec:Conc}. Appendix \ref{sec:appA} contains some
intermediate calculations which are needed for the final expressions of the
expectation values in the exterior region in the cases of nonzero and zero
chemical potentials. An alternative derivation of the zero temperature
limits of the expectation values of the charge and current densities is
presented in Appendix \ref{sec:AppB}.

\section{Problem setup and fermionic modes}

\label{sec:modes}

In this section we present the complete set of fermionic mode functions for
a conical geometry in the presence of a circular boundary. We begin with the
(3+1)-dimensional flat spacetime line element written in terms of spherical
coordinates $\left( r,\theta ,\tilde{\phi}\right) $ with a fixed value of
the polar coordinate $\theta =\theta _{0}$. It has the form $%
ds^{2}=dt^{2}-dr^{2}-r^{2}\sin ^{2}\theta _{0}d\tilde{\phi}^{2}$ where $r$
takes non-negative values and $0\leq \tilde{\phi}<2\pi $. Introducing a new
azimuthal coordinate $\phi =\tilde{\phi}\sin \theta _{0}$ with the variation
range $0\leq \phi <2\pi \sin \theta _{0}\equiv \phi _{0}$, the expression
for the line element reads
\begin{equation}
ds^{2}=g_{\mu \nu }dx^{\mu }dx^{\nu }=dt^{2}-dr^{2}-r^{2}d\phi ^{2}.
\label{linel}
\end{equation}%
In general, this corresponds to a conical spacetime described in terms of
spatial polar coordinates $\left( r,\phi \right) $. The point $r=0$
corresponds to the apex of a cone with a planar angle deficit $2\pi -\phi
_{0}$. In the special case $\phi _{0}=2\pi $ the bulk coincides with the
(2+1)-dimensional Minkowski spacetime.

We consider a spinor field $\psi (x)$ in the presence of an external
electromagnetic field with the vector potential $A_{\mu }=(0,0,A)$ having a
constant angular component $A_{2}=A$. This means that the physical component
of the vector potential is expressed as $A_{\phi }=-A/r$. The latter
corresponds to an infinitely thin magnetic flux $\Phi =-\phi _{0}A$ located
at $r=0$. The spinor field operator obeys the Dirac equation
\begin{equation}
\left( i\gamma ^{\mu }D_{\mu }-sm\right) \psi (x)=0.  \label{Dirac}
\end{equation}%
Here, we use the units $\hbar =c=1$, $m$ is the field mass and the parameter
$s=\pm 1$ describes two inequivalent irreducible representations of the
Clifford algebra (see the discussion in Section \ref{sec:Rep2} below). The
gauge covariant derivative in the Dirac equation is defined as $D_{\mu
}=\partial _{\mu }+\Gamma _{\mu }+ieA_{\mu }$, where $\Gamma _{\mu }$ is the
spin connection and $e$ is the charge of the field quantum. The spin
connection has the form
\begin{equation}
\Gamma _{\mu }=\frac{1}{4}\gamma ^{(a)}\gamma ^{(b)}e_{(a)}^{\nu }e_{(b)\nu
;\mu },\;a,b=0,1,2,  \label{Gam}
\end{equation}%
where the flat spacetime Dirac matrices $\gamma ^{(a)}$ are expressed in
terms of the Pauli matrices $\sigma _{\mu }$ as
\begin{equation}
\gamma ^{(0)}=\sigma _{3},\;\gamma ^{(1)}=i\sigma _{1},\;\gamma
^{(2)}=i\sigma _{2}.  \label{gamM}
\end{equation}%
A convenient form for the basis tetrads $e_{(a)}^{\mu }$ is given by
\begin{equation}
e_{\left( 0\right) }^{\mu }=\delta _{0}^{\mu },\;e_{\left( 1\right) }^{\mu
}=\left( 0,\cos \left( q\phi \right) ,-\frac{1}{r}\sin \left( q\phi \right)
\right) ,e_{\left( 2\right) }^{\mu }=\left( 0,\sin \left( q\phi \right) ,%
\frac{1}{r}\cos \left( q\phi \right) \right) ,  \label{tet}
\end{equation}%
where we have introduced the parameter $q=2\pi /\phi _{0}\geq 1$. For the
completeness of discussion we present the form of the curved spacetime Dirac
matrices $\gamma ^{\mu }=e_{(a)}^{\mu }\gamma ^{(a)}$ as well:
\begin{equation}
\gamma ^{1}=i\left(
\begin{array}{cc}
0 & e^{-iq\phi } \\
e^{iq\phi } & 0%
\end{array}%
\right) ,\;\gamma ^{2}=\frac{1}{r}\left(
\begin{array}{cc}
0 & e^{-iq\phi } \\
-e^{iq\phi } & 0%
\end{array}%
\right) ,  \label{gamma}
\end{equation}%
and $\gamma ^{0}=\sigma _{3}$.

\begin{figure}[tbph]
\begin{center}
\epsfig{figure=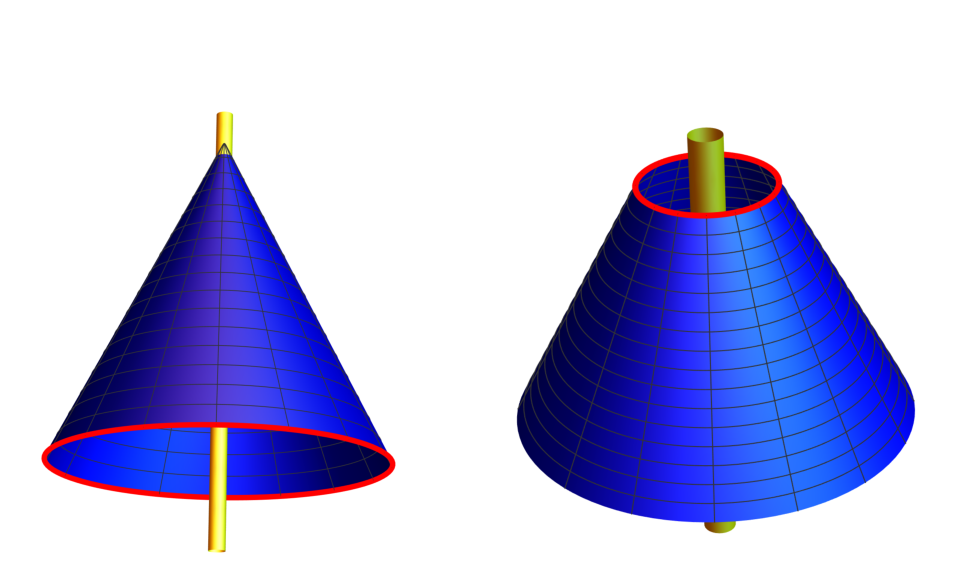,width=8cm,height=5cm}
\end{center}
\caption{The I- and E-regions of a conical space with a circular edge.}
\label{fig1}
\end{figure}

Additionally, we assume the presence of a circular boundary located at $r=a$
which divides the space into two separate regions. The region $r\leq a$
inside the boundary will be called as an interior region (I-region).
Similarly, the region $r\geq a$ outside the circle will be referred to as
the exterior region (E-region). These regions with circular boundaries,
embedded in three-dimensional Euclidean space with a magnetic flux, are
depicted in Figure \ref{fig1}.
We will denote the unit normal to the boundary having an inward direction by
$n_{\mu }$. Thus, we have $n_{\mu }=\delta ^{\mathrm{(J)}}\delta _{\mu }^{1}$%
, where $\mathrm{J}=\mathrm{I}$ and $\mathrm{J}=\mathrm{E}$ for the I- and
E-regions, respectively, and%
\begin{equation}
\delta ^{\mathrm{(J)}}=\left\{
\begin{array}{cc}
1, & \mathrm{J}=\mathrm{I} \\
-1, & \mathrm{J}=\mathrm{E}%
\end{array}%
\right. .  \label{delJ}
\end{equation}%
Further, it is assumed that at $r=a$ the field operator obeys the MIT bag
boundary condition
\begin{equation}
\left( 1+in_{\mu }\gamma ^{\mu }\right) \psi (x)=0,\;r=a.  \label{BCMIT}
\end{equation}%
This presents the most frequently used condition for confining fermions. The
charge and current densities for the boundary condition that differs from (%
\ref{BCMIT}) by the sign in front of the term involving the normal are
discussed at the end of Section \ref{sec:Inter}. Note that the radius of the
bounding circle in the embedding 3D Euclidean space is given by $a_{\mathrm{e%
}}=a/q$. In the part of the problem considering the expectation values in
the E-region the radius $a_{\mathrm{m}}$ of the location of magnetic flux
can be arbitrary in the range $a_{\mathrm{m}}<a_{\mathrm{e}}$.

Let $\{\psi _{\sigma }^{(+)}(x),\psi _{\sigma }^{(-)}(x)\}$ be the complete
normalized set of the positive- and negative-energy mode functions obeying
the equation (\ref{Dirac}). Here $\sigma $ is the set of quantum numbers
specifying the modes. The expansion of the field operator in terms of those
functions has the form
\begin{equation}
\psi =\sum_{\sigma }\left[ a_{\sigma }\psi _{\sigma }^{(+)}+b_{\sigma
}^{\dagger }\psi _{\sigma }^{(-)}\right] ,  \label{mexp}
\end{equation}%
where the creation and annihilation operators obey the standard
anticommutation relations.

The mode functions $\psi _{\sigma }^{(\pm )}(x)$ for the physical system
under consideration can be found, for example, in \cite{Saha19} and here we
present them for further references. Those functions are specified by the
quantum numbers $(\gamma ,j)$, with $j=\pm 1/2,\pm 3/2,\ldots $, and have
the form%
\begin{equation}
\psi _{\sigma }^{(\pm )}(x)=c^{(\pm )}e^{iqj\phi \mp iEt}\left(
\begin{array}{c}
Z_{\beta _{j}}^{(\pm )}(\gamma r)e^{-iq\phi /2} \\
\epsilon _{j}\frac{\gamma e^{iq\phi /2}}{\pm E+sm}Z_{\beta _{j}+\epsilon
_{j}}^{(\pm )}(\gamma r)%
\end{array}%
\right) ,  \label{psiIE}
\end{equation}%
where $E=E_{\sigma }=\sqrt{\gamma ^{2}+m^{2}}$ stands for the energy and $%
Z_{\beta _{j}}^{(\pm )}(\gamma r)$ is a cylindrical function of the order
\begin{equation}
\beta _{j}=\alpha _{j}-\epsilon _{j}/2,\;\alpha _{j}=q|j+\alpha |.
\label{betj}
\end{equation}%
Here we use the notations $\epsilon _{j}=\mathrm{sgn\,}(j+\alpha )$ and
\begin{equation}
\alpha =\frac{eA}{q}=-\frac{e\Phi }{2\pi }.  \label{alphan}
\end{equation}%
By taking into account that $\Phi _{0}=2\pi /e$ is the quantum of magnetic
flux, the parameter $\alpha $ is interpreted in terms of the ratio of the
magnetic flux and flux quantum. In the discussion below it will be shown
that, the expectation values of the charge and current densities depend only
on the fractional part of $\alpha $. This is a characteristic feature of the
Aharonov-Bohm type effects.\

The radial functions in the I- and E-regions are given by the expressions%
\begin{equation}
Z_{\nu }^{(\pm )}(\gamma r)=\left\{
\begin{array}{cc}
J_{\nu }(\gamma r), & r<a \\
g_{\beta _{j},\nu }^{(\pm )}(\gamma a,\gamma r), & r>a%
\end{array}%
\right. ,  \label{Znu}
\end{equation}%
where we have defined the function%
\begin{equation}
g_{\beta _{j},\nu }^{(\pm )}(\gamma a,\gamma r)=\bar{Y}_{\beta _{j}}^{(\pm
)}(\gamma a)J_{\nu }(\gamma r)-\bar{J}_{\beta _{j}}^{(\pm )}(\gamma a)Y_{\nu
}(\gamma r).  \label{gbet}
\end{equation}%
Here $J_{\nu }(x)$ and $Y_{\nu }(x)$ are the Bessel and Neumann functions
and the notation with the bar is defined as
\begin{align}
\bar{F}_{\beta _{j}}^{(\pm )}(z)& =zF_{\beta _{j}}^{\prime }(z)-\left[
\epsilon _{j}\beta _{j}-\delta ^{\mathrm{(J)}}(sm_{a}\pm \sqrt{%
z^{2}+m_{a}^{2}})\right] F_{\beta _{j}}(z)  \notag \\
& =-\epsilon _{j}zF_{\beta _{j}+\epsilon _{j}}(z)+\delta ^{\mathrm{(J)}%
}(sm_{a}\pm \sqrt{z^{2}+m_{a}^{2}})F_{\beta _{j}}(z),\;F=J,Y,  \label{Fbar}
\end{align}%
and $m_{a}=ma$. In the E-region $\mathrm{J}=\mathrm{E}$ and the notation for
the I-region with $\mathrm{J}=\mathrm{I}$ is used below. Note that one has%
\begin{equation}
g_{\beta _{j},\beta _{j}}^{(\pm )}(\gamma a,\gamma a)=\frac{2}{\pi }%
,\;g_{\beta _{j},\beta _{j}+\epsilon _{j}}^{(\pm )}(\gamma a,\gamma
a)=\delta ^{\mathrm{(J)}}\frac{2\epsilon _{j}}{\pi \gamma }\left( sm\pm
\sqrt{\gamma ^{2}+m^{2}}\right) .  \label{gaa}
\end{equation}

In the E-region the mode functions (\ref{psiIE}) with $Z_{\nu }^{(\pm
)}(\gamma r)=g_{\beta _{j},\nu }^{(\pm )}(\gamma a,\gamma r)$ obey the
boundary condition (\ref{BCMIT}) and the spectrum of the quantum number $%
\gamma $ is continuous, $0\leq \gamma <\infty $. For the I-region the
eigenvalues of $\gamma $ are quantized by the boundary condition. They are
solutions of the equation%
\begin{equation}
\bar{J}_{\beta _{j}}^{(\pm )}(\gamma a)=0.  \label{modesi}
\end{equation}%
We denote the positive roots of this equation by $\gamma a=\gamma
_{j,l}^{(\pm )}$, $l=1,2,\ldots $. Hence, in the I-region the modes are
specified by the set $\sigma =(l,j)$.

The mode functions (\ref{psiIE}) are normalized by the condition
\begin{equation}
\int_{0}^{\phi _{0}}d\phi \int dr\,r\,\bar{\psi}_{\sigma }^{(\pm )}\gamma
^{0}\psi _{\sigma ^{\prime }}^{(\pm )}=\delta _{\sigma \sigma ^{\prime }},
\label{NC}
\end{equation}%
where $\bar{\psi}=\psi ^{\dagger }\gamma ^{0}$ is the Dirac adjoint and the
radial integration goes over $[0,a]$ and $[a,\infty )$ for the I- and
E-regions. One has $\delta _{\sigma \sigma ^{\prime }}=\delta _{ll^{\prime
}}\delta _{jj^{\prime }}$ and $\delta _{\sigma \sigma ^{\prime }}=\delta
(\gamma -\gamma ^{\prime })\delta _{jj^{\prime }}$ in those regions,
respectively. For the normalization coefficient in the I-region one finds
\begin{equation}
|c^{(\pm )}|^{2}\equiv |c_{\mathrm{i}}^{(\pm )}|^{2}=\frac{\gamma }{2\phi
_{0}a}\frac{E\pm sm}{E}T_{\beta _{j}}^{(\pm )}(\gamma a),  \label{ci}
\end{equation}%
where we have defined
\begin{equation}
T_{\beta _{j}}^{(\pm )}(z)=\frac{zJ_{\beta _{j}}^{-2}(z)}{a\left( E\pm
sm\right) \left[ aE\mp q\left( j+\alpha _{0}\right) +\frac{sm}{2E}\right] },
\label{Tpm}
\end{equation}%
with $E=\sqrt{z^{2}/a^{2}+m^{2}}$ and $z=\gamma _{j,l}^{(\pm )}$. For the
E-region we get%
\begin{equation}
|c^{(\pm )}|^{2}\equiv |c_{\mathrm{e}}^{(\pm )}|^{2}=\frac{\gamma }{2\phi
_{0}E}\frac{E\pm sm}{\bar{J}_{\beta _{j}}^{(\pm )2}(\gamma a)+\bar{Y}_{\beta
_{j}}^{(\pm )2}(\gamma a)}.  \label{ce}
\end{equation}%
Note that the mode functions in the boundary-free conical space with $0\leq
r<\infty $ are given by (\ref{psiIE}) with $Z_{\nu }^{(\pm )}(\gamma
r)=J_{\nu }(\gamma r)$. The corresponding normalization coefficient is found
from (\ref{NC}) with the radial integration over $[0,\infty )$. This gives%
\begin{equation}
|c^{(\pm )}|^{2}\equiv |c_{0}^{(\pm )}|^{2}=\gamma \frac{E\pm sm}{2\phi _{0}E%
}.  \label{c0}
\end{equation}

We turn to the evaluation of the expectation values of the charge and
current densities for the field $\psi (x)$ in thermal equilibrium at
temperature $T$. They are expressed in terms of the density matrix $\hat{\rho%
}=e^{-\beta (\hat{H}-\mu ^{\prime }\hat{Q})}/Z$, with $\beta =1/T$, by the
formula
\begin{equation}
\left\langle j^{\nu }\right\rangle =e\,\mathrm{tr}\left[ \hat{\rho}\bar{\psi}%
\gamma ^{\nu }\psi \right] ,\;\nu =0,1,2,  \label{jnu}
\end{equation}%
where the angular brackets stand for the ensemble average. Here, $\hat{H}%
=\sum_{\sigma }E_{\sigma }(a_{\sigma }^{\dagger }a_{\sigma }-b_{\sigma
}b_{\sigma }^{\dagger })$ is the Hamilton operator, $\hat{Q}=e\sum_{\sigma
}(a_{\sigma }^{\dagger }a_{\sigma }+b_{\sigma }b_{\sigma }^{\dagger })$ is
the charge operator and $Z=\mathrm{tr\,}[e^{-\beta (\hat{H}-\mu ^{\prime }%
\hat{Q})}]$ is the partition function with the related chemical potential $%
\mu ^{\prime }$.

Plugging in (\ref{jnu}) the expansion (\ref{mexp}) of the operator $\psi (x)$
and the corresponding expansion for the Dirac adjoint, the mean current
density is presented in the form
\begin{equation}
\langle j^{\nu }\rangle =\langle j^{\nu }\rangle _{\mathrm{vac}}+\langle
j^{\nu }\rangle _{T+}+\langle j^{\nu }\rangle _{T-},  \label{jnudec}
\end{equation}%
where $\langle j^{\nu }\rangle _{\mathrm{vac}}$ is the vacuum expectation
value (VEV), $\langle j^{\nu }\rangle _{T+}$ and $\langle j^{\nu }\rangle
_{T-}$ are the contributions coming from particles and antiparticles,
respectively. Introducing the notation $\mu =e\mu ^{\prime }$, the separate
contributions are given by the following expansions over the mode functions:%
\begin{align}
\langle j^{\nu }\rangle _{\mathrm{vac}}& =-\frac{e}{2}\sum_{\sigma
}\sum_{\lambda =\pm }\lambda \bar{\psi}_{\sigma }^{(\lambda )}(x)\gamma
^{\nu }\psi _{\sigma }^{(\lambda )}(x),  \label{jnuvac} \\
\langle j^{\nu }\rangle _{T\pm }& =\pm e\sum_{\sigma }\frac{\bar{\psi}%
_{\sigma }^{(\pm )}(x)\gamma ^{\nu }\psi _{\sigma }^{(\pm )}(x)}{e^{\beta
(E_{\sigma }\mp \mu )}+1}.  \label{jnupm}
\end{align}%
Here, $\sum_{\sigma }=\sum_{j}\sum_{l=1}^{\infty }$ in the I-region and $%
\sum_{\sigma }=\sum_{j}\int_{0}^{\infty }d\gamma $ for the E-region with $%
\sum_{j}=\sum_{j=\pm 1/2,\pm 3/2,\cdots }$. We emphasize again that the part
(\ref{jnuvac}) corresponds to the VEV in the geometry with the boundary at $%
r=a$. It has been investigated in \cite{Beze10} (see also \cite{Bell20} for
a conical ring with two circular boundaries). The thermal contributions (\ref%
{jnupm}) generally vanishes in the limit $T\rightarrow 0$ except when $|\mu
|>m$ in which case $\langle j^{\nu }\rangle _{T+}$ (for $\mu >0$) or $%
\langle j^{\nu }\rangle _{T-}$ (for $\mu <0$) do not vanish at $T=0$ as will
be discussed later. If we present the parameter $\alpha $ in the form
\begin{equation}
\alpha =\alpha _{0}+n_{0},\;|\alpha _{0}|<1/2,  \label{alpha}
\end{equation}%
with $n_{0}$ being an integer, then, redefining $j+n_{0}\rightarrow j$ in
the series over $j$, we see that the expectation values (\ref{jnuvac}) and (%
\ref{jnupm}) do not depend on the integer part. In terms of the magnetic
flux this means that only the fractional part of the ratio of the magnetic
flux and flux quantum is physically relevant.

Hence, in the discussion below, without loss of generality, we can take $%
\alpha _{j}=q|j+\alpha _{0}|$. It can be checked that under the replacements
$\alpha _{0}\rightarrow -\alpha _{0}$ and $j\rightarrow -j$ we have $\beta
_{j}\rightleftarrows \beta _{j}+\epsilon _{j}$ and%
\begin{equation}
\bar{F}_{\beta _{j}}^{(\pm )}(\gamma a)\rightarrow b(\gamma )\bar{F}_{\beta
_{j}}^{(\mp )}(z),\;g_{\beta _{j},\beta _{j}}^{(\pm )}(\gamma a,\gamma
r)\rightarrow b(\gamma )g_{\beta _{j},\beta _{j}+\epsilon _{j}}^{(\mp
)}(\gamma a,\gamma r),  \label{Replalf}
\end{equation}%
with $b(\gamma )=-\delta ^{\mathrm{(J)}}\epsilon _{j}(sm\pm E)/\gamma $.
From the second relation in (\ref{Replalf}) it follows that $g_{\beta
_{j},\beta _{j}+\epsilon _{j}}^{(\pm )}(\gamma a,\gamma r)\rightarrow
b(\gamma )g_{\beta _{j},\beta _{j}}^{(\mp )}(\gamma a,\gamma r)$. For the
radial modes in the I-region one has $\gamma a=\gamma _{j,l}^{(\pm )}(\alpha
_{0})$. Now, by using the first relation in (\ref{Replalf}), we see that $%
\gamma _{-j,l}^{(-)}(-\alpha _{0})=\gamma _{j,l}^{(+)}(\alpha _{0})$. This
gives the relation between the eigenmodes for particles and antiparticles.

\section{Vacuum expectation values and thermal contributions in a conical
space without boundary}

\label{sec:BF}

For further convenience, in this section we summarize the results for the
VEVs in a conical space with a circular boundary and for the finite
temperature charge and current densities in a conical space with no
boundaries. The total current density in the problem under consideration is
presented in the form%
\begin{equation}
\langle j^{\nu }\rangle =\langle j^{\nu }\rangle ^{(0)}+\langle j^{\nu
}\rangle ^{\mathrm{(b)}},  \label{jnudec2}
\end{equation}%
where $\langle j^{\nu }\rangle ^{(0)}$ is the term independent of the
boundary at $r=a$. This term is decomposed into the vacuum and thermal
contributions as%
\begin{equation}
\langle j^{\nu }\rangle ^{(0)}=\langle j^{\nu }\rangle _{\mathrm{vac}%
}^{(0)}+\langle j^{\nu }\rangle _{T}^{(0)}.  \label{jnu0dec}
\end{equation}%
The terms with superscript (b) depend explicitly on the boundary at $r=a$
and are decomposed as%
\begin{equation}
\langle j^{\nu }\rangle ^{\mathrm{(b)}}=\langle j^{\nu }\rangle _{\mathrm{vac%
}}^{\mathrm{(b)}}+\langle j^{\nu }\rangle _{T}^{\mathrm{(b)}}.  \label{jnube}
\end{equation}%
This Section computes $\langle j^{\nu }\rangle _{\mathrm{vac}}^{(0)}$, $%
\langle j^{\nu }\rangle _{\mathrm{vac}}^{\mathrm{(b)}}$, and $\langle j^{\nu
}\rangle _{T}^{(0)}$ in that order. Section \ref{sec:Exter} computes $%
\langle j^{\nu }\rangle _{T}^{\mathrm{(b)}}$ in the region $r>a$. The part $%
\langle j^{\nu }\rangle _{T}^{\mathrm{(b)}}$ in the region $r<a$ is
evaluated in Section \ref{sec:Inter}. The radial current density always
vanishes: $\langle j^{1}\rangle =0$ and the remainder of the text will only
concern $\nu =0,2$.

The boundary-independent part of the VEV for $\nu =0,2$ is given by
\begin{align}
\left\langle j^{\nu }\right\rangle _{\mathrm{vac}}^{\left( 0\right) }=& -%
\frac{e}{2\pi r}\Big\{\sideset{}{'}{\sum}_{l=1}^{[q/2]}(-1)^{l}\sin (2\pi
l\alpha _{0})f_{\nu }\left( 2mr\sin (\pi l/q)\right)  \notag \\
& -\frac{q}{\pi }\int_{0}^{\infty }dy\frac{f_{\nu }\left( 2mr\cosh y\right)
f(q,\alpha _{0},y)}{\cosh (2qy)-\cos (q\pi )}\Big\},  \label{jm02}
\end{align}%
where the functions for the charge and azimuthal current densities are
defined as%
\begin{align}
f_{0}(z)& =sme^{-z},\;f_{2}(z)=2m^{2}e^{-z}\frac{1+z}{z^{2}},  \notag \\
f(q,\alpha _{0},y)& =\sum_{\chi =\pm 1}\chi \cos \left[ q\pi \left( 1/2-\chi
\alpha _{0}\right) \right] \cosh \left[ q\left( 1+2\chi \alpha _{0}\right) y%
\right] .  \label{f02}
\end{align}%
The square brackets in the upper limit of the summation in (\ref{jm02}) mean
the integer part of the enclosed expression and the prime means that for
even $q$ the term with $l=q/2$ should be taken with a coefficient 1/2. Note
that in \cite{Beze10} the negative energy modes have been used with $\alpha $
replaced by $-\alpha $ and as a consequence of that the VEV (\ref{jm02})
differ from the corresponding formula in \cite{Beze10} by the sign (see also
\cite{Bell20}). For the magnetic flux equal to an integer number of magnetic
flux quanta we have $\alpha _{0}=0$ and both the charge and current
densities in (\ref{jm02}) vanish. The VEV $\left\langle j^{\nu
}\right\rangle _{\mathrm{vac}}^{\left( 0\right) }$ is an odd periodic
function of the magnetic flux $\Phi $ with the period of flux quantum. From (%
\ref{jm02}) it follows that \cite{Beze10}%
\begin{align}
\lim_{\alpha _{0}\rightarrow \pm 1/2}\langle j^{0}\rangle _{\mathrm{vac}%
}^{\left( 0\right) }& =\pm \frac{eqm}{2\pi ^{2}r}K_{0}(2mr),\;  \notag \\
\lim_{\alpha _{0}\rightarrow \pm 1/2}\langle j^{2}\rangle _{\mathrm{vac}%
}^{\left( 0\right) }& =\pm \frac{eqm}{2\pi ^{2}r^{2}}K_{1}(2mr).
\label{j020alf}
\end{align}%
This shows that the VEVs $\left\langle j^{\nu }\right\rangle _{\mathrm{vac}%
}^{\left( 0\right) }$, considered as functions of $\alpha $ are
discontinuous at the points $\alpha =n_{0}+1/2$ with $n_{0}$ being an
integer. For a massless field the charge density $\left\langle
j^{0}\right\rangle _{\mathrm{vac}}^{\left( 0\right) }$ vanishes and the
expression for the current density $\left\langle j^{2}\right\rangle _{%
\mathrm{vac}}^{\left( 0\right) }$ is obtained from (\ref{jm02}) by the
replacement $f_{2}\left( 2mrb\right) \rightarrow 1/(2r^{2}b^{2})$.

For a conical space with a circular edge the boundary-induced contributions
in the VEVs are expressed as \cite{Bell20,Beze10}
\begin{align}
\left\langle j^{\nu }\right\rangle _{\mathrm{vac}}^{\mathrm{(b)}} &=\frac{e}{%
\pi \phi _{0}}\sum_{j}\int_{m}^{\infty }dx\,x\,\mathrm{Re}\left[ \frac{\bar{I%
}_{\beta _{j}}(ax)}{\bar{K}_{\beta _{j}}(ax)}\frac{U_{\nu ,\beta
_{j}}^{K}(rx)}{\sqrt{x^{2}-m^{2}}}\right] ,\;r>a,  \notag \\
\left\langle j^{\nu }\right\rangle _{\mathrm{vac}}^{\mathrm{(b)}} &=\frac{e}{%
\pi \phi _{0}}\sum_{j}\int_{m}^{\infty }dx\,x\,\mathrm{Re}\left[ \frac{\bar{K%
}_{\beta _{j}}(ax)}{\bar{I}_{\beta _{j}}(ax)}\frac{U_{\nu ,\beta
_{j}}^{I}(rx)}{\sqrt{x^{2}-m^{2}}}\right] ,\;r<a,  \label{jbvac}
\end{align}%
with $\nu =0$ and $\nu =2$ for the charge and azimuthal current densities%
\footnote{%
As it has been mentioned in \cite{Bell20}, comparing the expressions for the
charge and current densities given here and in \cite{Bell20} with the
corresponding expressions in \cite{Beze10}, the parameters $\alpha $ and $%
\alpha _{0}$ should be replaced by $-\alpha $ and $-\alpha _{0}$,
respectively. The change of the signs is related to the fact that the
negative energy mode functions used in \cite{Beze10} differ from those here
by the sign of $\alpha $.}. Here, for the modified Bessel functions $%
I_{\beta _{j}}(z)$ and $K_{\beta _{j}}(z)$ we use the notation
\begin{align}
\bar{F}_{\beta _{j}}(u)& =uF_{\beta _{j}}^{\prime }(u)-\left[ \epsilon
_{j}\beta _{j}-\delta ^{\mathrm{(J)}}\left( sm_{a}+i\sqrt{u^{2}-m_{a}^{2}}%
\right) \right] F_{\beta _{j}}(u)  \notag \\
& =\delta _{F}uF_{\beta _{j}+\epsilon _{j}}(u)+\delta ^{\mathrm{(J)}}\left(
sm_{a}+i\sqrt{u^{2}-m_{a}^{2}}\right) F_{\beta _{j}}(u),  \label{FbarIK}
\end{align}%
where $F=I,K$, $\delta _{I}=-\delta _{K}=1$ and $\delta ^{\mathrm{(J)}}$ is
defined by (\ref{delJ}). In (\ref{jbvac}) we have defined
\begin{align}
U_{0,\beta _{j}}^{F}(rx) &=\sum_{\chi =\pm 1}\left( sm+\chi i\sqrt{%
x^{2}-m^{2}}\right) F_{\alpha _{j}-\chi \epsilon _{j}/2}^{2}(rx),  \notag \\
U_{2,\beta _{j}}^{F}(rx) &=-\delta _{F}\frac{2x}{r}F_{\beta
_{j}}(rx)F_{\beta _{j}+\epsilon _{j}}(rx),  \label{U2}
\end{align}%
for the modified Bessel functions $F=I$ and $F=K$. It can be seen that under
the replacement $\alpha _{0}\rightarrow -\alpha _{0}$ and $j\rightarrow -j$
we have
\begin{equation}
\frac{\bar{I}_{\beta _{j}}(ax)}{\bar{K}_{\beta _{j}}(ax)}\rightarrow -\left[
\frac{\bar{I}_{\beta _{j}}(ax)}{\bar{K}_{\beta _{j}}(ax)}\right] ^{\ast
},\;U_{2,\beta _{j}}^{F}(rx)\rightarrow U_{2,\beta _{j}}^{F\ast }(rx),
\label{Replbar}
\end{equation}%
where the star stands for the complex conjugate. From these relations it
follows that%
\begin{equation}
\left\langle j^{\nu }\right\rangle _{\mathrm{vac}}^{\mathrm{(b)}}(-\alpha
_{0})=-\left\langle j^{\nu }\right\rangle _{\mathrm{vac}}^{\mathrm{(b)}%
}(\alpha _{0}).  \label{jbvacodd}
\end{equation}%
In particular, for a magnetic flux taking integer multiple values of flux
quantum one has $\alpha _{0}=0$ and the boundary-induced VEVs vanish, $%
\left\langle j^{\nu }\right\rangle _{\mathrm{vac}}^{\mathrm{(b)}}=0$. Note
that the VEV (\ref{jm02}) in the geometry without boundary is also an odd
function of $\alpha _{0}$.

In a conical space without boundaries the current density is decomposed as (%
\ref{jnu0dec}) where the vacuum part is given by (\ref{jm02}) and for the
thermal part one has%
\begin{align}
\langle j^{0}\rangle _{T}^{(0)}& =\frac{e}{2\phi _{0}}\int_{0}^{\infty
}d\gamma \gamma \sum_{\chi =\pm 1}\frac{\sinh (\beta \mu )+\chi sm\cosh
(\beta \mu )/E}{\cosh (\beta E)+\cosh (\beta \mu )}\sum_{j}J_{\alpha
_{j}-\chi \epsilon _{j}/2}^{2}(\gamma r),  \notag \\
\langle j^{2}\rangle _{T}^{(0)}& =\frac{e}{r\phi _{0}}\cosh (\beta \mu
)\int_{0}^{\infty }d\gamma \,\frac{\gamma ^{2}}{E}\frac{\sum_{j}\epsilon
_{j}J_{\beta _{j}}(\gamma r)J_{\beta _{j}+\epsilon _{j}}(\gamma r)}{\cosh
(\beta E)+\cosh (\beta \mu )}.  \label{j02T}
\end{align}%
These formulas are obtained by the transformations of the corresponding
expressions from \cite{Bell16T}. Alternative representations for the thermal
parts $\langle j^{\nu }\rangle _{T}^{(0)}$ having a structure similar to (%
\ref{jm02}) can be found in \cite{Bell16T}. Note that the physical azimuthal
component of the current density is expressed in terms of the contravariant
component $\left\langle j^{2}\right\rangle $ by the relation
\begin{equation}
\left\langle j_{\phi }\right\rangle =r\left\langle j^{2}\right\rangle .
\label{jphys}
\end{equation}
In all formulas below the index $\nu =0,2$ of the current density $j^{\nu }$
stands for the contravariant components. From (\ref{j02T}) we can see that
\begin{equation}
\langle j^{\nu }\rangle _{T}^{(0)}(-\alpha _{0},-\mu )=-\langle j^{\nu
}\rangle _{T}^{(0)}(\alpha _{0},\mu ).  \label{j0odd}
\end{equation}%
Now, by taking into account that the azimuthal component is an even function
of the chemical potential, we conclude that for the magnetic flux equal to
an integer multiple of flux quantum ($\alpha _{0}=0$) one gets $\langle
j^{2}\rangle _{T}^{(0)}=0$. The expression for the charge density in this
case is simplified to%
\begin{equation}
\langle j^{0}\rangle _{T}^{(0)}=\frac{e}{\phi _{0}}\sinh (\beta \mu
)\sum_{j>0}\int_{0}^{\infty }d\gamma \gamma \frac{J_{qj-1/2}^{2}(\gamma
r)+J_{qj+1/2}^{2}(\gamma r)}{\cosh (\beta E)+\cosh (\beta \mu )},
\label{j00T}
\end{equation}%
and it vanishes for a zero chemical potential. In the special case $\phi
_{0}=2\pi $, the formulas given in this section describe the combined
effects of the finite temperature, magnetic flux and concentric circular
edge on the mean charge and current densities in (2+1)-dimensional Minkowski
spacetime. In the simplest case of the absence of the magnetic flux and
boundary the current density vanishes and the distribution of the charge is
uniform with
\begin{equation}
\langle j^{0}\rangle ^{(0)}=\langle j^{0}\rangle _{T}^{(0)}=\frac{e}{2\pi }%
\int_{m}^{\infty }dx\frac{\sinh (\beta \mu )\,x}{\cosh (\beta x)+\cosh
(\beta \mu )}.  \label{j00T0}
\end{equation}%
In the zero temperature limit this gives%
\begin{equation}
\lim_{T\rightarrow 0}\langle j^{0}\rangle ^{(0)}=\mathrm{sgn\,}(\mu )e\frac{%
\mu ^{2}-m^{2}}{4\pi }\theta \left( |\mu |-m\right) ,  \label{j00T00}
\end{equation}%
with $\theta (x)$ being the Heaviside step function.

In the discussion below we are interested in the finite temperature parts $%
\langle j^{\nu }\rangle _{T\pm }$ for the geometry of a conical space with a
circular edge. Those parts in the E- and I-regions will be considered
separately.

\section{Charge and current densities in the E-region}

\label{sec:Exter}

This section will compute $\langle j^{\nu }\rangle _{T}^{\mathrm{(b)}}$ in
the region $r>a$. The corresponding mode functions are given by (\ref{psiIE}%
) with $Z_{\nu }^{(\pm )}(\gamma r)=g_{\beta _{j},\nu }^{(\pm )}(\gamma
a,\gamma r)$. With those modes the thermal contributions to the charge
density and azimuthal current density are given by
\begin{align}
\left\langle j^{0}\right\rangle _{T\pm }& =\frac{\pm e}{2\phi _{0}}%
\sum_{j}\int_{0}^{\infty }d\gamma \,\frac{\gamma }{E}\frac{\sum_{\chi =\pm
1}\left( E\pm \chi sm\right) g_{\beta _{j},\alpha _{j}-\chi \epsilon
_{j}/2}^{\left( \pm \right) 2}\left( \gamma a,\gamma r\right) }{\left[
e^{\beta \left( E\mp \mu \right) }+1\right] \left[ \bar{J}_{\beta
_{j}}^{\left( \pm \right) 2}\left( \gamma a\right) +\bar{Y}_{\beta
_{j}}^{\left( \pm \right) 2}\left( \gamma a\right) \right] },  \notag \\
\left\langle j^{2}\right\rangle _{T\pm }& =\frac{e}{\phi _{0}r}%
\sum_{j}\int_{0}^{\infty }d\gamma \,\frac{\epsilon _{j}\gamma ^{2}/E}{%
e^{\beta \left( E\mp \mu \right) }+1}\frac{g_{\beta _{j},\beta _{j}}^{\left(
\pm \right) }\left( \gamma a,\gamma r\right) g_{\beta _{j},\beta
_{j}+\epsilon _{j}}^{\left( \pm \right) }\left( \gamma a,\gamma r\right) }{%
\bar{J}_{\beta _{j}}^{\left( \pm \right) 2}\left( \gamma a\right) +\bar{Y}%
_{\beta _{j}}^{\left( \pm \right) 2}\left( \gamma a\right) }.  \label{jnupme}
\end{align}%
The radial component of the current density vanishes. As expected, the ratio
$\left\langle j^{0}\right\rangle _{T\pm }/e$ is positive for particles and
negative for antiparticles. Considering the expectation values as functions
of the parameters $\alpha _{0}$ and $\mu $, $\left\langle j^{\nu
}\right\rangle _{T\pm }=\left\langle j^{\nu }\right\rangle _{T\pm }(\alpha
_{0},\mu )$, and by using the transformations (\ref{Replalf}) under the
replacements $\alpha _{0}\rightarrow -\alpha _{0}$ and $j\rightarrow -j$,
the following relation is obtained between the current densities of
particles and antiparticles:%
\begin{equation}
\left\langle j^{\nu }\right\rangle _{T\pm }\left( -\alpha _{0},-\mu \right)
=-\left\langle j^{\nu }\right\rangle _{T\mp }\left( \alpha _{0},\mu \right) .
\label{jnualf}
\end{equation}%
From this relation, for the thermal part of the total current density we get%
\begin{equation}
\left\langle j^{\nu }\right\rangle _{T}\left( -\alpha _{0},-\mu \right)
=-\left\langle j^{\nu }\right\rangle _{T}\left( \alpha _{0},\mu \right) .
\label{jnuTodd}
\end{equation}%
As it has been discussed in the previous section, the same relation takes
place for the VEVs. Another simple relation between the physical components
on the boundary $r=a$ follows from Eq. (\ref{gaa}):%
\begin{equation}
\left\langle j^{0}\right\rangle _{T\pm }=-\left\langle j_{\phi
}\right\rangle _{T\pm }=\frac{4e}{\pi ^{2}\phi _{0}}\sum_{j}\int_{0}^{\infty
}d\gamma \,\frac{\gamma \left( sm/E\pm 1\right) }{e^{\beta \left( E\mp \mu
\right) }+1}\frac{1}{\bar{J}_{\beta _{j}}^{\left( \pm \right) 2}\left(
\gamma a\right) +\bar{Y}_{\beta _{j}}^{\left( \pm \right) 2}\left( \gamma
a\right) }.  \label{jon}
\end{equation}%
This relation for the charge and current densities on the edge is a
consequence of the boundary condition (\ref{BCMIT}) and has been mentioned
for the VEVs in other geometries in \cite{Bell20,Bene12,Bell16}.

The finite temperature charge and current densities in the boundary-free
geometry were studied in \cite{Bell16T} and here we are interested in the
boundary-induced effects. In order to extract those contributions we
subtract from the expectation values (\ref{jnupme}) the corresponding
quantities in the boundary-free conical space, denoted here by $\langle
j^{\nu }\rangle _{T\pm }^{(0)}$. The expressions for the latter are obtained
from (\ref{jnupme}) by the replacement%
\begin{equation}
\frac{g_{\beta _{j},\rho }^{\left( \pm \right) }\left( \gamma a,\gamma
r\right) }{\sqrt{\bar{J}_{\beta _{j}}^{\left( \pm \right) 2}\left( \gamma
a\right) +\bar{Y}_{\beta _{j}}^{\left( \pm \right) 2}\left( \gamma a\right) }%
}\rightarrow J_{\rho }(\gamma r),  \label{Repl}
\end{equation}%
with $\rho =\beta _{j},\beta _{j}+\epsilon _{j}$.

For the further transformation of the boundary-induced part $\langle j^{\nu
}\rangle _{T\pm }^{\mathrm{(b)}}=\langle j^{\nu }\rangle _{T\pm }-\langle
j^{\nu }\rangle _{T\pm }^{(0)}$, we use the identity
\begin{equation}
\frac{g_{\beta _{j},\mu }^{\left( \pm \right) }\left( x,y\right) g_{\beta
_{j},\rho }^{\left( \pm \right) }\left( x,y\right) }{\bar{J}_{\beta
_{j}}^{\left( \pm \right) 2}\left( x\right) +\bar{Y}_{\beta _{j}}^{\left(
\pm \right) 2}\left( x\right) }-J_{\mu }\left( y\right) J_{\rho }\left(
y\right) =-\frac{1}{2}\sum_{l=1,2}\frac{\bar{J}_{\beta _{j}}^{\left( \pm
\right) }\left( x\right) }{\bar{H}_{\beta _{j}}^{\left( \pm ,l\right)
}\left( x\right) }H_{\mu }^{\left( l\right) }\left( y\right) H_{\rho
}^{\left( l\right) }\left( y\right) ,  \label{ident1}
\end{equation}%
where $H_{\mu }^{(l)}(x)$ are the Hankel functions and $\mu ,\rho =\beta
_{j},\beta _{j}+\epsilon _{j}$. Here, the notation with the bar for the
Hankel functions is defined in accordance with (\ref{Fbar}). The
boundary-induced contributions are presented in the form
\begin{align}
\langle j^{0}\rangle _{T\lambda }^{\mathrm{(b)}}& =-\frac{\lambda e}{4\phi
_{0}}\sum_{j}\sum_{l=1,2}\int_{0}^{\infty }d\gamma \,\frac{\gamma \bar{J}%
_{\beta _{j}}^{\left( \lambda \right) }\left( \gamma a\right) }{E\bar{H}%
_{\beta _{j}}^{\left( l,\lambda \right) }\left( \gamma a\right) }\frac{%
\sum_{\chi =\pm 1}\left( E+\lambda \chi sm\right) H_{\alpha _{j}-\chi
\epsilon _{j}/2}^{\left( l\right) 2}\left( \gamma r\right) }{e^{\beta \left(
E-\lambda \mu \right) }+1},  \notag \\
\langle j^{2}\rangle _{T\lambda }^{\mathrm{(b)}}& =-\frac{e}{2\phi _{0}r}%
\sum_{j}\sum_{l=1,2}\int_{0}^{\infty }d\gamma \,\frac{\epsilon _{j}\gamma
^{2}\bar{J}_{\beta _{j}}^{\left( \lambda \right) }\left( \gamma a\right) }{E%
\bar{H}_{\beta _{j}}^{\left( l,\lambda \right) }\left( \gamma a\right) }%
\frac{H_{\beta _{j}}^{\left( l\right) }\left( \gamma r\right) H_{\beta
_{j}+\epsilon _{j}}^{\left( l\right) }\left( \gamma r\right) }{e^{\beta
\left( E-\lambda \mu \right) }+1},  \label{jnupmb1}
\end{align}%
with $\lambda =\pm $. The cases of $\mu \neq 0$ and $\mu =0$ will be
discussed separately.

For the case of nonzero chemical potential, after transformations presented
in Appendix \ref{sec:appA}, for the expectation values of the charge and
current densities coming from particles and antiparticles we get
\begin{align}
\left\langle j^{0}\right\rangle _{T\lambda }^{\mathrm{(b)}}& =\frac{e}{\pi
\phi _{0}}\sum_{j}\left\{ -\int_{m}^{\infty }dx\,x\,\mathrm{Re}\left[ \frac{%
\bar{I}_{\beta _{j}}\left( xa\right) }{\bar{K}_{\beta _{j}}\left( xa\right) }%
\frac{U_{0,\beta _{j}}^{K}(xr)}{\sqrt{x^{2}-m^{2}}}\frac{1}{e^{\lambda \beta
(i\sqrt{x^{2}-m^{2}}-\mu )}+1}\right] \right.  \notag \\
& \left. +2\pi T\,\theta \left( \lambda \mu \right) \sum_{n=0}^{\infty }%
\mathrm{Re}\left[ \frac{\bar{I}_{\beta _{j}}\left( u_{n}a\right) }{\bar{K}%
_{\beta _{j}}\left( u_{n}a\right) }\sum_{\chi =\pm 1}\left[ sm+\chi \mu
+\chi i\pi \left( 2n+1\right) T\right] K_{\alpha _{j}-\chi \epsilon
_{j}/2}^{2}\left( u_{n}r\right) \right] \right\} ,  \notag \\
\left\langle j^{2}\right\rangle _{T\lambda }^{\mathrm{(b)}}& =\frac{e}{\pi
\phi _{0}}\sum_{j}\left\{ -\int_{m}^{\infty }dx\,\frac{xU_{2,\beta
_{j}}^{K}(xr)}{\sqrt{x^{2}-m^{2}}}\mathrm{Re}\left[ \frac{\bar{I}_{\beta
_{j}}\left( xa\right) }{\bar{K}_{\beta _{j}}\left( xa\right) }\frac{1}{%
e^{\lambda \beta (i\sqrt{x^{2}-m^{2}}-\mu )}+1}\right] \right.  \notag \\
& \left. +2\pi T\,\theta \left( \lambda \mu \right) \sum_{n=0}^{\infty }%
\mathrm{Re}\left[ \frac{\bar{I}_{\beta _{j}}\left( u_{n}a\right) }{\bar{K}%
_{\beta _{j}}\left( u_{n}a\right) }U_{2,\beta _{j}}^{K}(u_{n}r)\right]
\right\} ,  \label{jnub3}
\end{align}%
where
\begin{equation}
u_{n}=\left\{ \left[ \pi \left( 2n+1\right) T-i\mu \right]
^{2}+m^{2}\right\} ^{1/2}.  \label{un}
\end{equation}%
We have also used the notation $\bar{F}_{\beta _{j}}(z)=\bar{F}_{\beta
_{j}}^{(+)}(z)$ for the modified Bessel functions in agreement with (\ref%
{FbarIK}) where $\delta ^{\mathrm{(J)}}=\delta ^{\mathrm{(E)}}=-1$.

By combining the contributions from the separate terms with $\lambda =+$ and
$\lambda =-$ and using the identity $\sum_{\lambda }1/(e^{\lambda z}+1)=1$,
we obtain the boundary-induced finite temperature contributions $\langle
j^{\nu }\rangle _{T}^{\mathrm{(b)}}=\sum_{\lambda =\pm }\langle j^{\nu
}\rangle _{T\lambda }^{\mathrm{(b)}}$. Afterwards, the total
boundary-induced contributions, given by the formula (\ref{jnube}) are
presented in the form
\begin{align}
\left\langle j^{0}\right\rangle ^{\mathrm{(b)}}& =\frac{2Te}{\phi _{0}}%
\sum_{j}\sum_{n=0}^{\infty }\mathrm{Re}\left\{ \frac{\bar{I}_{\beta
_{j}}\left( u_{n}a\right) }{\bar{K}_{\beta _{j}}\left( u_{n}a\right) }%
\sum_{\chi =\pm 1}\left[ sm+\chi \mu +\chi i\pi \left( 2n+1\right) T\right]
K_{\alpha _{j}-\chi \epsilon _{j}/2}^{2}\left( u_{n}r\right) \right\} ,
\notag \\
\left\langle j^{2}\right\rangle ^{\mathrm{(b)}}& =\frac{2Te}{\phi _{0}}%
\sum_{j}\sum_{n=0}^{\infty }\mathrm{Re}\left[ \frac{\bar{I}_{\beta
_{j}}\left( u_{n}a\right) }{\bar{K}_{\beta _{j}}\left( u_{n}a\right) }%
U_{2,\beta _{j}}^{K}(u_{n}r)\right] ,  \label{jnub4}
\end{align}%
where the function $U_{2,\beta _{j}}^{K}(u_{n}r)$ is given by (\ref{U2})
with $F=K$ and the modified Bessel functions with bar are defined as (\ref%
{FbarIK}) with $\mathrm{J}=\mathrm{E}$, $\delta ^{\mathrm{(E)}}=-1$.
Introducing the functions%
\begin{equation}
W_{j}^{\mathrm{(J)}}(z)=\sum_{\chi =\pm 1}\chi \left[ I_{\alpha _{j}-\chi
\epsilon _{j}/2}(z)+\delta ^{\mathrm{(J)}}\frac{sm_{a}}{z}I_{\alpha
_{j}+\chi \epsilon _{j}/2}(z)\right] K_{\alpha _{j}-\chi \epsilon _{j}/2}(z),
\label{Wj}
\end{equation}%
with $\mathrm{J}=\mathrm{I},\mathrm{E}$ and $\delta ^{\mathrm{(J)}}$ defined
as (\ref{delJ}), the real parts in (\ref{jnub200}) are explicitly separated
by using the relation
\begin{equation}
\frac{\bar{I}_{\beta _{j}}(z)}{\bar{K}_{\beta _{j}}(z)}=\frac{W_{j}^{\mathrm{%
(E)}}(z)+\left[ \mu +i\pi \left( 2n+1\right) T\right] a/z^{2}}{K_{\beta
_{j}+\epsilon _{j}}^{2}(z)+K_{\beta _{j}}^{2}(z)+2sm_{a}K_{\beta
_{j}}(z)K_{\beta _{j}+\epsilon _{j}}(z)/z}.  \label{IKrat}
\end{equation}%
The function $W_{j}^{\mathrm{(I)}}(z)$ will appear in the formulas below for
the I-region. Note that under the replacement $\alpha _{0}\rightarrow
-\alpha _{0}$, $j\rightarrow -j$ we have $W_{j}^{\mathrm{(J)}}(z)\rightarrow
-W_{j}^{\mathrm{(J)}}(z)$.

In the case of zero chemical potential, $\mu =0$, the poles of the integrand
in (\ref{jnupmb1}) are located on the imaginary axis and the procedure for
the transformation is different from that we have described above. This case
is considered in Appendix \ref{sec:appA} and the corresponding result is
obtained from (\ref{jnub4}) by taking the limit $\mu \rightarrow 0$. For $%
\mu =0$ in the arguments of the modified Bessel functions we have $%
u_{n}=u_{0n}$, with
\begin{equation}
u_{0n}=\sqrt{\left[ \pi \left( 2n+1\right) T\right] ^{2}+m^{2}},  \label{u0n}
\end{equation}%
and the arguments are real. By taking into account (\ref{IKrat}) the real
parts are explicitly separated and we get
\begin{align}
\left\langle j^{0}\right\rangle ^{\mathrm{(b)}}& =\frac{2Te}{a\phi _{0}}%
\sum_{j}\sum_{n=0}^{\infty }\sum_{\chi =\pm 1}\frac{\left[ sm_{a}W_{j}^{%
\mathrm{(E)}}(z)-\chi \left( 1-\frac{m_{a}^{2}}{z^{2}}\right) \right]
K_{\alpha _{j}-\chi \epsilon _{j}/2}^{2}\left( zr/a\right) }{K_{\beta
_{j}+\epsilon _{j}}^{2}(z)+K_{\beta _{j}}^{2}(z)+2sm_{a}K_{\beta
_{j}}(z)K_{\beta _{j}+\epsilon _{j}}(z)/z},  \notag \\
\left\langle j^{2}\right\rangle ^{\mathrm{(b)}}& =\frac{2Te}{\phi _{0}}%
\sum_{j}\sum_{n=0}^{\infty }\frac{W_{j}^{\mathrm{(E)}}(z)U_{2,\beta
_{j}}^{K}(zr/a)}{K_{\beta _{j}+\epsilon _{j}}^{2}(z)+K_{\beta
_{j}}^{2}(z)+2sm_{a}K_{\beta _{j}}(z)K_{\beta _{j}+\epsilon _{j}}(z)/z},
\label{jnub0}
\end{align}%
where $z=au_{0n}$. These expressions are further simplified for a massless
field:
\begin{align}
\left\langle j^{0}\right\rangle ^{\mathrm{(b)}}& =\frac{2Te}{a\phi _{0}}%
\sum_{j}\sum_{n=0}^{\infty }\frac{K_{\beta _{j}+\epsilon _{j}}^{2}\left(
zr/a\right) -K_{\beta _{j}}^{2}\left( zr/a\right) }{K_{\beta _{j}+\epsilon
_{j}}^{2}\left( z\right) +K_{\beta _{j}}^{2}\left( z\right) },  \notag \\
\left\langle j^{2}\right\rangle ^{\mathrm{(b)}}& =\frac{2Te}{\phi _{0}}%
\sum_{j}\sum_{n=0}^{\infty }\sum_{\chi =\pm 1}\frac{\chi I_{\alpha _{j}-\chi
\epsilon _{j}/2}(z)K_{\alpha _{j}-\chi \epsilon _{j}/2}(z)}{K_{\beta
_{j}+\epsilon _{j}}^{2}\left( z\right) +K_{\beta _{j}}^{2}\left( z\right) }%
U_{2,\beta _{j}}^{K}(zr/a),  \label{jnubm0}
\end{align}%
with $z=\left( 2n+1\right) \pi aT$. In this case the expectation values do
not depend on the parameter $s$, as anticipated.

For $\mu =0$ the zero temperature limit of the charge and current densities
is directly obtained from (\ref{jnub0}) (or from (\ref{jnub4}) with $%
u_{n}=u_{0n}$) by taking into account that for small temperatures the
dominant contribution comes from the terms with large $n$ and one can
replace the summation over $n$ by the integration. This is reduced to the
replacement
\begin{equation}
\sum_{n=0}^{\infty }f(u_{0n})\rightarrow \frac{1}{2\pi T}\int_{m}^{\infty }dx%
\frac{xf(x)}{\sqrt{x^{2}-m^{2}}},  \label{ReplT0}
\end{equation}%
and we can see that the result (\ref{jbvac}) for $r>a$ is obtained.

Now, let us consider the charge and current densities in the zero
temperature limit for $\mu \neq 0$. When discussing that limit, it is more
convenient to use the representations (\ref{jnupme}). For $|\mu |<m$ we have
$E>|\mu |$ in the entire range of $\gamma $-integration and the thermal
contributions $\left\langle j^{\nu }\right\rangle _{T\pm }$ in the
expectation values tend to zero. Consequently, the expectation values are
reduced to the corresponding VEVs:%
\begin{equation}
\langle j^{\nu }\rangle _{T=0}=\langle j^{\nu }\rangle _{\mathrm{vac}%
},\;|\mu |<m.  \label{jnubT0mu}
\end{equation}%
The zero temperature limit is qualitatively different in the case $|\mu |>m$%
. In this range, the contributions from particles/antiparticles survive in
the limit $T\rightarrow 0$. Those contributions come from the integration
range $\gamma \in \lbrack 0,\gamma _{\mathrm{F}}]$ with $\gamma _{\mathrm{F}%
}=\sqrt{\mu ^{2}-m^{2}}$. The zero temperature mean charge and current
densities are presented as%
\begin{align}
\left\langle j^{0}\right\rangle _{T=0}& =\left\langle j^{0}\right\rangle _{%
\mathrm{vac}}+\frac{e}{2\phi _{0}}\sum_{j}\int_{0}^{\gamma _{\mathrm{F}%
}}d\gamma \,\frac{\gamma }{E}\frac{\sum_{\chi =\pm 1}\left( \lambda E+\chi
sm\right) g_{\beta _{j},\alpha _{j}-\chi \epsilon _{j}/2}^{\left( \lambda
\right) 2}\left( \gamma a,\gamma r\right) }{\bar{J}_{\beta _{j}}^{\left(
\lambda \right) 2}\left( \gamma a\right) +\bar{Y}_{\beta _{j}}^{\left(
\lambda \right) 2}\left( \gamma a\right) },  \notag \\
\left\langle j^{2}\right\rangle _{T=0}& =\left\langle j^{2}\right\rangle _{%
\mathrm{vac}}+\frac{e}{r\phi _{0}}\sum_{j}\epsilon _{j}\int_{0}^{\gamma _{%
\mathrm{F}}}d\gamma \,\frac{\gamma ^{2}}{E}\frac{g_{\beta _{j},\beta
_{j}}^{\left( \lambda \right) }\left( \gamma a,\gamma r\right) g_{\beta
_{j},\beta _{j}+\epsilon _{j}}^{\left( \lambda \right) }\left( \gamma
a,\gamma r\right) }{\bar{J}_{\beta _{j}}^{\left( \lambda \right) 2}\left(
\gamma a\right) +\bar{Y}_{\beta _{j}}^{\left( \lambda \right) 2}\left(
\gamma a\right) },  \label{jnuT0}
\end{align}%
where $\lambda $ is determined by the relation $\mu =\lambda |\mu |$. Hence,
the zero temperature state contains particles for $\mu >m$ ($\lambda =+$)
and antiparticles for $\mu <-m$ ($\lambda =-$). As expected, the sign of the
ratio $(\left\langle j^{0}\right\rangle _{T=0}-\left\langle
j^{0}\right\rangle _{\mathrm{vac}})/e$ coincides with $\lambda $. Similarly,
the zero temperature limit in the boundary-free geometry for the case $|\mu
|>m$ is reduced to%
\begin{align}
\langle j^{0}\rangle _{T=0}^{(0)} &=\left\langle j^{0}\right\rangle _{%
\mathrm{vac}}^{\left( 0\right) }+\frac{e}{2\phi _{0}}\int_{0}^{\gamma _{%
\mathrm{F}}}d\gamma \,\frac{\gamma }{E}\sum_{\chi =\pm 1}\left( \lambda
E+\chi sm\right) \sum_{j}J_{\alpha _{j}-\chi \epsilon _{j}/2}^{2}(\gamma r),
\notag \\
\langle j^{2}\rangle _{T=0}^{(0)} &=\left\langle j^{2}\right\rangle _{%
\mathrm{vac}}^{\left( 0\right) }+\frac{e}{r\phi _{0}}\int_{0}^{\gamma _{%
\mathrm{F}}}d\gamma \,\frac{\gamma ^{2}}{E}\sum_{j}\epsilon _{j}J_{\beta
_{j}}(\gamma r)J_{\beta _{j}+\epsilon _{j}}(\gamma r),  \label{jnuT0i2}
\end{align}%
with the same choice of $\lambda $. The part of the integral in the
expression (\ref{jnuT0i2}) for the charge density containing $\lambda $ is
evaluated by using the formula from \cite{Prud2}. Note that in (\ref{jnuT0i2}%
) the azimuthal current density is the same for particles and antiparticles.
In Appendix \ref{sec:AppB} we show that the same zero temperature limit (\ref%
{jnuT0}) is obtained starting from (\ref{jnub4}).

\section{Charge and current densities in the I-region}

\label{sec:Inter}

In this section we consider the charge and current densities in the interior
region, $r\leq a$. The substitution of mode functions (\ref{psiIE}) into the
formula (\ref{jnupm}) gives
\begin{equation}
\left\langle j^{\nu }\right\rangle _{T\lambda }=\frac{\lambda e}{2\phi _{0}a}%
\sum_{j}\sum_{l=1}^{\infty }\frac{T_{\beta _{j}}^{\left( \lambda \right)
}\left( z\right) g^{\left( \nu \right) }\left( z/a\right) }{e^{\beta \left(
E-\lambda \mu \right) }+1},  \label{jnuTi}
\end{equation}%
with $\nu =0,2$, $z=\gamma _{j,l}^{\left( \lambda \right) }$ being the
positive roots of the equation (\ref{modesi}), and $E=\sqrt{z^{2}/a^{2}+m^{2}%
}$. As before, the expectation values with $\lambda =+$ and $\lambda =-$
present the contributions of the positive and negative energy modes. We have
also introduced the functions
\begin{align}
g^{\left( 0\right) }\left( \gamma \right) & =\gamma \sum_{\chi =\pm 1}\left(
1+\frac{\chi \lambda sm}{\sqrt{\gamma ^{2}+m^{2}}}\right) J_{\alpha
_{j}-\chi \epsilon _{j}/2}^{2}\left( \gamma r\right) ,  \notag \\
g^{\left( 2\right) }\left( \gamma \right) & =\lambda \epsilon _{j}\gamma ^{2}%
\frac{2J_{\beta _{j}}\left( \gamma r\right) J_{\beta _{j}+\epsilon
_{j}}\left( \gamma r\right) }{r\sqrt{\gamma ^{2}+m^{2}}}.  \label{gzi}
\end{align}%
for the charge and azimuthal current densities. The radial component of the
current density vanishes. From (\ref{ci}) it follows that $T_{\beta
_{j}}^{\left( \lambda \right) }\left( z\right) >0$ and, hence, the sign of
the ratio $\left\langle j^{0}\right\rangle _{T\lambda }/e$ coincides with $%
\lambda $. Of course, this agrees with the interpretation of $\left\langle
j^{\nu }\right\rangle _{T\lambda }$ as the contribution from particles for $%
\lambda =+$ and from antiparticles for $\lambda =-$. By using the fact that
the eigenvalues $\gamma _{j,l}^{\left( \pm \right) }$ are the roots of the
equation (\ref{modesi}), it can be seen that under the replacements $\alpha
_{0}\rightarrow -\alpha _{0}$ and $j\rightarrow -j$ one has $T_{\beta
_{j}}^{\left( \pm \right) }(\gamma _{j,l}^{\left( \pm \right) })\rightarrow
T_{\beta _{j}}^{\left( \mp \right) }(\gamma _{j,l}^{\left( \mp \right) })$.
In combination with (\ref{gzi}), this shows that the charge and current
densities obey the relations (\ref{jnualf}) and (\ref{jnuTodd}). In
addition, the following relation takes place on the boundary $r=a$ (compare
with (\ref{jon}) in the E-region):%
\begin{equation}
\left\langle j^{0}\right\rangle _{T\lambda }=\left\langle j_{\phi
}\right\rangle _{T\lambda }=\frac{\lambda e}{\phi _{0}a^{2}}%
\sum_{j}\sum_{l=1}^{\infty }\frac{uT_{\beta _{j}}^{\left( \lambda \right)
}\left( z\right) J_{\beta _{j}}^{2}\left( z\right) }{e^{\beta \left(
E-\lambda \mu \right) }+1}\left( 1+\frac{\lambda sm}{E}\right) ,
\label{joni}
\end{equation}%
with $z=\gamma _{j,l}^{\left( \lambda \right) }$.

The expressions for $\left\langle j^{\nu }\right\rangle _{T\lambda }$ given
by (\ref{jnuTi}) are not convenient for numerical analysis since the zeros $%
\gamma _{j,l}^{(\lambda )}$ are given implicitly. In addition, the terms
with large values of $l$ are highly oscillatory. We will transform the
series over $l$ in (\ref{jnuTi}). That is done by using the generalized
Abel-Plana formula \cite{Saharev,Saharev2} for series $\sum_{l=1}^{\infty
}T_{\beta _{j}}(\gamma _{j,l}^{(\lambda )})f^{(\nu )}(\gamma
_{j,l}^{(\lambda )})$ with the function $f^{(\nu )}(z)$ given by
\begin{equation}
f^{\left( \nu \right) }\left( z\right) =\frac{g^{\left( \nu \right) }\left(
z/a\right) }{e^{\beta (\sqrt{z^{2}/a^{2}+m^{2}}-\lambda \mu )}+1},
\label{fz}
\end{equation}%
for $\nu =0,2$.

Firstly, we discuss the case of $\mu \neq 0$. The functions (\ref{fz}) have
branch points $z=\pm im_{a}$, corresponding to $E=0$, and simple poles $%
z_{n}^{(\lambda )}=a\gamma _{n}^{(\lambda )}$, $n=0,\pm 1,\pm 2,\ldots $, at
the zeros of the function $e^{\beta \left( E-\lambda \mu \right) }+1$. These
zeros correspond to the values of energy $E=E_{n}^{(\lambda )}$ with
\begin{equation}
E_{n}^{(\lambda )}=\lambda \mu +i\pi \left( 2n+1\right) T,  \label{En}
\end{equation}%
and for them we have
\begin{equation}
\gamma _{n}^{(\lambda )2}=E_{n}^{(\lambda )2}-m^{2},  \label{gamn}
\end{equation}%
with $n=0,\pm 1,\pm 2,\ldots $. One has $\mathrm{Im}\,(z_{n}^{(\lambda )})>0$
for $n=0,1,2,\ldots $, and $\mathrm{Im}\,(z_{n}^{(\lambda )})<0$ for $%
n=\ldots ,-2,-1$. For the real parts we have $\,\mathrm{sgn}(\lambda \mu )\,%
\mathrm{Re}\,(z_{n}^{(\lambda )})>0$. In addition, the relations $%
E_{n}^{(\lambda )}=E_{-n-1}^{(\lambda )\ast }$ and $\gamma _{n}^{(\lambda
)}=\gamma _{-n-1}^{(\lambda )\ast }$, $n=\ldots ,-2,-1$, take place for the
poles in the lower and upper half-planes. For $\lambda \mu >0$ both
functions $f^{(\nu )}(z)$ have simple poles in the right half-plane, $%
z=z_{n}^{(\lambda )}\equiv iau_{n}^{\left( \lambda \right) }$, $n=0,\pm
1,\pm 2,\ldots $, with
\begin{equation}
u_{n}^{(\lambda )}=\left\{ \left[ \pi \left( 2n+1\right) T-i\lambda \mu %
\right] ^{2}+m^{2}\right\} ^{1/2}  \label{unlam}
\end{equation}%
and $\mathrm{Re\,}(au_{n}^{(\lambda )})>0$. Under these conditions we have
the following summation formula \cite{Saha19}
\begin{align}
\sum_{l=1}^{\infty }T_{\beta _{j}}^{\left( \lambda \right) }(\gamma
_{j,l}^{\left( \lambda \right) })f^{(\nu )}(\gamma _{j,l}^{\left( \lambda
\right) })& =\int_{0}^{\infty }dx\,f^{(\nu )}\left( x\right) +\frac{\pi }{2}%
\underset{z=0}{\mathrm{Res}}\frac{\tilde{Y}_{\beta _{j}}^{\left( \lambda
\right) }\left( z\right) }{\tilde{J}_{\beta _{j}}^{\left( \lambda \right)
}\left( z\right) }f^{(\nu )}\left( z\right)  \notag \\
& -4\sum_{n=0}^{\infty }\mathrm{Re}\left[ e^{-i\pi \beta _{j}}\frac{\tilde{K}%
_{\beta _{j}}^{\left( \lambda \right) }(u_{n}^{\left( \lambda \right) }a)}{%
\tilde{I}_{\beta _{j}}^{\left( \lambda \right) }(u_{n}^{\left( \lambda
\right) }a)}\underset{z=z_{n}^{\left( \lambda \right) }}{\mathrm{Res}}%
f^{(\nu )}\left( z\right) \right]  \notag \\
& -\frac{2}{\pi }\int_{0}^{\infty }dx\,\mathrm{Re}\left[ e^{-i\pi \beta
_{j}}f^{(\nu )}\left( xe^{\pi i/2}\right) \frac{\tilde{K}_{\beta
_{j}}^{\left( \lambda \right) }\left( x\right) }{\tilde{I}_{\beta
_{j}}^{\left( \lambda \right) }\left( x\right) }\right] ,  \label{SumAP}
\end{align}%
with the notation
\begin{equation}
\tilde{F}_{\beta _{j}}^{(\lambda )}(x)=xF_{\beta _{j}}^{\prime }(x)+[\delta
^{\mathrm{(J)}}(sm_{a}+\lambda \sqrt{\left( e^{\pi i/2}x\right)
^{2}+m_{a}^{2}})-\epsilon _{j}\beta _{j}]F_{\beta _{j}}(x),\;F=I,K,
\label{Ftild}
\end{equation}%
for the modified Bessel functions. We consider the I-region and in (\ref%
{Ftild}) $\delta ^{\mathrm{(J)}}=\delta ^{\mathrm{(I)}}=1$. The notation
with $\mathrm{J}=\mathrm{E}$ is used in the transformation of the
expectation values for the E-region, given in Appendix \ref{sec:appA}. The
contribution of the first term in the right-hand side of (\ref{SumAP}) gives
the expectation value in the boundary-free conical space. Denoting the
latter for the modes with $\lambda =+$ and $\lambda =-$ by $\langle j^{\nu
}\rangle _{T\lambda }^{(0)}$ and introducing a new integration variable $%
\gamma =x/a$, we have%
\begin{equation}
\langle j^{\nu }\rangle _{T\lambda }^{(0)}=\frac{\lambda e}{2\phi _{0}}%
\sum_{j}\int_{0}^{\infty }d\gamma \,\frac{g^{\left( \nu \right) }\left(
\gamma \right) }{e^{\beta (\sqrt{\gamma ^{2}+m^{2}}-\lambda \mu )}+1}.
\label{jnu0T}
\end{equation}%
Of course, this part does not depend on $a$. As it has been already
mentioned before, the expression (\ref{jnu0T}) for the boundary-free
geometry is obtained from the expressions (\ref{jnupme}) for the E-region by
the replacement (\ref{Repl}). It can be checked that the combined thermal
current density $\sum_{\lambda =\pm }$ $\langle j^{\nu }\rangle _{T\lambda
}^{(0)}$, with $\langle j^{\nu }\rangle _{T\lambda }^{(0)}$ from (\ref{jnu0T}%
), is transformed to (\ref{j02T}).

For the function (\ref{fz}) the term in (\ref{SumAP}) with the residue at $%
z=0$ and the part of the last integral over the region $x\in \lbrack 0,ma]$
become zero. Note that in the region $x\in \lbrack ma,\infty )$ we have
\begin{equation}
\tilde{F}_{\beta _{j}}^{(+)}(x)=[\tilde{F}_{\beta _{j}}^{(-)}(x)]^{\ast }=%
\bar{F}_{\beta _{j}}(x).  \label{RelFtild}
\end{equation}%
By using this property, the thermal contributions are presented in the form
\begin{equation}
\langle j^{\nu }\rangle _{T\lambda }=\langle j^{\nu }\rangle _{T\lambda
}^{(0)}+\langle j^{\nu }\rangle _{T\lambda }^{\mathrm{(b)}},
\end{equation}%
where the boundary-induced parts are expressed as
\begin{align}
\left\langle j^{0}\right\rangle _{T\lambda }^{\mathrm{(b)}}& =\frac{e}{\pi
\phi _{0}}\sum_{j}\sum_{\chi =\pm 1}\left\{ -\int_{m}^{\infty }dx\,x\,%
\mathrm{Re}\left[ \frac{\tilde{K}_{\beta _{j}}^{\left( \lambda \right)
}\left( xa\right) }{\tilde{I}_{\beta _{j}}^{\left( \lambda \right) }\left(
xa\right) }\frac{sm+\lambda \chi i\sqrt{x^{2}-m^{2}}}{e^{\beta (i\sqrt{%
x^{2}-m^{2}}-\lambda \mu )}+1}\frac{I_{\alpha _{j}-\chi \epsilon
_{j}/2}^{2}\left( xr\right) }{\sqrt{x^{2}-m^{2}}}\right] \right.  \notag \\
& \left. +2\pi T\,\theta \left( \lambda \mu \right) \sum_{n=0}^{\infty }%
\mathrm{Re}\left[ \frac{\tilde{K}_{\beta _{j}}^{\left( \lambda \right)
}(u_{n}^{\left( \lambda \right) }a)}{\tilde{I}_{\beta _{j}}^{\left( \lambda
\right) }(u_{n}^{\left( \lambda \right) }a)}\left[ sm+\chi \mu +\lambda
i\chi \pi \left( 2n+1\right) T\right] I_{\alpha _{j}-\chi \epsilon
_{j}/2}^{2}\left( u_{n}^{\left( \lambda \right) }r\right) \right] \right\} ,
\notag \\
\left\langle j^{2}\right\rangle _{T\lambda }^{\mathrm{(b)}}& =\frac{e}{\pi
\phi _{0}}\sum_{j}\left\{ -\int_{m}^{\infty }dx\,\frac{xU_{2,\beta
_{j}}^{I}(xr)}{\sqrt{x^{2}-m^{2}}}\,\mathrm{Re}\left[ \frac{\tilde{K}_{\beta
_{j}}^{\left( \lambda \right) }\left( xa\right) }{\tilde{I}_{\beta
_{j}}^{\left( \lambda \right) }\left( xa\right) }\frac{1}{e^{\beta (i\sqrt{%
x^{2}-m^{2}}-\lambda \mu )}+1}\right] \right.  \notag \\
& \left. +2\pi T\theta \left( \lambda \mu \right) \sum_{n=0}^{\infty }%
\mathrm{Re}\left[ \frac{\tilde{K}_{\beta _{j}}^{\left( \lambda \right)
}(u_{n}^{\left( \lambda \right) }a)}{\tilde{I}_{\beta _{j}}^{\left( \lambda
\right) }(u_{n}^{\left( \lambda \right) }a)}U_{2,\beta
_{j}}^{I}(u_{n}^{\left( \lambda \right) }r)\right] \right\} .  \label{jnub2i}
\end{align}%
By using the relations (\ref{RelFtild}) these expressions are transformed to
\begin{align}
\left\langle j^{0}\right\rangle _{T\lambda }^{\mathrm{(b)}}& =\frac{e}{\pi
\phi _{0}}\sum_{j}\left\{ -\int_{m}^{\infty }dx\,x\,\mathrm{Re}\left[ \frac{%
\bar{K}_{\beta _{j}}\left( xa\right) }{\bar{I}_{\beta _{j}}\left( xa\right) }%
\frac{U_{0,\beta _{j}}^{I}(xr)}{\sqrt{x^{2}-m^{2}}}\frac{1}{e^{\lambda \beta
(i\sqrt{x^{2}-m^{2}}-\mu )}+1}\right] \right.  \notag \\
& \left. +2\pi T\theta \left( \lambda \mu \right) \sum_{n=0}^{\infty }%
\mathrm{Re}\left[ \frac{\bar{K}_{\beta _{j}}\left( u_{n}a\right) }{\bar{I}%
_{\beta _{j}}\left( u_{n}a\right) }\sum_{\chi =\pm 1}\left[ sm+\chi \mu
+i\chi \pi \left( 2n+1\right) T\right] I_{\alpha _{j}-\chi \epsilon
_{j}/2}^{2}\left( u_{n}r\right) \right] \right\} ,  \notag \\
\left\langle j^{2}\right\rangle _{T\lambda }^{\mathrm{(b)}}& =-\frac{e}{\pi
\phi _{0}}\sum_{j}\left\{ \int_{m}^{\infty }dx\,\frac{x}{\sqrt{x^{2}-m^{2}}}%
\mathrm{Re}\left[ \frac{\bar{K}_{\beta _{j}}\left( xa\right) }{\bar{I}%
_{\beta _{j}}\left( xa\right) }\frac{U_{2,\beta _{j}}^{I}(xr)}{e^{\lambda
\beta \left( i\sqrt{x^{2}-m^{2}}-\mu \right) }+1}\right] \right.  \notag \\
& \left. -2\pi T\,\theta \left( \lambda \mu \right) \sum_{n=0}^{\infty }%
\mathrm{Re}\left[ \frac{\bar{K}_{\beta _{j}}\left( u_{n}a\right) }{\bar{I}%
_{\beta _{j}}\left( u_{n}a\right) }U_{2,\beta _{j}}^{I}(u_{n}r)\right]
\right\} ,  \label{jnub3i}
\end{align}%
where the notation with bar is defined by (\ref{FbarIK}) with $\mathrm{J}=%
\mathrm{I}$ and $\delta ^{\mathrm{(I)}}=1$ for the I-region.

By taking the sum of the currents for $\lambda =+$ and $\lambda =-$ and,
again, using the relation $\sum_{\lambda }1/(e^{\lambda z}+1)=1$, we see
that the sum of the first terms in the figure braces of (\ref{jnub3i}) gives
$-\langle j^{\nu }\rangle _{\mathrm{vac}}^{\mathrm{(b)}}$ for both
expectation values (see Eq. (\ref{jbvac})). As a consequence, the
boundary-induced contributions $\langle j^{\nu }\rangle ^{\mathrm{(b)}}$,
given by (\ref{jnube}), take the form
\begin{align}
\left\langle j^{0}\right\rangle ^{\mathrm{(b)}}& =\frac{2eT}{\phi _{0}}%
\sum_{j}\sum_{n=0}^{\infty }\mathrm{Re}\left[ \frac{\bar{K}_{\beta
_{j}}\left( u_{n}a\right) }{\bar{I}_{\beta _{j}}\left( u_{n}a\right) }%
\sum_{\chi =\pm 1}\left[ sm+\chi \mu +\chi i\pi \left( 2n+1\right) T\right]
I_{\alpha _{j}-\chi \epsilon _{j}/2}^{2}\left( u_{n}r\right) \right] ,
\notag \\
\left\langle j^{2}\right\rangle ^{\mathrm{(b)}}& =\frac{2eT}{\phi _{0}}%
\sum_{j}\sum_{n=0}^{\infty }\mathrm{Re}\left[ \frac{\bar{K}_{\beta
_{j}}\left( u_{n}a\right) }{\bar{I}_{\beta _{j}}\left( u_{n}a\right) }%
U_{2,\beta _{j}}^{I}(u_{n}r)\right] ,  \label{jnubTi}
\end{align}%
where $u_{n}$ is defined by (\ref{un}). The total expectation values are
presented as (\ref{jnudec2}) with $\langle j^{\nu }\rangle ^{(0)}$ having
the form (\ref{jnu0dec}). For the ratio of the combinations of the modified
Bessel functions in (\ref{jnubTi}) we have the representation
\begin{equation}
\frac{\bar{K}_{\beta _{j}}(z)}{\bar{I}_{\beta _{j}}(z)}=\frac{W_{j}^{\mathrm{%
(I)}}(z)+\left[ \mu +i\pi \left( 2n+1\right) T\right] a/z^{2}}{I_{\beta
_{j}}^{2}(z)+I_{\beta _{j}+\epsilon _{j}}^{2}(z)+2sm_{a}I_{\beta
_{j}}(z)I_{\beta _{j}+\epsilon _{j}}(z)/z},  \label{KIrat}
\end{equation}%
where the function $W_{j}^{\mathrm{(I)}}(z)$ is defined by (\ref{Wj}) with $%
\mathrm{J}=\mathrm{I}$ and $\delta ^{\mathrm{(I)}}=1$.

Now we consider the transformation of the thermal contributions $%
\left\langle j^{\nu }\right\rangle _{T\lambda }$ to the expectation values
coming from particles and antiparticles given by (\ref{jnuTi}) for the case
of $\mu =0$. The corresponding procedure for the evaluation of the
boundary-induced part $\langle j^{\nu }\rangle ^{\mathrm{(b)}}$ differs from
that we have described above for $\mu \neq 0$. Now, the poles of the
functions $f^{(\nu )}(z)$, given by (\ref{fz}), are located on the imaginary
axis. They are given as $z=\pm iau_{0n}$, $n=0,1,2,\ldots $, where $u_{0n}$
is defined in accordance with (\ref{u0n}). The Abel-Plana type summation
formula adapted for this case is given in \cite{Saha19}. It is obtained from
(\ref{SumAP}) by the replacement%
\begin{equation}
\underset{z=z_{n}^{\left( \lambda \right) }}{\mathrm{Res}}f^{(\nu )}\left(
z\right) \rightarrow \frac{1}{2}\underset{z=iau_{0n}}{\mathrm{Res}}f^{(\nu
)}\left( z\right) ,  \label{ReplAP}
\end{equation}%
and by the replacement $\int_{0}^{\infty }dx\rightarrow \mathrm{p.v.}%
\int_{0}^{\infty }dx$ in the last integral. Here, $\mathrm{p.v.}$ stands for
the principal value of the integral. For $\mu =0$ the term coming from the
poles $iau_{0n}$ is present for both $\lambda =+$ and $\lambda =-$, whereas
in (\ref{SumAP}) the pole term is present only in the case of $\lambda \mu
>0 $. Further transformations of the expectation values are similar to those
for the E-region. In the region $r<a$, the expressions for $\langle
j^{0}\rangle _{T\lambda }^{\mathrm{(b)}}$ and $\langle j^{0}\rangle _{T}^{%
\mathrm{(b)}}$ are obtained from (\ref{jnub20}) and (\ref{jnub200}) by the
replacements $I\rightleftarrows K$. For the expressions of $\langle
j^{2}\rangle _{T\lambda }^{\mathrm{(b)}}$ and $\langle j^{2}\rangle _{T}^{%
\mathrm{(b)}}$ in the I-region, in addition to those replacements, the sign
is changed. As a result, for the boundary-induced contributions we obtain%
\begin{align}
\left\langle j^{0}\right\rangle ^{\mathrm{(b)}}& =\frac{2eT}{a\phi _{0}}%
\sum_{j}\sum_{n=0}^{\infty }\sum_{\chi =\pm 1}\frac{\left[ sm_{a}W_{j}^{%
\mathrm{(I)}}(z)-\chi \left( 1-\frac{m_{a}^{2}}{z^{2}}\right) \right]
I_{\alpha _{j}-\chi \epsilon _{j}/2}^{2}\left( zr/a\right) }{I_{\beta
_{j}}^{2}(z)+I_{\beta _{j}+\epsilon _{j}}^{2}(z)+2sm_{a}I_{\beta
_{j}}(z)I_{\beta _{j}+\epsilon _{j}}(z)/z},  \notag \\
\left\langle j^{2}\right\rangle ^{\mathrm{(b)}}& =\frac{2eT}{\phi _{0}}%
\sum_{j}\sum_{n=0}^{\infty }\,\frac{W_{j}^{\mathrm{(I)}}(z)U_{2,\beta
_{j}}^{I}(zr/a)}{I_{\beta _{j}}^{2}(z)+I_{\beta _{j}+\epsilon
_{j}}^{2}(z)+2sm_{a}I_{\beta _{j}}(z)I_{\beta _{j}+\epsilon _{j}}(z)/z}%
,\;z=au_{0n}.  \label{jnubtoti0}
\end{align}%
We see that the expressions (\ref{jnubtoti0}) for $\langle j^{\nu }\rangle ^{%
\mathrm{(b)}}$ are also directly obtained from (\ref{jnubTi}) in the limit $%
\mu \rightarrow 0$. Note that for a massless field the expressions for the
boundary-induced contributions in the expectation values are reduced to
\begin{align}
\left\langle j^{0}\right\rangle ^{\mathrm{(b)}}& =\frac{2eT}{\phi _{0}a}%
\sum_{j}\sum_{n=0}^{\infty }\frac{I_{\beta _{j}+\epsilon _{j}}^{2}\left(
zr/a\right) -I_{\beta _{j}}^{2}\left( zr/a\right) }{I_{\beta _{j}+\epsilon
_{j}}^{2}\left( z\right) +I_{\beta _{j}}^{2}\left( z\right) },  \notag \\
\left\langle j^{2}\right\rangle ^{\mathrm{(b)}}& =\frac{2eT}{\phi _{0}}%
\sum_{j}\sum_{n=0}^{\infty }\sum_{\chi =\pm 1}\frac{\chi I_{\alpha _{j}-\chi
\epsilon _{j}/2}(z)K_{\alpha _{j}-\chi \epsilon _{j}/2}(z)}{I_{\beta
_{j}}^{2}\left( z\right) +I_{\beta _{j}+\epsilon _{j}}^{2}\left( z\right) }%
U_{2,\beta _{j}}^{I}(zr/a),  \label{jnubm0i}
\end{align}%
where $z=\left( 2n+1\right) \pi aT$.

Now we turn to the zero temperature limit for the I-region. In the case of
zero chemical potential that limit is directly obtained from (\ref{jnubtoti0}%
) (or from (\ref{jnubTi}) with $u_{n}=u_{n0}$) by the replacement (\ref%
{ReplT0}) and we get the VEVs given by (\ref{jbvac}) in the region $r<a$.
For $\mu \neq 0$, as a starting point it is convenient to use the
representation (\ref{jnuTi}). For the chemical potential in the range $\mu
^{2}<\gamma _{j,1}^{\left( \lambda \right) 2}/a^{2}+m^{2}$ the relation (\ref%
{jnubT0mu}) takes place and the zero temperature expectation values coincide
with the corresponding VEVs. In the case $\mu ^{2}>\gamma _{j,1}^{\left(
\lambda \right) 2}/a^{2}+m^{2}$ one has $\lim_{T\rightarrow 0}[e^{\beta
\left( E-\lambda \mu \right) }+1]=1$ for the modes with $E<\lambda \mu $ and
$\lim_{T\rightarrow 0}[e^{\beta \left( E-\lambda \mu \right) }+1]=2$ for $%
E=\lambda \mu $. For the zero temperature limit of the expectation values
from (\ref{jnuTi}) we get%
\begin{equation}
\left\langle j^{\nu }\right\rangle _{T=0}=\left\langle j^{\nu }\right\rangle
_{\mathrm{vac}}+\frac{\lambda e}{2\phi _{0}a}\sum_{j}\sideset{}{'}%
\sum_{l=1}^{l_{m}}T_{\beta _{j}}^{\left( \lambda \right) }(\gamma
_{j,l}^{\left( \lambda \right) })g^{\left( \nu \right) }(\gamma
_{j,l}^{\left( \lambda \right) }/a),  \label{jnuT0i}
\end{equation}%
where $\lambda $ and $l_{m}$ are defined by $\mu =\lambda |\mu |$ and
\begin{equation}
\gamma _{j,l_{m}}^{(\lambda )}\leq a\gamma _{\mathrm{F}}<\gamma
_{j,l_{m}+1}^{(\lambda )}.  \label{poles}
\end{equation}%
The prime on the summation over $l$ means that in case of $\gamma
_{j,l_{m}}^{\left( \lambda \right) }=a\gamma _{\mathrm{F}}$ the term with $%
l=l_{m}$ must be taken with an additional factor 1/2. The last term in (\ref%
{jnuT0i}) comes from particles for $\mu >0$ and antiparticles for $\mu <0$.
The zero temperature limit of the expectation values in the geometry without
boundary is given by (\ref{jnuT0i2}). In Appendix \ref{sec:AppB} we show
that the same result (\ref{jnuT0i}) is obtained from the representation (\ref%
{jnubTi}).

The MIT bag boundary condition (\ref{BCMIT}) provides zero normal fermionic
current on the boundary. The same is done by the boundary condition that
differs from (\ref{BCMIT}) by the sign of the term involving the normal to
the boundary (see, e.g., \cite{Berr87}). We can combine both the conditions
in
\begin{equation}
\left( 1+\eta in_{\mu }\gamma ^{\mu }\right) \psi (x)=0,\;r=a,  \label{BCeta}
\end{equation}%
introducing the parameter $\eta =\pm 1$. It can be shown that the results
for the boundary condition with $\eta =-1$ are obtained from the
corresponding formulas for the condition (\ref{BCMIT}) ($\eta =+1$) by the
replacement $\delta ^{\mathrm{(J)}}\rightarrow -\delta ^{\mathrm{(J)}}$ in
the definitions (\ref{Fbar}), (\ref{FbarIK}), and (\ref{Ftild}).
Equivalently, we can generalize the expressions for the edge induced
expectation values in the case of the condition (\ref{BCeta}) by making the
replacement $\delta ^{\mathrm{(J)}}\rightarrow \eta \delta ^{\mathrm{(J)}}$.
The corresponding problem is specified by the set of parameters $(s,\mu
,\eta )$. From the definitions (\ref{FbarIK}) and (\ref{un}) it is seen that%
\begin{equation}
\bar{F}_{\beta _{j}}\left( u_{n}a\right) |_{(s,\mu ,\eta )}=\left[ \bar{F}%
_{\beta _{j}}\left( u_{n}a\right) |_{(-s,-\mu ,-\eta )}\right] ^{\ast }.
\label{Rel2sets}
\end{equation}%
Now, from (\ref{jnub4}) and (\ref{jnubTi}) we get the following relations
between the edge induced expectation values in two different problems with
the sets $(s,\mu ,\eta )$ and $(-s,-\mu ,-\eta )$:%
\begin{equation}
\left\langle j^{\nu }\right\rangle _{(s,\mu ,\eta )}^{\mathrm{(b)}%
}=(-1)^{1-\nu /2}\left\langle j^{\nu }\right\rangle _{(-s,-\mu ,-\eta )}^{%
\mathrm{(b)}},  \label{jnurel}
\end{equation}%
for $\nu =0,2$. Hence, the expectation values for the condition (\ref{BCeta}%
) with $\eta =-1$ are obtained from those in the case $\eta =+1$ (given
above) by the replacements $s\rightarrow -s$ and $\mu \rightarrow -\mu $. In
a similar way, from the formulas in Section \ref{sec:BF} it can be seen that%
\begin{equation}
\langle j^{\nu }\rangle _{(s,\mu )}^{(0)}=(-1)^{1-\nu /2}\langle j^{\nu
}\rangle _{(-s,-\mu )}^{(0)}.  \label{jnu0rel}
\end{equation}%
This shows that the relation (\ref{jnurel}) takes place for the total
expectation values $\left\langle j^{\nu }\right\rangle _{(s,\mu ,\eta )}$ as
well.

The mode functions used above in the investigations of the charge and
current densities in the E- and I-regions are periodic functions of the
angular coordinate $\phi $ with the period $\phi _{0}$. We could impose more
general periodicity condition with a nontrivial phase, given by $\psi
(t,r,\phi +\phi _{0})=e^{2\pi i\varsigma }\psi (t,r,\phi )$, where $%
\varsigma $ is a constant. It can be checked that the fermionic modes in
this case are obtained from those discussed above by the shift $j\rightarrow
j+\varsigma $. The corresponding expressions for the expectation values of
the charge and current densities are given by the formulas given above
making the replacement $\alpha \rightarrow eA/q+\varsigma $. This shows that
the phase $\varsigma $ can be interpreted in terms of the Aharonov-Bohm type
vector potential and vice versa.

\section{Asymptotic and numerical analysis}

\label{sec:Num}

In this section we investigate the asymptotics for the boundary-induced
expectation values of the charge and current densities corresponding to the
limiting values of radial coordinate $r$, temperature $T$ and parameters $a$%
, $\alpha _{0}$.

\subsection{Radial asymptotics}

Firstly, we consider the expectation values in the E-region at large
distances from the circular boundary by fixing the location of the boundary,
the chemical potential and the mass. By taking into account that for large $%
|z|$ one has $K_{\beta }(z)\sim \sqrt{\pi /2z}e^{-z}$, from (\ref{jnub4}) we
find%
\begin{equation}
\left\langle j^{\nu }\right\rangle ^{\mathrm{(b)}}\approx \frac{2\pi eT}{%
r^{2}\phi _{0}}\sum_{j}\sum_{n=0}^{\infty }\mathrm{Re}\left[ \frac{\bar{I}%
_{\beta _{j}}\left( u_{n}a\right) }{\bar{K}_{\beta _{j}}\left( u_{n}a\right)
}\left( \frac{smr}{u_{n}}\right) ^{1-\nu /2}e^{-2ru_{n}}\right] .
\label{jnur}
\end{equation}%
For temperatures $T\gtrsim m,|\mu |$ the series over $n$ is dominated by the
first term and one gets $\left\langle j^{\nu }\right\rangle ^{\mathrm{(b)}%
}\propto e^{-2ru_{0}}$. In the case $\mu =0$ the large distance asymptotic
is further simplified with the current density
\begin{equation}
\left\langle j^{2}\right\rangle ^{\mathrm{(b)}}\approx \frac{2\pi eT}{\phi
_{0}r^{2}}\sum_{j}\frac{W_{j}^{\mathrm{(E)}}(z)e^{-2zr/a}}{K_{\beta
_{j}+\epsilon _{j}}^{2}(z)+K_{\beta _{j}}^{2}(z)+2sm_{a}K_{\beta
_{j}}(z)K_{\beta _{j}+\epsilon _{j}}(z)/z},  \label{j2br}
\end{equation}%
where $z=a\sqrt{\pi ^{2}T^{2}+m^{2}}$. The charge density is expressed as%
\begin{equation}
\left\langle j^{0}\right\rangle ^{\mathrm{(b)}}\approx \frac{sm\left\langle
j_{\phi }\right\rangle ^{\mathrm{(b)}}}{\sqrt{\pi ^{2}T^{2}+m^{2}}}.
\label{j0br}
\end{equation}%
For a massless field the leading term for the charge density in (\ref{j0br})
vanishes and keeping the next term in the expansion one finds%
\begin{equation}
\left\langle j^{0}\right\rangle ^{\mathrm{(b)}}\approx \frac{4e}{\phi
_{0}^{2}Tar^{2}}\sum_{j}\frac{(j+\alpha _{0})e^{-2\pi Tr}}{K_{\beta
_{j}+\epsilon _{j}}^{2}(\pi Ta)+K_{\beta _{j}}^{2}(\pi Ta)}.  \label{j0brm0}
\end{equation}

It is of interest to compare the large distance asymptotics with those for
the expectation value $\langle j^{\nu }\rangle ^{(0)}$ in the boundary-free
conical space. The leading term for the charge density $\langle j^{0}\rangle
^{(0)}$ coincides with the expectation value $\langle j^{0}\rangle ^{\mathrm{%
(M)}}$ in the Minkowski spacetime with $\alpha _{0}=0$, given by (\ref{j00T0}%
). The latter does not depend on the radial coordinate. The topological
contribution induced by a boundary-free conical geometry and magnetic flux
is described by the difference $\langle j^{\nu }\rangle _{\mathrm{t}%
}^{(0)}=\langle j^{\nu }\rangle ^{(0)}-\langle j^{\nu }\rangle ^{\mathrm{(M)}%
}$, with $\langle j^{2}\rangle ^{\mathrm{(M)}}=0$. For $\phi _{0}>\pi $ the
topological part $\langle j^{\nu }\rangle _{\mathrm{t}}^{(0)}$ is suppressed
by the factor $e^{-2ru_{0}}$. In the case $\phi _{0}<\pi $ the suppression
of the expectation value $\langle j^{\nu }\rangle _{\mathrm{t}}^{(0)}$ is
weaker, by the factor $e^{-2ru_{0}\sin (\phi _{0}/2)}$. For a massive field
the asymptotics of the VEVs in a boundary-free conical space are described by%
\begin{align}
\left\langle j^{\nu }\right\rangle _{\mathrm{vac}}^{\left( 0\right) }
&\propto e^{-2mr},\;mr\gg 1,\;\phi _{0}>\pi ,  \notag \\
\left\langle j^{\nu }\right\rangle _{\mathrm{vac}}^{\left( 0\right) }
&\propto e^{-2mr\sin (\phi _{0}/2)},\;mr\gg 1,\;\phi _{0}<\pi .
\label{jvacr}
\end{align}%
For a massless field the charge density vanishes, $\left\langle
j^{0}\right\rangle _{\mathrm{vac}}^{\left( 0\right) }=0$, and the decay of
the current density follows a power-law, like $\left\langle
j^{2}\right\rangle _{\mathrm{vac}}^{\left( 0\right) }\propto 1/r^{2}$. At
large distances the boundary-induced contributions in the VEVs behave like $%
\left\langle j^{0}\right\rangle _{\mathrm{vac}}^{\mathrm{(b)}},\left\langle
j_{\phi }\right\rangle _{\mathrm{vac}}^{\mathrm{(b)}}\propto
e^{-2mr}/r^{3/2} $, $mr\gg 1$, for a massive field and as $\left\langle
j^{0}\right\rangle _{\mathrm{vac}}^{\mathrm{(b)}}\propto 1/r^{2\rho +2}$ in
the case of a massless field. Here, we have defined%
\begin{equation}
\rho =q\left( 1/2-|\alpha _{0}|\right) .  \label{ro}
\end{equation}%
In the case of a massless field for the boundary-induced VEV of the current
density one has $\left\langle j_{\phi }\right\rangle _{\mathrm{vac}}^{%
\mathrm{(b)}}\propto 1/r^{2\rho +3}$ for $\rho >1/2$ and $\left\langle
j_{\phi }\right\rangle _{\mathrm{vac}}^{\mathrm{(b)}}\propto 1/r^{4\rho +2}$
for $\rho <1/2$.

\begin{figure}[tbph]
\begin{centering}
\begin{tabular}{cc}
\epsfig{figure=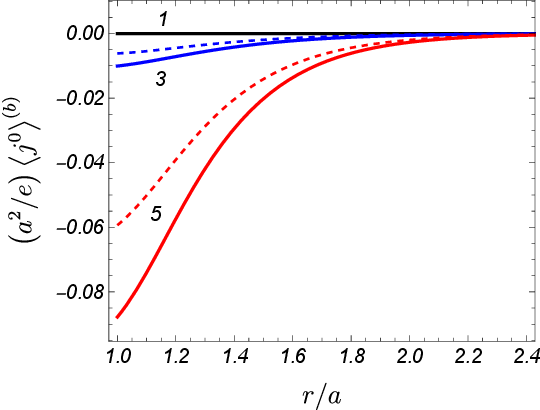,width=7.5cm,height=6.cm} & \quad{}\epsfig{figure=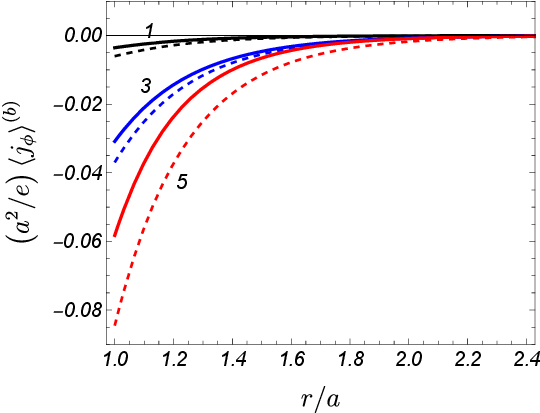,width=7.5cm,height=6.cm}\tabularnewline
\end{tabular}
\par\end{centering}
\caption{Expectation values of the charge (left panel) and current (right
panel) densities in the E-region for a massless field with zero chemical
potential as functions of the radial coordinate. The graphs are plotted for $%
Ta=0.5$ and the full/dashed curves correspond to $\protect\alpha _{0}=0.2$/$%
\protect\alpha _{0}=0.4$. The numbers near the curves are the values of $q$.}
\label{fig2}
\end{figure}

In Figure \ref{fig2} we have plotted the radial dependence of the
boundary-induced contributions in the charge and current densities in the
E-region for $m=\mu =0$ and for fixed temperature corresponding to $Ta=0.5$.
The full and dashed curves correspond to $\alpha _{0}=0.2$ and $\alpha
_{0}=0.4$, respectively, and the numbers near the curves present the values
of the parameter $q$. An important feature for the charge and current
densities is their finiteness in the limit $r\rightarrow a$. This is in
contrast to the behavior of the fermion condensate that diverges on the edge
(see \cite{Saha19}).

The difference in the near-boundary behavior of the fermion condensate from
one side and the charge and current densities from the other side can also
be understood by general arguments of quantum field theory in external
fields. In quantum field theory with boundaries the divergences in the
expectation values of local physical observables, bilinear in the field
operator, are divided into bulk and boundary divergences. For a given
quantum field in an external gauge field $A_{\mu }(x)$, the bulk divergences
are uniquely determined by the local geometrical characteristics of the
spacetime, constructed from the Riemann tensor, and by the gauge field
strength tensor $F_{\mu \nu }$. For smooth boundaries, the boundary
divergences are specified by the extrinsic curvature tensor of the boundary
\cite{Deut79,Kenn80} (see also \cite{Bord09,Casi11} for discussions in the
context of the Casimir effect). An important point is that the boundary
divergences are entirely due to the vacuum parts in the expectation values.
In the problem under consideration the field tensor $F_{\mu \nu }$ is zero
and the background geometry is flat except the point corresponding to the
cone apex $r=0$ in the I-region. Consequently, the renormalization of the
expectation values of local observables for $r\neq 0$ and $r\neq a$ is
realized by the subtraction of the corresponding quantities for the
boundary-free Minkowski spacetime in the absence of the gauge field. The
vacuum contributions in the expectation values of the charge and current
densities are odd functions of the parameter $\alpha $, determined by
magnetic flux in accordance with (\ref{alphan}). From here it follows that
in the problem with zero gauge field ($A_{\mu }=0$ and $\alpha =0$) the
renormalized charge and current densities become zero everywhere including
the points on the boundary. Now, when we add an external constant gauge
field, the bulk and boundary geometries are not changed and the field tensor
remains equal to zero. Thus, the divergences in the problems without and
with constant gauge fields are the same. In particular, the turning on of the
constant gauge field does not add boundary divergences in the expectation
values of the charge and current densities. Note that the given argument
does not work for the fermion condensate. The latter is an even function of $%
\alpha $ and it does vanish for $\alpha =0$. The vacuum fermion condensate
diverges on the boundary and the diverging contribution does not depend on
the parameter $\alpha $.

\begin{figure}[tbph]
\begin{centering}
\epsfig{figure=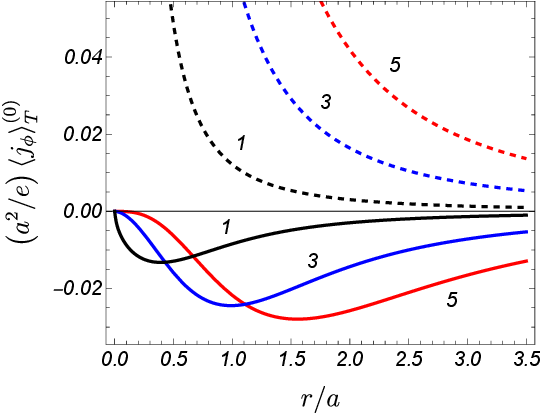,width=7.5cm,height=6.cm}
\par\end{centering}
\caption{Vacuum (dashed curves) and thermal (full curves) current densities
for a massless field with zero chemical potential in the boundary-free
geometry. For the parameters we have taken $Ta=0.5$, $\protect\alpha %
_{0}=0.2 $, and the numbers near the curves are the values of $q$. }
\label{fig3}
\end{figure}

For a massless field with zero chemical potential the charge density in the
boundary-free geometry vanishes, $\langle j^{0}\rangle ^{(0)}=0$, and the
boundary induced contribution in the charge density plotted in Figure \ref%
{fig2} coincides with the total charge density $\left\langle
j^{0}\right\rangle $. That is not the case for the current density. For $%
m=\mu =0$ the current density in a conical space without boundary is
expressed as \cite{Bell16T}%
\begin{align}
\langle j_{\phi }\rangle ^{(0)}& =-\frac{4eT^{3}r}{\pi }\sideset{}{'}{\sum}%
_{n=0}^{\infty }(-1)^{n}\left\{ \sideset{}{'}{\sum}_{l=1}^{[q/2]}\frac{%
(-1)^{l}\sin (\pi l/q)\sin (2\pi l\alpha _{0})}{\left[ n^{2}+\left( 2Tr\sin
(\pi l/q)\right) ^{2}\right] ^{\frac{3}{2}}}\right.  \notag \\
& \left. -\frac{q}{\pi }\int_{0}^{\infty }dy\,\frac{f(q,\alpha _{0},y)\cosh y%
}{\cosh (2qy)-\cos (q\pi )}\frac{1}{\left[ n^{2}+(2Tr\cosh y)^{2}\right] ^{%
\frac{3}{2}}}\right\} ,  \label{jphi0}
\end{align}%
where the function $f(q,\alpha _{0},y)$ is defined by Eq. (\ref{f02}) and
the prime on the sum over $n$ means that the term $n=0$ is taken with the
coefficient 1/2 (for the prime on the sum over $l$ see (\ref{jm02})). That
term corresponds to the vacuum current density $\langle j_{\phi }\rangle _{%
\mathrm{vac}}^{(0)}$ and the remaining part presents the thermal
contribution. In the massless case the vacuum current density behaves like $%
\langle j_{\phi }\rangle _{\mathrm{vac}}^{(0)}\propto 1/r^{2}$. For
comparison with the contributions induced by the boundary, Figure \ref{fig3}
shows the radial dependence of the thermal contribution to the current
density, $\langle j_{\phi }\rangle _{T}^{(0)}$, for a massless field with
zero chemical potential in the boundary-free geometry (full curves) for $%
Ta=0.5$ and $\alpha _{0}=0.2$. The numbers near the curves present the
corresponding values of $q$. The dashed curves describe the radial
dependence of the vacuum current density, $\langle j_{\phi }\rangle _{%
\mathrm{vac}}^{(0)}$, in the same geometry for $\alpha _{0}=0.2$. Although
the problem corresponding to Figure \ref{fig3} does not contain the
parameter $a$, the quantities on the horizontal and vertical axes correspond
to the radial distance and the current density measured in units of $a$ and $%
e/a^{2}$, respectively. This choice of the measurement units makes it easier
to compare with data in figures for the geometry with the boundary.

Now we consider the asymptotic of the boundary-induced expectation values (%
\ref{jnubTi}) in the I-region for points near the cone apex. In the limit of
small values of $r$ the series over $j$ is dominated by the contribution of
the mode with $j=-\mathrm{sgn}(\alpha _{0})/2$ and, by using the small
argument asymptotic for the function $I_{\beta }(z)$, the leading order
terms read
\begin{align}
\left\langle j^{0}\right\rangle ^{\mathrm{(b)}}& \approx \frac{2\mathrm{sgn}%
\left( \alpha _{0}\right) eT\left( r/2\right) ^{2\rho -1}}{\phi _{0}\Gamma
^{2}\left( \rho +1/2\right) }\sum_{n=0}^{\infty }\mathrm{Re}\left\{
u_{n}^{2\rho -1}\right.  \notag \\
& \left. \times \frac{\bar{K}_{\beta _{j}}\left( u_{n}a\right) }{\bar{I}%
_{\beta _{j}}\left( u_{n}a\right) }\left[ \pi \left( 2n+1\right) T-i\mu +i\,%
\mathrm{sgn}\left( \alpha _{0}\right) sm\right] \right\} ,  \notag \\
\left\langle j^{2}\right\rangle ^{\mathrm{(b)}}& \approx \frac{-4eT\left(
r/2\right) ^{2\rho -1}}{\phi _{0}\left( 2\rho +1\right) \Gamma ^{2}\left(
\rho +1/2\right) }\sum_{n=0}^{\infty }\mathrm{Re}\left[ \frac{\bar{K}_{\beta
_{j}}\left( u_{n}a\right) }{\bar{I}_{\beta _{j}}\left( u_{n}a\right) }%
u_{n}^{2\rho +1}\right] .  \label{jnuapex}
\end{align}%
As seen, the physical component $\left\langle j_{\phi }\right\rangle ^{%
\mathrm{(b)}}$ of the boundary-induced current density vanishes on the cone
apex. According to (\ref{jnuapex}), the boundary-induced charge density
vanishes on the cone apex for $\rho >1/2$ and diverges for $\rho <1/2$. For $%
\rho =1/2$ it takes a finite nonzero limiting value.

\begin{figure}[tbph]
\begin{centering}
\begin{tabular}{cc}
\epsfig{figure=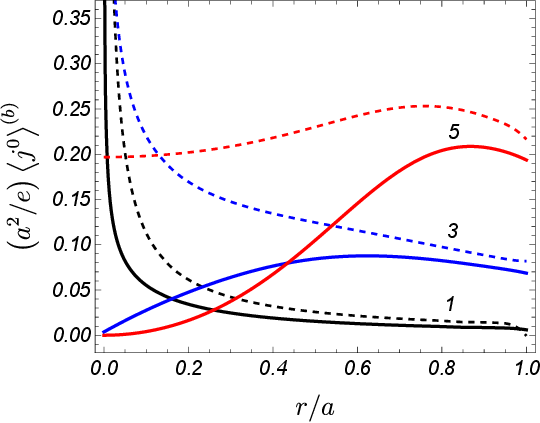,width=7.5cm,height=6.cm} & \quad{}\epsfig{figure=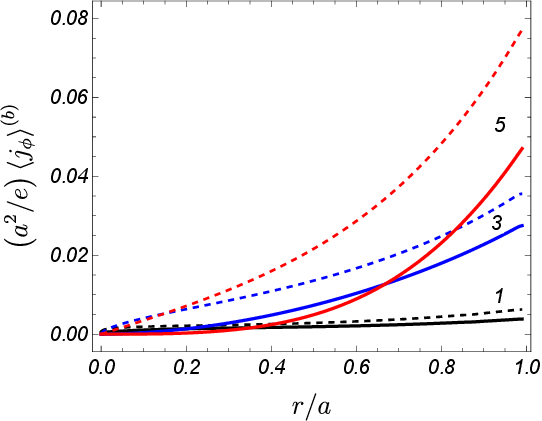,width=7.5cm,height=6.cm}\tabularnewline
\end{tabular}
\par\end{centering}
\caption{The same as in Figure \protect\ref{fig2} for the I-region.}
\label{fig4}
\end{figure}

In the limit $r\rightarrow 0$, the VEVs in a boundary-free conical space
behave as $\langle j^{\nu }\rangle _{\mathrm{vac}}^{(0)}\propto 1/r^{1+\nu }$
(see (\ref{jm02})). In the case of a massless field the corresponding charge
density vanishes. The near-apex asymptotic of the thermal part of the
expectation value in the boundary-free geometry, $\langle j^{\nu }\rangle
_{T}^{(0)}=\langle j^{\nu }\rangle ^{(0)}-\langle j^{\nu }\rangle _{\mathrm{%
vac}}^{(0)}$, is given by $\langle j^{\nu }\rangle _{T}^{(0)}\propto
r^{2\rho -1}$ for $\rho <1/2$. For the values of the parameters in the
region $\rho >1/2$ the thermal expectation value $\langle j^{\nu }\rangle
_{T}^{(0)}$ tends to a finite nonzero value in the limit $r\rightarrow 0$.
We see that near the apex the expectation values of the charge and current
densities are dominated by the vacuum parts independent of $a$. The radial
dependence of the boundary-induced charge and current densities in the
I-region is plotted in Figure \ref{fig4} for the same values of the
parameters as in Figure \ref{fig2}. As follows from the asymptotic analysis,
at the cone apex the boundary-induced expectation value of the current
density $\left\langle j_{\phi }\right\rangle ^{\mathrm{(b)}}$ always
vanishes, whereas the boundary-induced expectation value of the charge
density diverges in case of $2|\alpha _{0}|>1-1/q$ and vanishes for $%
2|\alpha _{0}|<1-1/q$. In the special case $2|\alpha _{0}|=1-1/q$ the charge
density tends to a finite nonzero value in the limit $r\rightarrow 0$. On
the left panel of Figure \ref{fig4} that case corresponds to the dashed
curve for $q=5$. In addition, for large values of $Tr$ the boundary-induced
expectation values are suppressed by the factor $e^{-2\pi T\left( r-a\right)
}$.

It is also of interest to consider the behavior of the boundary-induced
expectation values for small values of the radius $a$ and for fixed $r$,
assuming that $Ta,ma\ll 1$. By using the asymptotic expressions for the
modified Bessel functions for small values of the argument, from (\ref{jnub4}%
) one can see that $\left\langle j^{0}\right\rangle ^{\mathrm{(b)}%
},\left\langle j_{\phi }\right\rangle ^{\mathrm{(b)}}\propto a^{2\rho }$ and
for $|\alpha _{0}|<1/2$ the boundary-induced contributions tend to zero in
the limit $a\rightarrow 0$. The expectation values of the charge and current
densities in the limit $|\alpha _{0}|\rightarrow 1/2$ are discussed in the
next subsection. This limit corresponds to a half-integer magnetic flux in
units of the flux quantum.

\subsection{Charge and current densities for half-integer fluxes}

The charge and current densities, in general, are discontinuous at
half-integer values of the parameter $\alpha $. By taking into account that
the expectation values are periodic functions of $\alpha $ with the period
equal to 1, the discontinuity at $\alpha =1/2$ can be expressed as%
\begin{equation}
\langle j^{\nu }\rangle |_{\alpha =1/2-0}^{\alpha =1/2+0}=\langle j^{\nu
}\rangle |_{\alpha _{0}=1/2-0}^{\alpha _{0}=-1/2+0}\equiv \langle j^{\nu
}\rangle |_{\alpha _{0}=1/2}^{\alpha _{0}=-1/2}.  \label{Disj}
\end{equation}%
Here and below, $\alpha _{0}=\pm 1/2$ is understood in the sense $\alpha
_{0}\rightarrow \pm \left( 1/2-0\right) $.

First we consider the expectation values in the boundary-free conical
geometry. By using the expressions $\langle j^{\nu }\rangle ^{(0)}$ from
\cite{Bell16T} we get%
\begin{equation}
\langle j^{\nu }\rangle ^{(0)}|_{\alpha _{0}=1/2}^{\alpha _{0}=-1/2}=-\frac{%
2^{7/2}em^{2}s}{\pi ^{3/2}\phi _{0}}\left( 2ms\right) ^{\frac{\nu }{2}}%
\sideset{}{'}{\sum}_{n=0}^{\infty }(-1)^{n}\cosh (n\mu /T)\int_{0}^{\infty
}dy\,h_{\frac{1+\nu }{2}}\left( c_{n}(y)\right) \cosh y,  \label{jnu0dis}
\end{equation}%
with the functions $h_{\beta }(x)=x^{-\beta }K_{\beta }(x)$ and $c_{n}(y)=m%
\sqrt{n^{2}/T^{2}+4r^{2}\cosh ^{2}y}$. Here, the prime on the summation sign
means that the term $n=0$ is taken with an additional coefficient 1/2. Note
that the current density $\langle j^{2}\rangle ^{(0)}$ is an even function
of the chemical potential (and, hence, an odd function of $\alpha $) and the
corresponding limiting values are expressed as $\langle j^{2}\rangle
^{(0)}|_{\alpha _{0}=\pm 1/2}=\mp \langle j^{2}\rangle ^{(0)}|_{\alpha
_{0}=1/2}^{\alpha _{0}=-1/2}/2$. Introducing a new integration variable $%
w=c_{n}(y)$, the integral in (\ref{jnu0dis}) is evaluated by using the
formula from \cite{Prud2} with the result%
\begin{equation}
\int_{0}^{\infty }dy\,\cosh yh_{\frac{1+\nu }{2}}\left( c_{n}(y)\right) =%
\frac{\sqrt{\pi /2}}{2mr}h_{\frac{\nu }{2}}(m\sqrt{n^{2}/T^{2}+4r^{2}}).
\label{IntForm}
\end{equation}%
This gives%
\begin{equation}
\langle j^{\nu }\rangle ^{(0)}|_{\alpha _{0}=1/2}^{\alpha _{0}=-1/2}=-\frac{%
4esm}{\pi \phi _{0}r}\left( 2ms\right) ^{\frac{\nu }{2}}\sideset{}{'}{\sum}%
_{n=0}^{\infty }(-1)^{n}\cosh (n\mu /T)h_{\frac{\nu }{2}}(m\sqrt{%
n^{2}/T^{2}+4r^{2}}).  \label{jnu0dis2}
\end{equation}%
In particular, for $\mu =0$ from here it follows that%
\begin{equation}
\langle j^{\nu }\rangle ^{(0)}|_{\alpha _{0}=\pm 1/2}=\pm \frac{2esm}{\pi
\phi _{0}r}\left( 2ms\right) ^{\frac{\nu }{2}}\sideset{}{'}{\sum}%
_{n=0}^{\infty }(-1)^{n}h_{\frac{\nu }{2}}(m\sqrt{n^{2}/T^{2}+4r^{2}}).
\label{jnu0dismu0}
\end{equation}

Now we turn to the boundary-induced contributions. By using the defintions
of $\beta _{j}$ and $\epsilon _{j}$, it can be seen that for a function $%
f(x,y)$ one has%
\begin{equation}
\sum_{j}f\left( \beta _{j},\beta _{j}+\epsilon _{j}\right) |_{\alpha
_{0}=\pm 1/2}=\sum_{l=1}^{\infty }\sum_{\chi =\pm 1}f\left( ql-\chi
/2,ql+\chi /2\right) +f\left( \pm 1/2,\mp 1/2\right) ,  \label{fsumj}
\end{equation}%
where the last term comes from the $j=\mp 1/2$ mode for $\alpha _{0}=\pm 1/2$%
. From here it follows that the discontinuity at $\alpha =1/2$ is determined
by%
\begin{equation}
\sum_{j}f\left( \beta _{j},\beta _{j}+\epsilon _{j}\right) |_{\alpha
_{0}=1/2}^{\alpha _{0}=-1/2}=f\left( -1/2,1/2\right) -f\left(
1/2,-1/2\right) ,  \label{Disf}
\end{equation}%
$\ $ and it comes from the mode $j=\mp 1/2$ for $\alpha _{0}=\pm 1/2$.

Taking the expressions of the function $f\left( \beta _{j},\beta
_{j}+\epsilon _{j}\right) $ from (\ref{jnub4}), for the expectation values
of the charge and current densities in the E-region we find%
\begin{equation}
\left\langle j^{\nu }\right\rangle ^{\mathrm{(b)}}|_{\alpha
_{0}=1/2}^{\alpha _{0}=-1/2}=\left( sm\right) ^{1-\frac{\nu }{2}}\frac{4Te}{%
\phi _{0}}\sum_{n=0}^{\infty }\mathrm{Re}\left[ \left( \frac{u_{n}}{r}%
\right) ^{\frac{\nu }{2}}\left[ I_{-1/2}(u_{n}a)-I_{1/2}(u_{n}a)\right]
\frac{K_{1/2}^{2}\left( u_{n}r\right) }{K_{1/2}(u_{n}a)}\right] ,
\label{jnudis}
\end{equation}%
with $\nu =0,2$. The combination of the modified Bessel functions is
simplified to $e^{-2u_{n}r}/(u_{n}r)$ and one gets%
\begin{equation}
\left\langle j^{\nu }\right\rangle ^{\mathrm{(b)}}|_{\alpha
_{0}=1/2}^{\alpha _{0}=-1/2}=\frac{2Te}{\phi _{0}r^{1+\frac{\nu }{2}}}%
\sum_{n=-\infty }^{+\infty }\left( \frac{sm}{u_{n}}\right) ^{1-\frac{\nu }{2}%
}e^{-2u_{n}r}.  \label{jnudis2}
\end{equation}%
Note that the right-hand side does not depend on the radius $a$ of the
boundary. In the special case of zero chemical potential, $\mu =0$, the
expectation values are odd functions of $\alpha _{0}$ and from here we get%
\begin{equation}
\left\langle j^{\nu }\right\rangle ^{\mathrm{(b)}}|_{\alpha _{0}=\pm
1/2}=\mp \frac{Te}{\phi _{0}r^{1+\frac{\nu }{2}}}\sum_{n=-\infty }^{+\infty
}\left( \frac{sm}{u_{0n}}\right) ^{1-\frac{\nu }{2}}e^{-2u_{0n}r}.
\label{jnudis2mu0}
\end{equation}%
In particular, for a massless field the charge density is continuous.

Alternative representations for the discontinuities in the E-region are
obtained from (\ref{jnudis2}) by applying the formula \cite{Bell09,Bell10}
\begin{equation}
\sum_{n=-\infty }^{+\infty }w_{n}^{2\beta -1}h_{\beta -1/2}(2rw_{n})=\frac{%
m^{2\beta }}{\sqrt{2\pi }T}\sum_{n=-\infty }^{+\infty }\cos (n\alpha
)h_{\beta }(m\sqrt{4r^{2}+n^{2}/T^{2}}),  \label{SF2}
\end{equation}%
with $w_{n}=\sqrt{(2\pi n+\alpha )^{2}T^{2}+m^{2}}$. For $\alpha =\pi -i\mu
/T$ the series in the left-hand side of (\ref{SF2}) coincides with the
series in (\ref{jnudis2}) for the charge density ($\nu =0$) in the case $%
\beta =0$ and for the current density ($\nu =2$) in the case $\beta =1$.
Applying the formula (\ref{SF2}) in these special cases, from (\ref{jnudis2}%
) one gets%
\begin{equation}
\left\langle j^{\nu }\right\rangle ^{\mathrm{(b)}}|_{\alpha
_{0}=1/2}^{\alpha _{0}=-1/2}=\frac{4esm}{\pi \phi _{0}r}\left( 2ms\right) ^{%
\frac{\nu }{2}}\sideset{}{'}{\sum}_{n=0}^{\infty }\left( -1\right) ^{n}\cosh
(n\mu /T)h_{\frac{\nu }{2}}(m\sqrt{n^{2}/T^{2}+4r^{2}}).  \label{jnudis3}
\end{equation}%
Combining this expression with the corresponding result (\ref{jnu0dis2}) for
the boundary-free geometry we see that in the E-region the total expectation
value (\ref{jnudec2}) is continuous at half-integer values of the parameter $%
\alpha $: $\left\langle j^{\nu }\right\rangle |_{\alpha _{0}=1/2}^{\alpha
_{0}=-1/2}=0$. The boundary conditions used above for the confinement of the
field separate the field fluctuations in the E-region from the cone apex and
magnetic flux. All the fermionic modes in that region are regular and the
total expectation values are continuous functions of the parameter $\alpha $.

\begin{figure}[tbph]
\begin{centering}
\begin{tabular}{cc}
\epsfig{figure=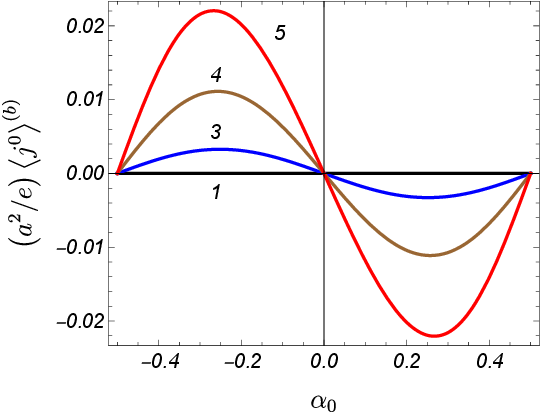,width=7.5cm,height=6.cm} & \quad{}\epsfig{figure=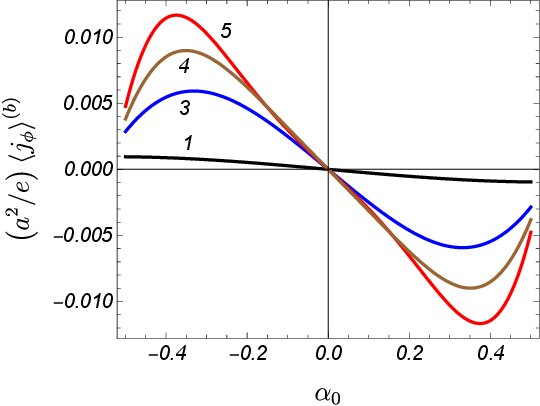,width=7.5cm,height=6.cm}\tabularnewline
\end{tabular}
\par\end{centering}
\caption{Expectation values of the charge (left panel) and current (right
panel) densities in the E-region for a massless field with a zero chemical
potential versus the parameter $\protect\alpha _{0}$. The graphs are plotted
for $r/a=1.5$, $Ta=0.5$ and the numbers near the curves correspond to the
values of $q$.}
\label{fig5}
\end{figure}

Figure \ref{fig5} displays the boundary-induced expectation values of the
charge and current densities for a massless field with zero chemical
potential in the E-region versus the parameter $\alpha _{0}$ for fixed
values of $r/a=1.5$ and $Ta=0.5$. The numbers near the curves correspond to
the values $q=1,3,4,5$ for the parameter describing the planar angle
deficit. For the considered example one has $\langle j^{0}\rangle ^{(0)}=0$
and both the total and boundary induced charge densities are continuous at
half integer values of the parameter $\alpha $ ($\langle j^{0}\rangle
_{\alpha _{0}=\pm 1/2}=\left\langle j^{0}\right\rangle _{\alpha _{0}=\pm
1/2}^{\mathrm{(b)}}=0$ for $\mu =0$). For the current density the part
independent of $a$ is given by (\ref{jphi0}) and it is discontinuous at half
integer values of $\alpha $. This discontinuity cancels the discontinuity of
the boundary-induced expectation value, depicted on the right panel of
Figure \ref{fig5}, and the total current density in the E-region is
continuous, in accordance with the asymptotic analysis given above.

\begin{figure}[tbph]
\begin{centering}
\begin{tabular}{cc}
\epsfig{figure=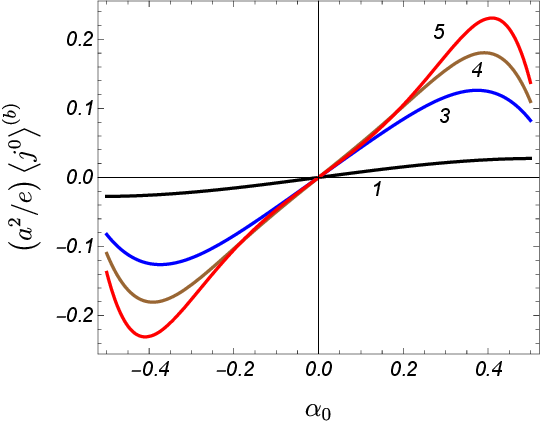,width=7.5cm,height=6.cm} & \quad{}\epsfig{figure=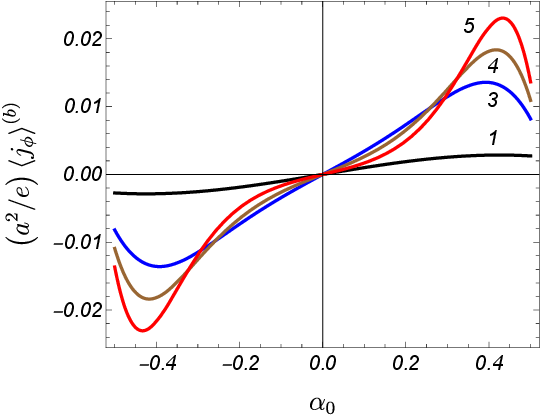,width=7.5cm,height=6.cm}\tabularnewline
\end{tabular}
\par\end{centering}
\caption{The same as in Figure \protect\ref{fig5} for the I-region. The
expectation values are evaluated for $r/a=0.5$ and for the same values of
the remaining parameters as in Figure \protect\ref{fig5}.}
\label{fig6}
\end{figure}

In the way similar to that for the E-region, the discontinuities of the
boundary induced expectation values in the I-region are expressed as%
\begin{align}
\left\langle j^{0}\right\rangle ^{\mathrm{(b)}}|_{\alpha _{0}=1/2}^{\alpha
_{0}=-1/2} &=\frac{8eT}{\phi _{0}r}\sum_{n=0}^{\infty }\mathrm{Re}\left[
\frac{sm\cosh \left( 2u_{n}r\right) -\left( u_{n}+sm\right) e^{2u_{n}a}}{%
u_{n}\left( \frac{u_{n}+sm}{u_{n}-sm}e^{4u_{n}a}+1\right) }\right] ,  \notag
\\
\left\langle j^{2}\right\rangle ^{\mathrm{(b)}}|_{\alpha _{0}=1/2}^{\alpha
_{0}=-1/2} &=-\frac{8eT}{\phi _{0}r^{2}}\sum_{n=0}^{\infty }\mathrm{Re}\left[
\frac{\sinh (2u_{n}r)}{\frac{u_{n}+sm}{u_{n}-sm}e^{4u_{n}a}+1}\right] .
\label{j02disI}
\end{align}%
For $\mu =0$ from here we find%
\begin{align}
\left\langle j^{0}\right\rangle ^{\mathrm{(b)}}|_{\alpha _{0}=\pm 1/2} &=\pm
\frac{4eT}{\phi _{0}r}\sum_{n=0}^{\infty }\frac{\left( u_{0n}+sm\right)
e^{2u_{0n}a}-sm\cosh \left( 2u_{0n}r\right) }{u_{0n}\left( \frac{u_{0n}+sm}{%
u_{0n}-sm}e^{4u_{0n}a}+1\right) },  \notag \\
\left\langle j^{2}\right\rangle ^{\mathrm{(b)}}|_{\alpha _{0}=\pm 1/2} &=\pm
\frac{4eT}{\phi _{0}r^{2}}\sum_{n=0}^{\infty }\frac{\sinh (2u_{0n}r)}{\frac{%
u_{0n}+sm}{u_{0n}-sm}e^{4u_{0n}a}+1}.  \label{j02Half}
\end{align}%
Unlike to the case of the E-region, the total expectation values in the
I-region are discontinuous at half-integer values of $\alpha $. By taking
into account that the expectation value $\langle j^{\nu }\rangle
^{(0)}|_{\alpha _{0}=1/2}^{\alpha _{0}=-1/2}$ is given by the right-hand
side of (\ref{jnudis2}) with the opposite sign, for the discontinuities in
the I-region we get%
\begin{align}
\left\langle j^{0}\right\rangle |_{\alpha _{0}=1/2}^{\alpha _{0}=-1/2} &=-%
\frac{4eT}{\phi _{0}r}\sum_{n=0}^{\infty }\mathrm{Re}\left[ \frac{sm\left(
\frac{u_{n}+sm}{u_{n}-sm}e^{2u_{n}(a-r)}-e^{-2u_{n}(a-r)}\right) +2\left(
u_{n}+sm\right) }{u_{n}\left( \frac{u_{n}+sm}{u_{n}-sm}%
e^{2u_{n}a}+e^{-2u_{n}a}\right) }\right] ,  \notag \\
\left\langle j^{2}\right\rangle |_{\alpha _{0}=1/2}^{\alpha _{0}=-1/2} &=-%
\frac{4eT}{\phi _{0}r^{2}}\sum_{n=0}^{\infty }\mathrm{Re}\left[ \frac{\frac{%
u_{n}+sm}{u_{n}-sm}e^{2u_{n}(a-r)}+e^{-2u_{n}(a-r)}}{\frac{u_{n}+sm}{u_{n}-sm%
}e^{2u_{n}a}+e^{-2u_{n}a}}\right] .  \label{j02tHalf}
\end{align}%
In particular, it can be checked that for $r=a$ one has $\left\langle
j^{0}\right\rangle |_{\alpha _{0}=1/2}^{\alpha _{0}=-1/2}=r\left\langle
j^{2}\right\rangle |_{\alpha _{0}=1/2}^{\alpha _{0}=-1/2}$, in accordance
with (\ref{joni}). The dependence of the boundary-induced charge and current
densities in the I region on the parameter $\alpha _{0}$ is presented in
Figure \ref{fig6} for $r/a=0.5$. The values of the remaining parameters are
the same as those for Figure \ref{fig5}. For the example presented in Figure %
\ref{fig6} we have taken $m=\mu =0$ and the charge density in the
boundary-free geometry vanishes. Hence, the graphs on the left panel also
present the total charge density. As explained above, both the boundary-free
and boundary induced contributions are discontinuous at half-integer values
of $\alpha $.

\subsection{Asymptotics and numerical analysis with respect to other
parameters}

\begin{figure}[tbph]
\begin{centering}
\begin{tabular}{cc}
\epsfig{figure=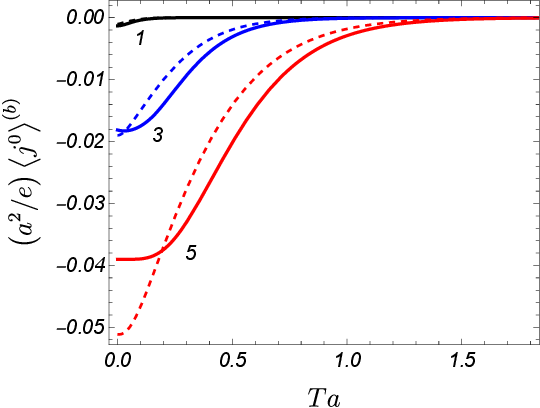,width=7.5cm,height=6.cm} & \quad{}\epsfig{figure=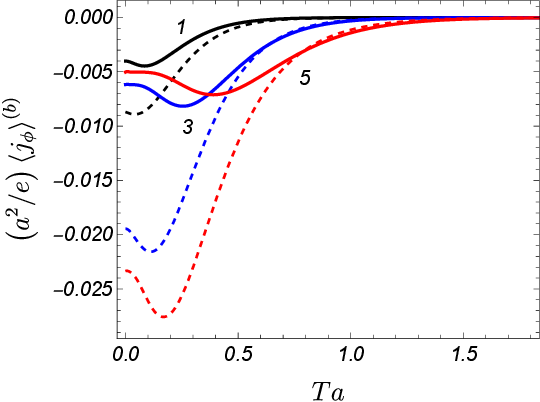,width=7.5cm,height=6.cm}\tabularnewline
\end{tabular}
\par\end{centering}
\caption{Expectation values of the charge (left panel) and current (right
panel) densities in the E-region for a massless field with zero chemical
potential versus the temperature. The graphs are plotted for $r/a=1.5$ and
the numbers near the curves correspond to the values of $q$. The full/dashed
curves correspond to $\protect\alpha _{0}=0.2$/$\protect\alpha _{0}=0.4$.}
\label{fig7}
\end{figure}

\begin{figure}[tbph]
\begin{centering}
\begin{tabular}{cc}
\epsfig{figure=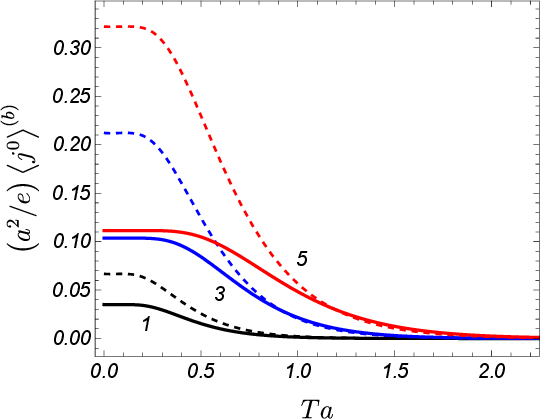,width=7.5cm,height=6.cm} & \quad{}\epsfig{figure=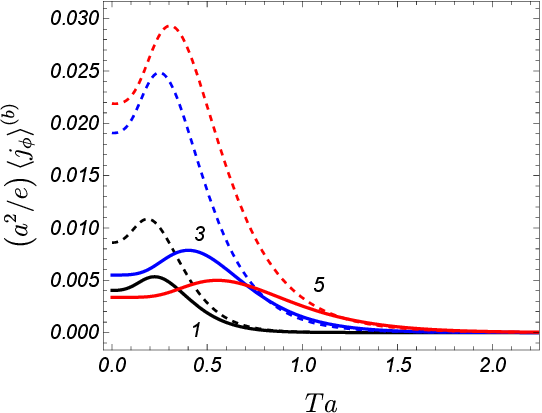,width=7.5cm,height=6.cm}\tabularnewline
\end{tabular}
\par\end{centering}
\caption{The same as in Figure \protect\ref{fig7} in the I-region for $%
r/a=0.5$.}
\label{fig8}
\end{figure}

In the high temperature limit, $T\gg m,|\mu |,1/(r-a)$, the contributions of
the $n=0$ terms dominate in (\ref{jnub4}) and (\ref{jnubTi}). In that limit
the boundary-induced expectation values of the charge and current densities
for a given $r$ are suppressed by the factor $e^{-2\pi T\left( r-a\right) }$%
. The corresponding asymptotics for the parts independent of $a$ have been
discussed in \cite{Bell16T}. For the topological part of the charge density
in a conical space without boundary one has $\left\langle j^{0}\right\rangle
_{\mathrm{t}}^{\left( 0\right) }\propto e^{-2\pi Tr\sin \left( \phi
_{0}/2\right) }$ for $\phi _{0}<\pi $. In the case $\phi _{0}>\pi $ the
decay is faster, like $\left\langle j^{0}\right\rangle _{\mathrm{t}}^{\left(
0\right) }\propto e^{-2\pi Tr}$. As a consequence, at high temperatures and
for points not too close to the boundary, the total charge density is
dominated by the Minkowskian part which behaves like $\left\langle
j^{0}\right\rangle _{\mathrm{M}}^{\left( 0\right) }\approx e\mu T\ln 2/\pi $%
, provided that the chemical potential is nonzero. In the case of $\mu =0$
the Minkowskian part vanishes. Similarly, for the current density
independent of $a$ one has $\left\langle j_{\phi }\right\rangle ^{\left(
0\right) }\propto e^{-2\pi Tr\sin \left( \phi _{0}/2\right) }$ for $\phi
_{0}<\pi $ and $\left\langle j_{\phi }\right\rangle ^{\left( 0\right)
}\propto e^{-2\pi Tr}$ in the region $\phi _{0}>\pi $. The zero temperature
limit of the expectation values is analyzed in appendix \ref{sec:AppB}. As
it has been shown, in the region $|\mu |>m$ for the chemical potential the
expectation values differ from the respective vacuum counterparts. The
corresponding asymptotic expressions are given by (\ref{jnuT0i2}) for
boundary-free geometry and by (\ref{jnuT0}) and (\ref{jnuT0i}) for the
boundary-induced contributions in the E- and I-regions, respectively. For a
massless field with zero chemical potential, the Figure \ref{fig7} displays
the expectation values of the charge (left panel) and current (right panel)
densities in the E-region versus the temperature. The graphs are plotted for
$r/a=1.5$ and for several values of the parameter $q$ indicated next to the
curves. The full and dashed curves correspond to $\alpha _{0}=0.2$ and $%
\alpha _{0}=0.4$, respectively. For $q=1$ the dependence of the charge
density on $\alpha _{0}$ is weak and for that case the full and dashed
curves nearly coincide with each other. In accordance with the asymptotic
analysis given above, the expectation values are suppressed exponentially at
high temperatures. The same dependence in the I-region is depicted in Figure %
\ref{fig8} for the value of the radial coordinate corresponding to $r/a=0.5$.

\begin{figure}[tbph]
\begin{centering}
\begin{tabular}{cc}
\epsfig{figure=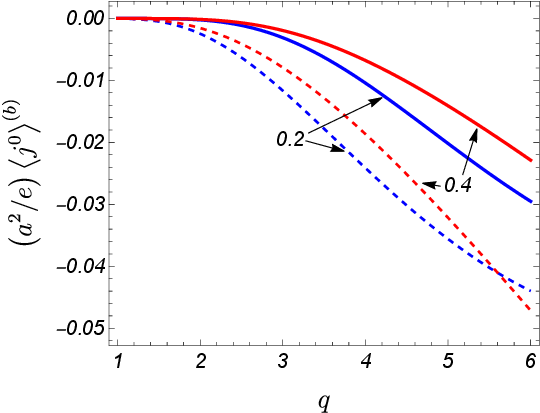,width=7.5cm,height=6.cm} & \quad{}\epsfig{figure=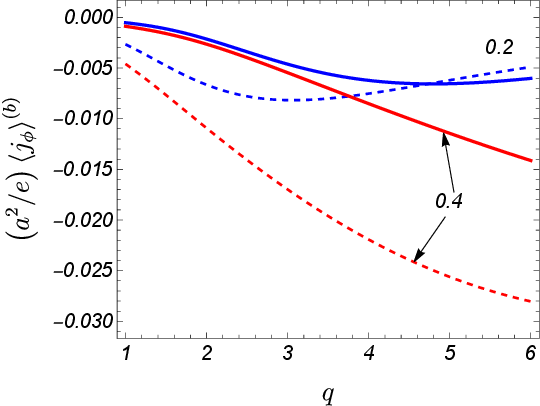,width=7.5cm,height=6.cm}\tabularnewline
\end{tabular}
\par\end{centering}
\caption{Expectation values of the charge (left panel) and current (right
panel) densities for a massless field with zero chemical potential as
functions of the parameter $q$. The graphs are plotted for $r/a=1.5$, $%
Ta=0.5 $ (full curves), $Ta=0.25$ (dashed curves) and the numbers near the
curves are the corresponding values of $\protect\alpha _{0}$.}
\label{fig9}
\end{figure}

\begin{figure}[tbph]
\begin{centering}
\begin{tabular}{cc}
\epsfig{figure=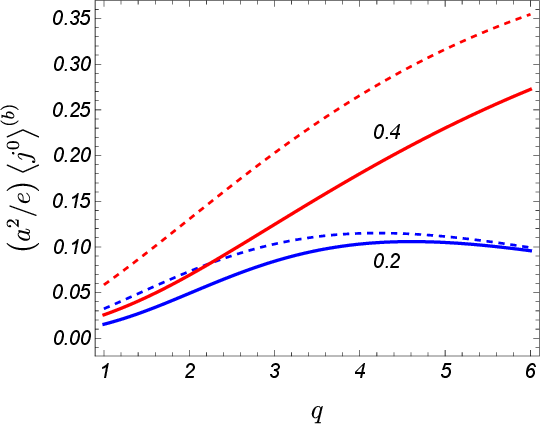,width=7.5cm,height=6.cm} & \quad{}\epsfig{figure=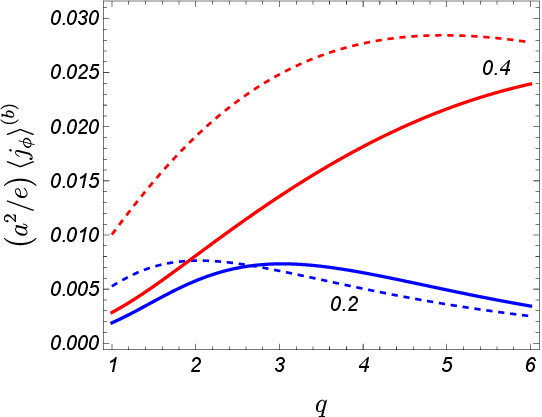,width=7.5cm,height=6.cm}\tabularnewline
\end{tabular}
\par\end{centering}
\caption{The same as in figure \protect\ref{fig9} for the I-region. For the
radial coordinate the value $r/a=0.5$ is taken.}
\label{fig10}
\end{figure}

Now we turn to the dependence on the planar angle deficit. For the
expectation values in the E-region it is displayed in Figure \ref{fig9} for
the parameters $r/a=1.5$, $Ta=0.5$ (full curves) and $Ta=0.25$ (dashed
curves). The numbers near the curves correspond to the values $\alpha
_{0}=0.2$, $0.4$ for the parameter describing the magnetic flux. The same
dependence in the I-region for the radial coordinate corresponding to $%
r/a=0.5$ is presented in Figure \ref{fig10}. Note that the behavior of the
boundary-induced expectation values as functions of $q$ essentially depends
on the value of $\alpha _{0}$.

\begin{figure}[tbph]
\begin{centering}
\begin{tabular}{cc}
\epsfig{figure=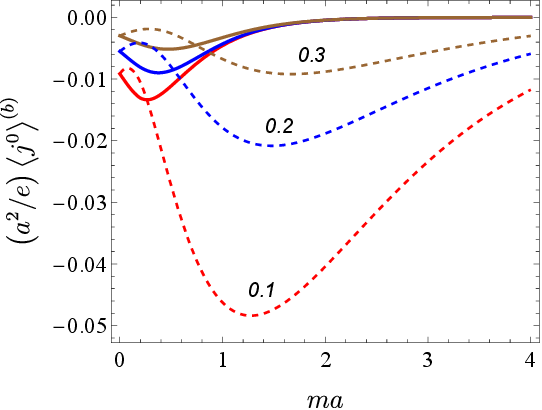,width=7.5cm,height=6.cm} & \quad{}\epsfig{figure=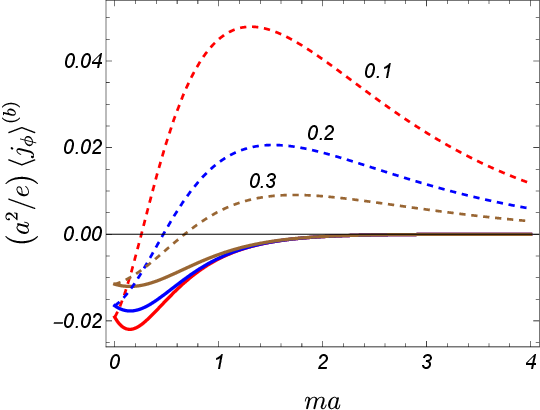,width=7.5cm,height=6.cm}\tabularnewline
\end{tabular}
\par\end{centering}
\caption{Boundary-induced expectation values of the charge (left panel) and
current (right panel) densities in the E-region as functions of the field
mass in the case of a zero chemical potential. The full and dashed curves
correspond to $s=1$ and $s=-1$, respectively. The numbers near the curves
are the values of $Ta$ and for remaining parameters we have taken $q=2.5$, $%
\protect\alpha _{0}=0.4$, and $r/a=1.5$.}
\label{fig11}
\end{figure}

\begin{figure}[tbph]
\begin{centering}
\begin{tabular}{cc}
\epsfig{figure=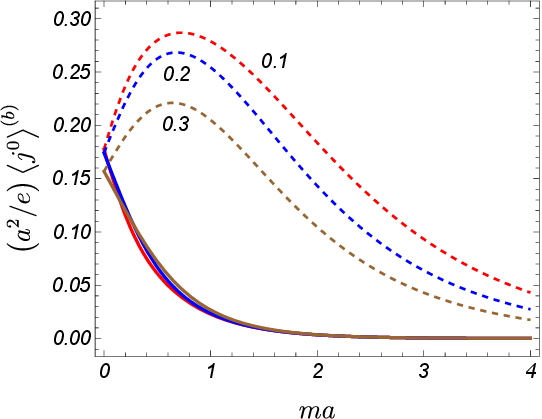,width=7.5cm,height=6.cm} & \quad{}\epsfig{figure=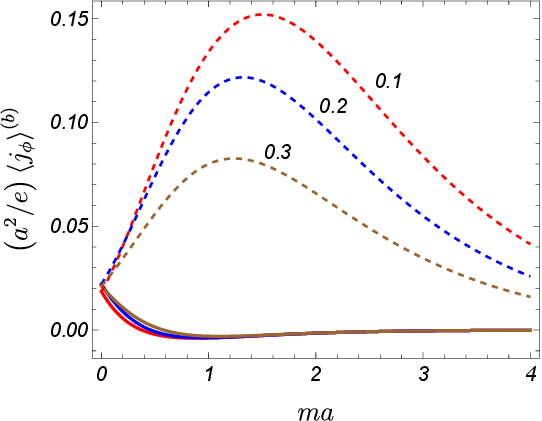,width=7.5cm,height=6.cm}\tabularnewline
\end{tabular}
\par\end{centering}
\caption{The same as in Figure \protect\ref{fig11} for the I-region with the
value of the radial coordinate corresponding to $r/a=0.5$.}
\label{fig12}
\end{figure}

The numerical examples presented in the discussion above are given for
massless fields with zero chemical potential. For those fields the
expectation values do not depend on the parameter $s$. In Figure \ref{fig11}%
, as another numerical example, we display the dependence of the
boundary-induced expectation values in the E-region on the mass for a field
with zero chemical potential. The graphs are plotted for $r/a=1.5$ and with
the parameters describing the deficit angle and the magnetic flux having the
values $q=2.5$ and $\alpha _{0}=0.4$. The full (dashed) curves present the
case $s=1$ ($s=-1$) and the numbers near the curves indicate the values of $%
Ta$. The corresponding graphs for the I-region, evaluated at $r/a=0.5$, are
depicted in Figure \ref{fig12}. As seen from the graphs the dependence on
the mass, in general, is not monotonic. For fields with $s=-1$ the
expectation values for massive fields can be essentially larger compared to
the massless case.

\section{Expectation values for the fields realizing two representations of
the Clifford algebra}

\label{sec:Rep2}

The inequivalent irreducible representations of the Clifford algebra in
3-dimensional spacetime described by (\ref{linel}) are realized by two sets
of $\gamma $-matrices specified by $s=+1$ and $s=-1$ given as $\gamma
_{(s)}^{\mu }=(\gamma ^{0},\gamma ^{1},\gamma _{(s)}^{2})$ with $\gamma
_{(s)}^{2}=-is\gamma ^{0}\gamma ^{1}/r$ in the coordinates $(t,r,\phi )$.
Note that the matrix $\gamma _{(+1)}^{2}$ coincides with $\gamma ^{2}$ used
in the discussion above. The spinor fields realizing the inequivalent
representations we will denote by $\psi _{(s)}$. The corresponding
Lagrangian densities read $L_{s}=\bar{\psi}_{(s)}(i\gamma _{(s)}^{\mu
}D_{\mu }-m)\psi _{(s)}$. For massive fields the model with given $s$ is not
invariant under the parity and time-reversal transformations. In the absence
of magnetic field we can construct fermionic models invariant under those
transformations combining the separate field in the Lagrangian density $%
L=\sum_{s=\pm 1}L_{s}$. The corresponding problem can be mapped to the
problem we have considered in the previous text. In order to realize that
let us introduce new fields $\psi _{(s)}^{\prime }$ in accordance with $\psi
_{(s)}^{\prime }=\gamma ^{0}\gamma ^{(1-s)/2}\psi _{(s)}$. In terms of those
fields the Lagrangian density is rewritten as $L=\sum_{s=\pm 1}\bar{\psi}%
_{(s)}^{\prime }(i\gamma ^{\mu }D_{\mu }-sm)\psi _{(s)}^{\prime }$, with the
same set of gamma matrices as in (\ref{Dirac}).

Let assume that the fields $\psi _{(s)}$ in the initial representation obey
the boundary conditions%
\begin{equation}
(1+\eta _{(s)}in_{\mu }\gamma _{(s)}^{\mu })\psi _{(s)}=0,\;r=a,  \label{BCs}
\end{equation}%
with $\eta _{(s)}=\pm 1$. The parameter $\eta _{(s)}$ may differ for the
fields $s=+1$ and $s=-1$. In terms of new fields $\psi _{(s)}^{\prime }$,
the boundary conditions are transformed to
\begin{equation}
\left( 1+s\eta _{(s)}in_{\mu }\gamma ^{\mu }\right) \psi _{(s)}^{\prime
}=0,\;r=a.  \label{BCsp}
\end{equation}%
Introducing the current densities for the fields $\psi _{(s)}$ and $\psi
_{(s)}^{\prime }$ by the standard formulas $j_{(s)}^{\nu }=e\bar{\psi}%
_{(s)}\gamma _{(s)}^{\nu }\psi _{(s)}$ and $j_{(s)}^{\prime \nu }=e\bar{\psi}%
_{(s)}^{\prime }\gamma _{(s)}^{\nu }\psi _{(s)}^{\prime }$, we obtain the
following relations for the expectation values of the charge and current
densities:%
\begin{equation}
\langle j_{(s)}^{\nu }\rangle =s^{\nu /2}\langle j_{(s)}^{\prime \nu
}\rangle ,  \label{relss}
\end{equation}%
for $\nu =0,2$. The fields $\psi _{(s)}^{\prime }$ obey the equation (\ref%
{Dirac}) and the boundary conditions (\ref{BCsp}).

From the discussion at the end of Section \ref{sec:Inter} it follows that
the expressions for the expectation values $\langle j_{(s)}^{\prime \nu
}\rangle $ are related to the charge and current densities presented in the
previous sections by the formula%
\begin{equation}
\langle j_{(s)}^{\prime \nu }\rangle =(s\eta _{(s)})^{1-\nu /2}\left\langle
j^{\nu }\right\rangle _{(\eta _{(s)},s\eta _{(s)}\mu ,+1)}.  \label{relsp}
\end{equation}%
By using the relation (\ref{relss}), the charge ($\nu =0$) and current ($\nu
=2$) densities for the initial fields $\psi _{(s)}$ are expressed as
\begin{equation}
\langle j_{(s)}^{\nu }\rangle =s\eta _{(s)}^{1-\nu /2}\left\langle j^{\nu
}\right\rangle _{(\eta _{(s)},s\eta _{(s)}\mu ,+1)}.  \label{rels}
\end{equation}%
For the total expectation values in models with Lagrangian density $%
L=\sum_{s=\pm 1}L_{s}$ one gets
\begin{equation}
\langle J^{\nu }\rangle =\sum_{s=\pm 1}s\eta _{(s)}^{1-\nu /2}\left\langle
j^{\nu }\right\rangle _{(\eta _{(s)},s\eta _{(s)}\mu ,+1)}.  \label{Jnu}
\end{equation}%
From here it follows that if the fields $\psi _{(s)}$ obey the same boundary
condition ($\eta _{(+1)}=\eta _{(-1)}$) then%
\begin{equation}
\langle J^{\nu }\rangle =\eta _{(+1)}^{1-\nu /2}\sum_{s=\pm 1}s\left\langle
j^{\nu }\right\rangle _{(\eta _{(+1)},s\eta _{(+1)}\mu ,+1)},
\label{Jnusame}
\end{equation}%
and the total charge and current densities are zero for $\mu =0$. In this
special case, the expectation values are odd functions of the chemical
potential. Now, by taking into account that $\left\langle j^{\nu
}\right\rangle \left( -\alpha _{0},-\mu \right) =-\left\langle j^{\nu
}\right\rangle \left( \alpha _{0},\mu \right) $, we conclude that the
expectation value $\langle J^{\nu }\rangle $ is an odd function of $\alpha
_{0}$. For the fields $\psi _{(s)}$ obeying different boundary conditions ($%
\eta _{(+1)}=-\eta _{(-1)}$, the fields $\psi _{(s)}^{\prime }$ obey the
same boundary conditions), the formula (\ref{Jnu}) is reduced to
\begin{equation}
\langle J^{\nu }\rangle =\eta _{(+1)}^{1-\nu /2}\left[ \left\langle j^{\nu
}\right\rangle _{(\eta _{(+1)},\eta _{(+1)}\mu ,+1)}-(-1)^{1-\nu
/2}\left\langle j^{\nu }\right\rangle _{(-\eta _{(+1)},\eta _{(+1)}\mu ,+1)}%
\right] .  \label{Jnudif}
\end{equation}%
For a massless field one has $\left\langle j^{\nu }\right\rangle _{(-\eta
_{(+1)},\eta _{(+1)}\mu ,+1)}=\left\langle j^{\nu }\right\rangle _{(\eta
_{(+1)},\eta _{(+1)}\mu ,+1)}$ and the total current density in (\ref{Jnudif}%
) vanishes, $\langle J^{2}\rangle =0$. For the charge density one gets $%
\langle J^{0}\rangle =2\eta _{(+1)}\left\langle j^{0}\right\rangle _{(\eta
_{(+1)},\eta _{(+1)}\mu ,+1)}$. Note that, in general, the fields $\psi
_{(+1)}$ and $\psi _{(-1)}$ may have different masses and in the
corresponding problems there will be no cancellations between the separate
contributions of those field into the total expaectation values.

The spinor fields $\psi _{(s)}$ appear in the Dirac model describing the
long wavelength properties of the electronic subsystem of graphene. The
parameter $s$ enumerates the valley degrees of freedom and corresponds to
the points $\mathbf{K}_{+}$ and $\mathbf{K}_{-}$ of the graphene Brillouin
zone. The spinors are presented as $\psi _{(s)}=(\psi _{s,AS},\psi
_{s,BS})^{T}$, where the upper and lower components are expressed in terms
of the electron wave functions on the A and B sites of the graphene lattice
and $S=+1$ and $S=-1$ correspond to additional degree of freedom related to
spin. The Dirac equation describing the dynamics of the subsystem of $\pi $%
-electrons can be expressed in terms of four-component spinors $\Psi
_{S}=(\psi _{(+1)},\psi _{(-1)})^{T}$ introducing the $4\times 4$ Dirac
matrices $\gamma _{(4)}^{\mu }=\sigma _{\mathrm{P}3}\otimes \gamma ^{\mu }$
and the related spin connection $\Gamma _{\mu }^{(4)}=I\otimes \Gamma _{\mu
} $ with $\sigma _{\mathrm{P}3}=\mathrm{diag}(1,-1)$ and $I=\mathrm{diag}%
(1,1)$. In the part containing the spatial derivatives, the speed of light
is replaced by the Fermi velocity of electrons $v_{\mathrm{F}}\approx
7.9\times 10^{7}$cm/s. The mass $m$ in the corresponding Dirac equation is
expressed in terms of the energy gap $\Delta $ by the relation $m=\Delta /v_{%
\mathrm{F}}^{2}$ and for the corresponding Compton wavelength one has (in
standard units) $\lambda _{\mathrm{C}}=\hbar v_{\mathrm{F}}/\Delta $.
Several mechanisms have been considered in the literature for the generation
of the gap in the range $1\,\mathrm{meV}\lesssim \Delta \lesssim 1\,\mathrm{%
eV}$. In translating the results given above for graphene nanocones the
replacement $m\rightarrow 1/\lambda _{\mathrm{C}}$ should be done and an
additional factor $v_{\mathrm{F}}$ appears in the definition of the spatial
components of the current density operator. In graphene nanocones the
possible values of the planar angle deficit are dictated by the symmetry of
the hexagonal lattice and are given by $2\pi -\phi _{0}=\pi n_{c}/3$ with $%
n_{c}=1,2,\ldots ,5$ (for effects of conical topology on the electronic
properties of graphene see, e.g., \cite{Lamm00}-\cite{Matt23} and references
therein). The graphitic cones with these values of opening angle have been
experimentally observed \cite{Kris97}.

\section{Conclusion}

\label{sec:Conc}

We have investigated the impact of a circular edge of a 2D conical space on
the expectation values of the charge and current densities for a massive
spinor field in thermal equilibrium at temperature $T$. In the presence of
an external gauge field the Dirac equation is presented as (\ref{Dirac}),
where the parameter $s$ in front of the mass term describes the two
inequivalent irreducible representations of the Clifford algebra. For
conical geometry, the general case of the planar angle deficit is
considered. The edge at the radial coordinate $r=a$ divides the space into
two causally separated regions: I- and E-regions for $r<a$ and $r>a$,
respectively. On the edge the field operator is constrained by the boundary
condition (\ref{BCMIT}). In the I-region that leads to the discretization of
the radial quantum number $\gamma $ with the eigenvalues being the roots of
the equation (\ref{modesi}). In the E-region the spectrum of $\gamma $ is
continuous. The complete set of spinor modes is given by (\ref{psiIE}) with
the radial functions expressed as (\ref{Znu}) in the I- and E-regions. The
expectation values of the charge and current densities are decomposed into
three parts corresponding to the VEVs and contributions coming from
particles and antiparticles (see (\ref{jnudec})). They are presented in the
form of sums over the spinorial modes. We have explicitly separated from
those mode sums the edge induced contributions. In the I-region that is done
by the application of the Abel-Plana type summation formula (\ref{SumAP}) to
the series over the eigenvalues of the radial quantum number. In the
corresponding integral representation of the edge-induced expectation values
the explicit knowledge of those eigenvalues is not required. Note that the
current density considered above is the analog of the persistent currents in
normal metal rings predicted in \cite{Butt83} and experimentally confirmed
in a number of papers (see, e.g., \cite{Bluh09,Bles09}). That type of
currents may appear also in a number of other mesoscopic condensed matter
system such as graphene and topological insulator rings (see, for example,
\cite{Lin98,Lati03,Rech07,Wu10,Mich11} and references given in \cite%
{Saha24b,Arau22}).

The mean radial current density vanishes and the finite temperature
boundary-induced contributions in the expectation values of the charge and
azimuthal current densities are given by the expressions (\ref{jnub4}) in
the E-region and by (\ref{jnubTi}) in the I-region, with $u_{n}$ from (\ref%
{un}). The influence of the magnetic flux on the expectation values is of
Aharonov-Bohm type effect and, as expected, they are periodic with respect
to the magnetic flux with the period of flux quantum. In addition, the
charge and current densities are odd functions under the reflection $(\alpha
_{0},\mu )\rightarrow $ $(-\alpha _{0},-\mu )$ in the space of the parameter
$\alpha _{0}$, determining the fractional part of the magnetic flux (in
units of flux quantum), and the chemical potential. In particular, for $%
\alpha _{0}=0$ the VEVs vanish and the net charge and current densities for
the nonzero chemical potential are purely finite temperature effect. For the
special case of zero chemical potential the arguments of the modified Bessel
functions in general formulas are real and the corresponding expressions are
reduced to (\ref{jnub0}) and (\ref{jnubtoti0}) in the E- and I-regions,
respectively. For a massless field the further simplifications lead to the
representations (\ref{jnubm0}) and (\ref{jnubm0i}). An important difference
from the fermionic condensate, considered in \cite{Saha19} (see also \cite%
{Saha19FC} for the corresponding problem in the case of two circular
boundaries at zero temperature), is that the charge and current densities
are finite on the boundary. The corresponding thermal contributions are
connected by the relations (\ref{jon}) and (\ref{joni}) for the E- and
I-regions, respectively. Having the expectation values for the boundary
condition (\ref{BCMIT}) one can obtain the corresponding expressions for the
charge and current densities in the case of the condition that differs from (%
\ref{BCMIT}) by the sign of the term containing the normal vector (the
boundary condition (\ref{BCeta}) with $\eta =-1$). The expectation values in
problems with two sets of parameters $(s,\mu ,\eta )$ and $(-s,-\mu ,-\eta )$
are connected by the relation (\ref{jnurel}). Another extension of the
obtained results corresponds to fields with more general periodicity
condition with respect to the angular coordinate $\phi $. As it has been
shown at the end of Section \ref{sec:Inter}, the nontrivial phase in the
respective quasiperiodicity condition can be interpreted in terms of the
effective magnetic flux and vice versa.

General formulas for the expectation values are rather complicated. To
clarify the behavior of the charge and current densities we have considered
different asymptotic regions of the parameters. In the E-region, at large
distances from the boundary and for a massive field the expectation value of
the charge density is dominated by the Minkowskian part $\left\langle
j^{0}\right\rangle _{\mathrm{M}}^{\left( 0\right) }$ which does not depend
on $r$. Contrary to that, the expectation value of the current density
decays exponentially at large distances. In the I-region and for small
values of the radial coordinate $r$ the boundary-induced current density
tends to zero, whereas the boundary-induced charge density tends to zero for
$2|\alpha _{0}|<1-1/q$ and diverges in the range $2|\alpha _{0}|>1-1/q$. In
the last case, near the cone apex, for a massive field the expectation
values of the charge and current densities in the boundary-free geometry are
dominated by the vacuum parts. If the radial coordinate is held constant and
the circular boundary is brought closer to the apex, then in the limit of
small values of $a$, the boundary-induced expectation values in the E-region
will tend to zero for $|\alpha _{0}|<1/2$.

The expectation values at half-integer values of the ratio of magnetic flux
to flux quantum (corresponding to half-integer values of the parameter $%
\alpha $) are obtained from the general formulas in the limit $\alpha
_{0}\rightarrow \pm 1/2$. The limiting values in the boundary-free geometry
are given by (\ref{jnu0dismu0}) and the corresponding charge and current
densities are discontinuous at half-integer values of $\alpha $. Similar
discontinuities are present also in the boundary-induced parts and for $%
\alpha _{0}=\pm 1/2$ they come from the contribution of the fermionic mode
with $j=\mp 1/2$. They are given by the formulas (\ref{jnudis2}) and (\ref%
{j02disI}) in the E- and I-regions, respectively. We have shown that in the
E-region the discontinuities in the parts $\langle j^{\nu }\rangle ^{(0)}$
and $\left\langle j^{\nu }\right\rangle ^{\mathrm{(b)}}$ are canceled out in
the total expectation value $\langle j^{\nu }\rangle =\langle j^{\nu
}\rangle ^{(0)}+\langle j^{\nu }\rangle ^{\mathrm{(b)}}$ and the latter is
continuous at half-integer values of $\alpha $. This is related to the fact
that all the spinorial modes in the E-region are regular. For zero chemical
potential the expectation values are odd functions of $\alpha _{0}$ and from
the expressions for discontinuities one can find the corresponding values at
$\alpha _{0}=\pm 1/2$ (see Eqs. (\ref{jnudis2mu0}) and (\ref{j02Half})).

At high temperatures, the boundary-induced expectation values of the charge
and current densities for a given $r$ are suppressed by the factor $e^{-2\pi
T\left( r-a\right) }$. In that limit and for nonzero chemical potential the
total charge density is dominated by the Minkowskian part with the high
temperature behavior $\left\langle j^{0}\right\rangle _{\mathrm{M}}^{\left(
0\right) }\propto \mu T$. For the current density independent of $a$ one has
$\left\langle j_{\phi }\right\rangle ^{\left( 0\right) }\propto e^{-2\pi
Tr\sin \left( \phi _{0}/2\right) }$ for $\phi _{0}<\pi $ and $\left\langle
j_{\phi }\right\rangle ^{\left( 0\right) }\propto e^{-2\pi Tr}$ in the
region $\phi _{0}>\pi $. The zero temperature limit of the expectation
values depends on the relative values of the mass and chemical potential. In
the range $|\mu |<m$ the expectation values tend to the corresponding VEVs
in the limit $T\rightarrow 0$. For the values of the chemical potential in
the range $|\mu |>m$ the contributions of particles/antiparticles survive in
the zero temperature limit. Those contributions come from particles for $\mu
>0$ and antiparticles for $\mu <0$. They are given by the expressions (\ref%
{jnuT0}) and (\ref{jnuT0i}).

In Section \ref{sec:Rep2} we have shown that the expectation values of the
charge and current densities for fields realizing the second inequivalent
irreducible representation of the Clifford algebra are obtained from the
formulas presented in the preceding sections. For fields with boundary
conditions (\ref{BCs}) the corresponding relation is given by (\ref{rels}).
In fermionic models invariant under parity and time-reversal transformations
(in the absence of magnetic fields), combining fields in two inequivalent
representations, the total expectation values are expressed by (\ref{Jnu}).
We have considered different combinations of boundary conditions for
separate fields. In the long wavelength description of the eletronic
subsystem in graphene, based on the Dirac model, the parameter $s$
corresponds to the valley degrees of freedom. At the end of Section \ref%
{sec:Rep2} we have discussed applications in graphene nanocones described by
the effective field-theoretical model.

\section*{Acknowledgments}

The work was supported by the grants No. 21AG-1C047 and 24FP-3B021 of the
Higher Education and Science Committee of the Ministry of Education,
Science, Culture and Sport RA.

\appendix

\section{Transformation of the expectation values in the E-region}

\label{sec:appA}

In this appendix we present the details of the transformations for the
thermal expectation values in the E-region. They will be presented
separately for nonzero and zero chemical potentials.

\subsection{Nonzero chemical potential}

For the case of nonzero chemical potential, $\mu \neq 0$, the integrands in (%
\ref{jnupmb1}) have branch points $\gamma =\pm im$ and simple poles $\gamma
=\gamma _{n}^{(\lambda )}$, $n=0,\pm 1,\pm 2,\ldots $, in the complex plane $%
\gamma $. The poles, given by (\ref{gamn}), are located in the right
half-plane for $\lambda \mu >0$ and in the left half-plane for $\lambda \mu
<0$. Other features of the locations for the poles are described in Section %
\ref{sec:Inter}.

By taking into account that for $r>a$ and $l=1$ ($l=2$) the integrands in (%
\ref{jnupmb1}) exponentially decrease in the upper (lower) half-plane for $%
|\gamma |\gg 1$, in the integrals over $\gamma $ we rotate the contour by
the angle $\pi /2$ ($-\pi /2$). The integrals over $\gamma \in \lbrack
0,\infty )$ are transformed to the integrals over the imaginary axis of the
complex plane $\gamma $ (over $\gamma \in \lbrack 0,i\infty )$ for $l=1$ and
$\gamma \in \lbrack 0,-i\infty )$ for $l=2$). For $\lambda \mu >0$ the parts
involving the residues at the poles $\gamma =\gamma _{n}^{(\lambda )}$ have
to be added. Using the relation $\sqrt{\left( \pm i\gamma \right) ^{2}+m^{2}}%
=\sqrt{m^{2}-\gamma ^{2}}$ for $0\leq \gamma \leq m$ and the definition (\ref%
{Fbar}), it can be seen that the integrals over the intervals $[0,im]$ and $%
[0,-im]$ cancel each other. Introducing the modified Bessel functions in the
integrals over the intervals $[im,i\infty )$ and $[-im,-i\infty )$, and also
in the residue terms, we obtain the representation%
\begin{align}
\left\langle j^{0}\right\rangle _{T\lambda }^{\mathrm{(b)}}& =\frac{e}{\pi
\phi _{0}}\sum_{j}\sum_{\chi =\pm 1}\left\{ -\int_{m}^{\infty }dx\,x\,%
\mathrm{Re}\left[ \frac{\tilde{I}_{\beta _{j}}^{\left( \lambda \right)
}\left( xa\right) }{\tilde{K}_{\beta _{j}}^{\left( \lambda \right) }\left(
xa\right) }\frac{K_{\alpha _{j}-\chi \epsilon _{j}/2}^{2}\left( xr\right) }{%
e^{\beta \left( i\sqrt{x^{2}-m^{2}}-\lambda \mu \right) }+1}\frac{sm+\lambda
\chi i\sqrt{x^{2}-m^{2}}}{\sqrt{x^{2}-m^{2}}}\right] \right.  \notag \\
& \left. +2\pi T\,\theta \left( \lambda \mu \right) \sum_{n=0}^{\infty }%
\mathrm{Re}\left[ \frac{\tilde{I}_{\beta _{j}}^{\left( \lambda \right)
}(u_{n}^{\left( \lambda \right) }a)}{\tilde{K}_{\beta _{j}}^{\left( \lambda
\right) }(u_{n}^{\left( \lambda \right) }a)}\left[ \chi \mu +sm+\lambda \chi
i\pi \left( 2n+1\right) T\right] K_{\alpha _{j}-\chi \epsilon
_{j}/2}^{2}(u_{n}^{\left( \lambda \right) }r)\right] \right\} ,
\label{jnub0T}
\end{align}%
for the boundary-induced contribution in the thermal part of the charge
density. Here, the notations (\ref{unlam}) and (\ref{Ftild}) have been used.
In the similar way, for the corresponding azimuthal current density one gets%
\begin{align}
\left\langle j^{2}\right\rangle _{T\lambda }^{\mathrm{(b)}}& =\frac{e}{\pi
\phi _{0}}\sum_{j}\left\{ -\int_{m}^{\infty }dx\,\frac{xU_{2,\beta
_{j}}^{K}\left( xr\right) }{\sqrt{x^{2}-m^{2}}}\mathrm{Re}\left[ \frac{%
\tilde{I}_{\beta _{j}}^{\left( \lambda \right) }\left( xa\right) }{\tilde{K}%
_{\beta _{j}}^{\left( \lambda \right) }\left( xa\right) }\frac{1}{e^{\beta
\left( i\sqrt{x^{2}-m^{2}}-\lambda \mu \right) }+1}\right] \right.  \notag \\
& \left. +2\pi T\,\theta \left( \lambda \mu \right) \sum_{n=0}^{\infty }%
\mathrm{Re}\left[ \frac{\tilde{I}_{\beta _{j}}^{\left( \lambda \right)
}(u_{n}^{\left( \lambda \right) }a)}{\tilde{K}_{\beta _{j}}^{\left( \lambda
\right) }(u_{n}^{\left( \lambda \right) }a)}U_{2,\beta
_{j}}^{K}(u_{n}^{\left( \lambda \right) }r)\right] \right\} .  \label{jnub2}
\end{align}%
Note that $\gamma _{n}^{(\lambda )}=iu_{n}^{(\lambda )}$ and $%
u_{n}^{(-)}=u_{n}^{(+)\ast }$. By taking into account the relations (\ref%
{RelFtild}), the expressions (\ref{jnub0T}) and (\ref{jnub2}) are
transformed to (\ref{jnub3}).

\subsection{Zero chemical potential}

In the case of zero chemical potential, $\mu =0$, for the poles of the
integrands in (\ref{jnupmb1}) located in the upper half-plane of complex
variable $\gamma $ we have $\gamma =\gamma _{n}=iu_{0n}$,$\;n=0,1,2,\ldots $%
, with $u_{0n}$ defined in (\ref{u0n}). The poles in the lower half-plane
are expressed as $\gamma _{n}=\gamma _{-n-1}^{\ast }$ with$\;n=\ldots ,-2,-1$%
. All the poles are located on the imaginary axis. Similar to the case $\mu
\neq 0$, in the separate integrals of (\ref{jnupmb1}) with $l=1,2$ we rotate
the contours by the angle $(-1)^{l-1}\pi /2$ bypassing the poles $\gamma
_{n} $ on the imaginary axis by semi-circular arcs $C_{\rho }(\gamma _{n})$
of small radius $\rho $. These arcs pass around the poles clockwise in the
upper half-plane and counter-clockwise in the lower half-plane. It can be
seen that the sum of the integrals along $C_{\rho }(\gamma _{n})$ with $%
n=\ldots ,-2,-1$ is the complex conjugate of the sum of the integrals with $%
n=0,1,2,\ldots $. In the limit $\rho \rightarrow 0$ the sum of the integrals
over the straight segments gives the principal values of the integrals over
the positive and negative imaginary semiaxes (denoted here as p.v.). The
integrals over the intervals $[0,im]$ and $[0,-im]$ cancel each other,
whereas the integral over $[-im,-i\infty )$ is the complex conjugate of the
integral over $[im,i\infty )$. Introducing the modified Bessel functions, we
get
\begin{align}
\left\langle j^{0}\right\rangle _{T\lambda }^{\mathrm{(b)}}& =\frac{e}{\pi
\phi _{0}}\sum_{j}\sum_{\chi =\pm 1}\left\{ -\mathrm{p.v.}\int_{m}^{\infty
}d\gamma \,\frac{\gamma K_{\alpha _{j}-\chi \epsilon _{j}/2}^{2}\left(
\gamma r\right) }{\sqrt{\gamma ^{2}-m^{2}}}\,\mathrm{Re}\left[ \frac{\bar{I}%
_{\beta _{j}}^{\left( \lambda \right) }\left( \gamma a\right) }{\bar{K}%
_{\beta _{j}}^{\left( \lambda \right) }\left( \gamma a\right) }\frac{%
sm+\lambda \chi i\sqrt{\gamma ^{2}-m^{2}}}{e^{i\beta \sqrt{\gamma ^{2}-m^{2}}%
}+1}\right] \right.  \notag \\
& +\pi T\sum_{n=0}^{\infty }\mathrm{Re}\left[ \frac{\bar{I}_{\beta
_{j}}^{\left( \lambda \right) }\left( u_{0n}a\right) }{\bar{K}_{\beta
_{j}}^{\left( \lambda \right) }\left( u_{0n}a\right) }\left[ sm+\lambda \chi
i\pi \left( 2n+1\right) T\right] K_{\alpha _{j}-\chi \epsilon
_{j}/2}^{2}\left( u_{0n}r\right) \right] ,  \notag \\
\left\langle j^{2}\right\rangle _{T\lambda }^{\mathrm{(b)}}& =\frac{e}{\pi
\phi _{0}}\sum_{j}\left\{ -\mathrm{p.v.}\int_{m}^{\infty }d\gamma \,\frac{%
\gamma U_{2,\beta _{j}}^{K}\left( \gamma r\right) }{\sqrt{\gamma ^{2}-m^{2}}}%
\mathrm{Re}\left[ \frac{\bar{I}_{\beta _{j}}^{\left( \lambda \right) }\left(
\gamma a\right) }{\bar{K}_{\beta _{j}}^{\left( \lambda \right) }\left(
\gamma a\right) }\frac{1}{e^{i\beta \sqrt{\gamma ^{2}-m^{2}}}+1}\right]
\right.  \notag \\
& \left. +\pi T\sum_{n=0}^{\infty }\mathrm{Re}\left[ \frac{\bar{I}_{\beta
_{j}}^{\left( \lambda \right) }\left( u_{0n}a\right) }{\bar{K}_{\beta
_{j}}^{\left( \lambda \right) }\left( u_{0n}a\right) }\right] U_{2,\beta
_{j}}^{K}\left( u_{0n}r\right) \right\} .  \label{jnub20}
\end{align}%
Here, the parts involving the series over $n$ come from the integrals along
the contours $C_{\rho }(\gamma _{n})$ and the parts with the integrals over $%
\gamma $ are the contributions from the integrals over the straight segments
on the imaginary axis. Now, we need to evaluate the thermal contribution in
the boundary-induced expectation values as the sum $\langle j^{\nu }\rangle
_{T}^{(b)}=\sum_{\lambda =\pm }\langle j^{\nu }\rangle _{T\lambda }^{(b)}$.
By using the relations $\bar{F}_{\beta _{j}}^{(-)}(z)=\bar{F}_{\beta
_{j}}^{(+)\ast }(z)$, $F=I,K$, we can see that the contributions coming from
the first terms in the right-hand side for each expectation value in (\ref%
{jnub20}) to the sum over $\lambda $ give $-\langle j^{\nu }\rangle _{%
\mathrm{vac}}^{(b)}$ which are given by (\ref{jbvac}). In this way we find
\begin{align}
\left\langle j^{0}\right\rangle _{T}^{\mathrm{(b)}} &=\frac{2eT}{\phi _{0}}%
\sum_{j}\sum_{n=0}^{\infty }\mathrm{Re}\left[ \frac{\bar{I}_{\beta
_{j}}\left( u_{0n}a\right) }{\bar{K}_{\beta _{j}}\left( u_{0n}a\right) }%
\sum_{\chi =\pm 1}\left[ sm+\chi i\pi \left( 2n+1\right) T\right] K_{\alpha
_{j}-\chi \epsilon _{j}/2}^{2}\left( u_{0n}r\right) \right] -\left\langle
j^{0}\right\rangle _{\mathrm{vac}}^{\mathrm{(b)}},  \notag \\
\left\langle j^{2}\right\rangle _{T}^{\mathrm{(b)}} &=\frac{2eT}{\phi _{0}}%
\sum_{j}\sum_{n=0}^{\infty }\mathrm{Re}\left[ \frac{\bar{I}_{\beta
_{j}}\left( u_{0n}a\right) }{\bar{K}_{\beta _{j}}\left( u_{0n}a\right) }%
\right] U_{2,\beta _{j}}^{K}\left( u_{0n}r\right) -\left\langle
j^{2}\right\rangle _{\mathrm{vac}}^{\mathrm{(b)}}.  \label{jnub200}
\end{align}%
Substituting (\ref{jnub200}) into (\ref{jnube}), for the total
boundary-induced expectation values $\langle j^{\nu }\rangle ^{\mathrm{(b)}}$
in the case of zero chemical potential we get (\ref{jnub0}). The
corresponding formulae are also obtained from (\ref{jnub4}) in the limit $%
\mu \rightarrow 0$.

\section{Zero temperature limit}

\label{sec:AppB}

In the main text we have considered the zero temperature limit of the charge
and current densities started from the initial representations (\ref{jnupme}%
) and (\ref{jnuTi}). In this appendix we show that the same results are
obtained started from (\ref{jnub4}) and (\ref{jnubTi}). To be short, the
presentation of the limiting transition will be given for the current
density only. The corresponding steps for the charge density are similar. In
the E-region and for small temperatures the series over $n$ in (\ref{jnub4})
is dominated by the contribution of the terms with large $n$ and we replace
the corresponding summation by the integration in accordance with $%
\sum_{n=0}^{\infty }f(u_{n})\rightarrow \frac{1}{2\pi T}\int_{-i\mu
}^{\infty -i\mu }dx\,f(u)$, where $u=(x^{2}+m^{2})^{1/2}$. The leading order
term, obtained in this way, does not depend on the temperature and one gets
\begin{equation}
\lim_{T\rightarrow 0}\left\langle j^{2}\right\rangle ^{\mathrm{(b)}}=\frac{e%
}{\pi \phi _{0}}\sum_{j}\mathrm{Re}\left[ \int_{-i\mu }^{\infty -i\mu }dx\,%
\frac{\bar{I}_{\beta _{j}}\left( ua\right) }{\bar{K}_{\beta _{j}}\left(
ua\right) }U_{2,\beta _{j}}^{K}(ur)\right] .  \label{jnubT01}
\end{equation}%
By taking into account that for $r>a$ the integrand is exponentially small
for large values of $a\,\mathrm{Re}\,u$, the integral in (\ref{jnubT01}) is
transformed to the sum of the integrals over the straight contours $[-i\mu
,0]$ and $[0,\infty )$. Comparing with (\ref{jbvac}), we see that the part
coming from the integral over $[0,\infty )$ coincides with the
boundary-induced vacuum current density $\left\langle j^{2}\right\rangle _{%
\mathrm{vac}}^{\mathrm{(b)}}$. The remaining part with the integral over $%
[-i\mu ,0]$ is the contribution from particles and antiparticles. In the
case $|\mu |<m$ we pass to the integral over $y=-ix$ with $\int_{-i\mu
}^{0}dx=i\int_{-\mu }^{0}dy$. In the new integral one has $u=\sqrt{%
m^{2}-y^{2}}$ and $0<u<m$. In this range, all the functions in the integrand
of (\ref{jnubT01}) are real and the real part of integral $i\int_{-\mu
}^{0}dy$ is zero. Hence, for $|\mu |<m$ we get $\lim_{T\rightarrow
0}\left\langle j^{2}\right\rangle ^{\mathrm{(b)}}=\left\langle
j^{2}\right\rangle _{\mathrm{vac}}^{\mathrm{(b)}}$, in agreement with (\ref%
{jnubT0mu}).

In the range $|\mu |>m$ for the chemical potential the integral over $[-i\mu
,0]$ is further decomposed into the sum of the integrals over $[-i\mu
,-\lambda im]$ and $[-\lambda im,0]$, with $\lambda =\pm $ defined in
accordance with $\mu =\lambda |\mu |$. By the arguments similar to those
used for the integral over $[-i\mu ,0]$ in the case $|\mu |<m$, we can see
that the real part of the integral over $[-\lambda im,0]$ is zero. In the
remaining integral over the region $[-i\mu ,-\lambda im]$ we introduce $%
y=e^{\lambda i\pi /2}x$ and then pass to the integration over $\gamma =\sqrt{%
y^{2}-m^{2}}$. As a result, the integral is transformed as%
\begin{equation}
\int_{-\lambda i|\mu |}^{-\lambda im}dx\,\,f(u)=\lambda i\int_{0}^{\gamma _{%
\mathrm{F}}}d\gamma \,\frac{\gamma }{E}\,f(e^{-\lambda i\pi /2}\gamma ),
\label{fuint}
\end{equation}%
where $f(u)$ is the integrand in (\ref{jnubT01}) and, as before, $E=\sqrt{%
\gamma ^{2}+m^{2}}$. Next, in the right-hand side of (\ref{fuint}) we return
to the Bessel and Hankel functions by using the relations%
\begin{equation}
\bar{I}_{\beta _{j}}(e^{-\lambda \pi i/2}u)=e^{-\lambda i\pi \beta _{j}/2}%
\bar{J}_{\beta _{j}}^{(\lambda )}(u),\;\bar{K}_{\beta _{j}}(e^{-\lambda \pi
i/2}u)=\frac{\lambda \pi i}{2}e^{\lambda i\pi \beta _{j}/2}\bar{H}_{\beta
_{j}}^{(l,\lambda )}(u),  \label{BesHan}
\end{equation}%
where $l=1$ ($l=2$) for $\lambda =+$ ($\lambda =-$) and the notation $\bar{F}%
_{\beta _{j}}^{(\pm )}(z)$ for the Bessel and Hankel functions is defined by
(\ref{Fbar}). Finally, the edge induced contribution in the E-region is
expressed as
\begin{equation}
\lim_{T\rightarrow 0}\left\langle j^{2}\right\rangle ^{\mathrm{(b)}%
}=\left\langle j^{2}\right\rangle _{\mathrm{vac}}^{\mathrm{(b)}}-\frac{e}{%
\phi _{0}r}\sum_{j}\epsilon _{j}\int_{0}^{\gamma _{\mathrm{F}}}d\gamma \,%
\frac{\gamma ^{2}}{E}\mathrm{Re}\left[ \frac{\bar{J}_{\beta _{j}}^{\left(
\lambda \right) }\left( \gamma a\right) }{\bar{H}_{\beta _{j}}^{\left(
l,\lambda \right) }\left( \gamma a\right) }H_{\beta _{j}}^{\left( l\right)
}\left( \gamma r\right) H_{\beta _{j}+\epsilon _{j}}^{\left( l\right)
}\left( \gamma r\right) \right] .  \label{jnu2bT0}
\end{equation}%
Note that in the integration range of (\ref{jnu2bT0}) the function $\bar{J}%
_{\beta _{j}}^{\left( \lambda \right) }\left( za\right) $ is real and the
real part in the integrand is the same for $l=1$ and $l=2$. Adding to (\ref%
{jnu2bT0}) the corresponding limit for the contribution independent of $a$,
expressed as (\ref{jnuT0i2}), and by using the identity (\ref{ident1}), we
obtain the result (\ref{jnuT0}), as was expected.

Now we turn to the zero temperature limit of the current density in the
I-region based on the representation (\ref{jnubTi}) for the boundary-induced
contribution. The leading order term in the corresponding asymptotic
expansion is obtained replacing the summation over $n$ by the integration.
That term does not depend on the temperature and gives the zero temperature
limit. Similar to (\ref{jnubT01}), the current density it is presented in
the form
\begin{equation}
\lim_{T\rightarrow 0}\left\langle j^{2}\right\rangle ^{\mathrm{(b)}}=\frac{e%
}{\pi \phi _{0}}\sum_{j}\mathrm{Re}\left[ \int_{-i\mu }^{\infty -i\mu }dx\,%
\frac{\bar{K}_{\beta _{j}}\left( ua\right) }{\bar{I}_{\beta _{j}}\left(
ua\right) }U_{2,\beta _{j}}^{I}(ur)\right] ,  \label{jnubT0i}
\end{equation}%
again, with $u=(x^{2}+m^{2})^{1/2}$. Now we deform the integration contour
in the way described above for the E-region. The part of the integral over $%
x\in \lbrack 0,\infty )$ gives the VEV $\left\langle j^{2}\right\rangle _{%
\mathrm{vac}}^{\mathrm{(b)}}$ (see (\ref{jbvac})). For $|\mu |<m$, we
introduce in the integral $\int_{-i\mu }^{0}dx$ new variable $y=-ix$ with $u=%
\sqrt{m^{2}-y^{2}}>0$. In this case the functions in the integrand of (\ref%
{jnubT0i}) are real and for the part of the integral $\int_{-i\mu
}^{0}dx=i\int_{-\mu }^{0}dy$ the real part is zero. The same arguments are
valid in the case $|\mu |>m$ for the part of the integral $\int_{-i\mu
}^{0}dx$ over the interval $[-\lambda im,0]$. By transformations similar to
those for (\ref{fuint}) we can pass from the integral $\int_{-\lambda i|\mu
|}^{-\lambda im}dx\,$\ to the integral $\int_{0}^{\gamma _{\mathrm{F}%
}}d\gamma $. Passing to the Bessel and Hankel functions by using the
relations (\ref{BesHan}), the function in the integrand is transformed as%
\begin{equation}
\int_{-\lambda i|\mu |}^{-\lambda im}dx\frac{\bar{K}_{\beta _{j}}\left(
ua\right) }{\bar{I}_{\beta _{j}}\left( ua\right) }U_{2,\beta
_{j}}^{I}(ur)=-\lambda \frac{\pi }{2}\int_{0}^{\gamma _{\mathrm{F}}}d\gamma
\,\frac{\bar{H}_{\beta _{j}}^{(l,\lambda )}(\gamma a)}{\bar{J}_{\beta
_{j}}^{(\lambda )}(\gamma a)}g^{\left( 2\right) }\left( \gamma \right) ,
\label{Trans2}
\end{equation}%
where $u=e^{-\lambda \pi i/2}\gamma $, the function $g^{\left( 2\right)
}\left( \gamma \right) $ is defined by (\ref{gzi}), and $l$ is the same as
in (\ref{BesHan}). An important difference from the E-region is that now the
integrand in (\ref{Trans2}) has poles $\gamma =\gamma _{j,l}^{\left( \lambda
\right) }/a$, $l=1,\ldots ,l_{m}$, with $l_{m}$ defined in accordance with (%
\ref{poles}). These poles have to be avoided by semicircles $C_{\rho
}(\gamma _{j,l}^{\left( \lambda \right) })$, with small radius $\rho $, in
the right half-plane. The integral in the right-hand side of (\ref{Trans2})
should be understood as
\begin{equation}
\int_{0}^{\gamma _{\mathrm{F}}}d\gamma =\mathrm{p.v.}\int_{0}^{\gamma _{%
\mathrm{F}}}d\gamma +\lim_{\rho \rightarrow 0}\sideset{}{'}%
\sum_{l=1}^{l_{m}}\int_{C_{\rho }(\gamma _{j,l}^{\left( \lambda \right)
})}d\gamma .  \label{Intgam}
\end{equation}%
Here the prime means that in case $\gamma _{j,l_{m}}^{(\lambda )}=a\gamma _{%
\mathrm{F}}$ the integral $\int_{C_{\rho }(\gamma _{j,l_{m}}^{\left( \lambda
\right) })}d\gamma $ is taken over the quarter circle $C_{\rho }(\gamma
_{j,l_{m}}^{\left( \lambda \right) })$ with its center located at $\gamma
=\gamma _{\mathrm{F}}$. The real part of the first integral in the
right-hand side of (\ref{Intgam}) gives $\int_{0}^{\gamma _{\mathrm{F}%
}}d\gamma \,g^{\left( 2\right) }\left( \gamma \right) $ and the
corresponding contribution to (\ref{jnubT0i}) is expressed as $\left\langle
j^{2}\right\rangle _{\mathrm{vac}}^{\left( 0\right) }-\langle j^{2}\rangle
_{T=0}^{(0)}$. The contours $C_{\rho }(\gamma _{j,l}^{\left( \lambda \right)
})$ in (\ref{Intgam}) are determined by the condition that the poles $%
x=-\lambda i\sqrt{\gamma _{j,l}^{\left( \lambda \right) 2}/a^{2}+m^{2}}$ in
the left-hand side of (\ref{Trans2}) are avoided by small semicircles in the
right half-plane of complex variable $x$. From that condition it follows
that in (\ref{Intgam}) $C_{\rho }(\gamma _{j,l}^{\left( \lambda \right) })$
is a semicircular contour in the upper/lower half-plane of complex variable $%
\gamma $, centered at $\gamma =\gamma _{j,l}^{\left( \lambda \right) }$,
with clockwise/counter-clockwise directions for $\lambda =+$/$\lambda =-$.
The corresponding integral is expressed in terms of the respective residue: $%
\int_{C_{\rho }(\gamma _{j,l}^{\left( \lambda \right) })}d\gamma =-\lambda
\pi i\,\mathrm{Res}_{\gamma =\gamma _{j,l}^{\left( \lambda \right) }/a}$. In
the evaluation of the residue with the integrand from the right-hand side of
(\ref{Trans2}) we use the relations%
\begin{equation}
\bar{J}_{\beta _{j}}^{(\lambda )\prime }(z)=\frac{-2}{T_{\beta _{j}}^{\left(
\lambda \right) }\left( z\right) J_{\beta _{j}}\left( z\right) },\;\bar{Y}%
_{\beta _{j}}\left( z\right) =\frac{2}{\pi J_{\beta _{j}}\left( z\right) },
\label{JbarDer}
\end{equation}%
for $z=\gamma _{j,l}^{\left( \lambda \right) }$. Collecting all the
contributions to the right-hand side of (\ref{jnubT0i}) we get%
\begin{equation}
\lim_{T\rightarrow 0}\left\langle j^{2}\right\rangle ^{\mathrm{(b)}%
}=\left\langle j^{2}\right\rangle _{\mathrm{vac}}-\langle j^{2}\rangle
_{T=0}^{(0)}+\frac{\lambda e}{2\phi _{0}a}\sum_{j}\sideset{}{'}%
\sum_{l=1}^{l_{m}}T_{\beta _{j}}^{\left( \lambda \right) }\left( a\gamma
_{j,l}^{\left( \lambda \right) }\right) g^{\left( 2\right) }\left( \gamma
_{j,l}^{\left( \lambda \right) }\right) ,  \label{jnuT0ic}
\end{equation}%
which is equivalent to (\ref{jnuT0i}).


\begin{thebibliography}{99}
\bibitem{Frad13} E. Fradkin, \textit{Field Theory of Condensed Matter Systems%
} (Cambridge University Press, Cambridge, 2013).

\bibitem{Naga99} N. Nagaosa, \textit{Quantum Field Theory in Condensed
Matter Physics and Quantum Field Theory in Strongly Correlated Electronic
Systems} (Springer, Berlin, 1999).

\bibitem{Jack85} R. Jackiw, Nucl. Phys. B \textbf{252}, 343 (1985).

\bibitem{Gusy07} V.P. Gusynin, S.G. Sharapov, and J.P. Carbotte, Int. J.
Mod. Phys. B \textbf{21}, 4611 (2007).

\bibitem{Cast09} A.H. Castro Neto, F. Guinea, N.M.R. Peres, K.S. Novoselov,
and A.K. Geim, Rev. Mod. Phys. \textbf{81}, 109 (2009).

\bibitem{Xiao11} X.-L. Qi and S.-C. Zhang, Rev. Mod. Phys. \textbf{83}, 1057
(2011).

\bibitem{Muno20} A. Mu\~{n}oz de las Heras, E. Macaluso, and I. Carusotto,
Phys. Rev. X \textbf{10}, 041058 (2020).

\bibitem{Most97} V.M. Mostepanenko and N.N. Trunov, \textit{The Casimir
Effect and Its Applications} (Clarendon, Oxford, 1997).

\bibitem{Milt02} K.A. Milton, \textit{The Casimir Effect: Physical
Manifestation of Zero-Point Energy} (World Scientific, Singapore, 2002).

\bibitem{Bord09} M. Bordag, G.L. Klimchitskaya, U. Mohideen, and V.M.
Mostepanenko, \textit{Advances in the Casimir Effect} (Oxford University
Press, New York, 2009).

\bibitem{Casi11} \textit{Casimir Physics}, edited by D. Dalvit, P. Milonni,
D. Roberts, and F. da Rosa, Lecture Notes in Physics Vol. 834
(Springer-Verlag, Berlin, 2011).

\bibitem{Bell06} S. Bellucci, J. Gonz\'{a}lez, P. Onorato, and E. Perfetto,
Phys. Rev. B \textbf{74}, 045427 (2006).

\bibitem{Kibb76} T.W. Kibble, J. Phys. A \textbf{9}, 1387 (1976).

\bibitem{Vile94} A. Vilenkin and E.P.S. Shellard, \textit{Cosmic Strings and
Other Topological Defects} (Cambridge University Press, Cambridge, England,
1994).

\bibitem{Ge94} M. Ge and K. Sattler, Chem. Phys. Lett. \textbf{220}, 192
(1994).

\bibitem{Kris97} A. Krishnan, et al, Nature \textbf{388}, 451 (1997).

\bibitem{Lamm00} P.E. Lammert and V.H. Crespi, Phys. Rev. Lett. \textbf{85},
5190 (2000).

\bibitem{Char01} J.-Ch. Charlier and G.-M. Rignanese, Phys. Rev. Lett.
\textbf{86}, 5970 (2001).

\bibitem{Osip01} V.A. Osipov and E.A. Kochetov, JETP Letters \textbf{73},
562 (2001).

\bibitem{Lamm04} P.E. Lammert and V.H. Crespi, Phys. Rev. B \textbf{69},
035406 (2004).

\bibitem{Cort07} A. Cortijo and M.A.H. Vozmediano, Nucl. Phys. B \textbf{763}%
, 293 (2007).

\bibitem{Furt08} C. Furtado, F. Moraes, and A.M.M. Carvalho, Phys. Lett. A
\textbf{372}, 5368 (2008).

\bibitem{Naes09} S.N. Naess, A. Elgsaeter, G. Helgesen, and K.D. Knudsen,
Sci. Technol. Adv. Mater. \textbf{10}, 065002 (2009).

\bibitem{Ullo13} P. Ulloa, A. Latg\'{e}, L. E. Oliveira, and M. Pacheco,
Nanoscale Research Letters \textbf{8}, 384 (2013).

\bibitem{Matt23} S. H. Mattoso, V. Brumas, S. Evangelisti, G. Fronzoni, T.
Leininger, M. Stener, J. Phys. Chem. A \textbf{127}, 9723 (2023).

\bibitem{Bell20} S. Bellucci, I. Brevik, A. A. Saharian, and H. G. Sargsyan,
Eur. Phys. J. C \textbf{80}, 281 (2020).

\bibitem{Srir01} L. Sriramkumar, Classical Quantum Gravity \textbf{18}, 1015
(2001).

\bibitem{Site08} Yu.A. Sitenko and N.D. Vlasii, J. Phys. A: Math. Theor.
\textbf{41}, 164034 (2008).

\bibitem{Site09} Yu.A. Sitenko and N.D. Vlasii, Classical Quantum Gravity
\textbf{26}, 195009 (2009).

\bibitem{Beze10a} E.R. Bezerra de Mello, Classical Quantum Gravity \textbf{27%
}, 095017 (2010).

\bibitem{Beze10} E.R. Bezerra de Mello, V. Bezerra, A.A. Saharian, and V.M.
Bardeghyan, Phys. Rev. D\textbf{\ 82}, 085033 (2010).

\bibitem{Moha15} A. Mohammadi, E.R. Bezerra de Mello, and A.A. Saharian, J.
Phys. A: Math. Theor. \textbf{48}, 185401 (2015).

\bibitem{Brag15} E.A.F. Bragan\c{c}a, H.F. Santana Mota, and E.R. Bezerra de
Mello, Int. J. Mod. Phys. D \textbf{24}, 1550055 (2015).

\bibitem{Bell16T} S. Bellucci, E.R. Bezerra de Mello, E. Bragan\c{c}a, and
A.A. Saharian, Eur. Phys. J. C \textbf{76}, 359 (2016).

\bibitem{Moha16} A. Mohammadi and E.R. Bezerra de Mello, Phys. Rev. D
\textbf{93}, 123521 (2016).

\bibitem{Site18} Yu.A. Sitenko and N.D. Vlasii, Universe \textbf{4}, 23
(2018).

\bibitem{Saha24} A.A. Saharian, Phys. Rev. D \textbf{110}, 065020 (2024).

\bibitem{Beze15} E.R. Bezerra de Mello, V.B. Bezerra, A.A. Saharian, and
H.H. Harutyunyan, Phys. Rev. D \textbf{91}, 064034 (2015).

\bibitem{Oliv19} W. Oliveira dos Santos, H.F. Santana Mota, and E.R. Bezerra
de Mello, Phys. Rev. D \textbf{99}, 045005 (2019).

\bibitem{Brag20} E.A.F. Bragan\c{c}a, E.R. Bezerra de Mello, and
A.Mohammadi, Int. J. Mod. Phys. D \textbf{29}, 2050103 (2020).

\bibitem{Bell20b} S. Bellucci, W. Oliveira Dos Santos, and E.R. Bezerra de
Mello, Eur. Phys. J. C \textbf{80}, 963 (2020).

\bibitem{Beze22} E.R. Bezerra de Mello, W. Oliveira dos Santos, and A.A.
Saharian, Phys. Rev. D \textbf{106}, 125009 (2022).

\bibitem{Oliv24} W. Oliveira dos Santos, H.F. Santana Mota, and E.R. Bezerra
de Mello, arXiv:2409.05691.

\bibitem{Saha24b} A.A. Saharian, Symmetry \textbf{16}, 92 (2024).

\bibitem{Saha19} A.A. Saharian, E.R. Bezerra de Mello and A.A. Saharyan,
Phys. Rev. D \textbf{100}, 105014 (2019).

\bibitem{Bene12} C.G. Beneventano and E.M. Santangelo, Int. J. Mod. Phys.:
Conf. Series \textbf{14}, 240 (2012).

\bibitem{Bell16} S. Bellucci, A.A. Saharian, and A.Kh. Grigoryan, Phys. Rev.
D \textbf{94}, 105007 (2016).

\bibitem{Prud2} A.P. Prudnikov, Yu.A. Brychkov, and O.I. Marichev, \textit{%
Integrals and Series} (Gordon and Breach, New York, 1986), Vol. 2.

\bibitem{Saharev} A.A. Saharian, Izv. Akad. Nauk. Arm. SSR. Mat. \textbf{22}%
, 166 (1987) {[}English translation: A. A. Saaryan, Sov. J. Contemp. Math.
Anal. \textbf{22}, 70 (1987){]}.

\bibitem{Saharev2} A.A. Saharian, \textit{The Generalized Abel-Plana Formula
with Applications to Bessel Functions and Casimir Effect} (Yerevan State
University Publishing House, Yerevan, 2008) (arXiv:0708.1187).

\bibitem{Berr87} M.V. Berry and R.J. Mondragon, Proc. R. Soc. A \textbf{412}%
, 53 (1987).

\bibitem{Deut79} D. Deutsch and P. Candelas, Phys. Rev. D \textbf{20}, 3063
(1979).

\bibitem{Kenn80} G. Kennedy, R. Critchley, and J.S. Dowker, Ann. Phys.
\textbf{125}, 346 (1980).

\bibitem{Bell09} S. Bellucci and A.A. Saharian, Phys. Rev. D \textbf{79},
085019 (2009).

\bibitem{Bell10} S. Bellucci, A.A. Saharian, and V.M. Bardeghyan, Phys. Rev.
D \textbf{82}, 065011 (2010).

\bibitem{Butt83} M. B\"{u}ttiker, Y. Imry, R. Landauer, Phys. Lett. A
\textbf{96}, 365 (1983).

\bibitem{Bluh09} H. Bluhm et al., Phys. Rev. Lett. \textbf{102}, 136802
(2009).

\bibitem{Bles09} A.C. Bleszynski-Jayich et al., Science \textbf{326}, 272
(2009).

\bibitem{Lin98} M.F. Lin and D.S. Chuu, Phys. Rev. B \textbf{57}, 6731
(1998).

\bibitem{Lati03} S. Latil, S. Roche, and A. Rubio,\ Phys. Rev. B \textbf{67}%
, 165420 (2003).

\bibitem{Rech07} P. Recher, B. Trauzettel, A. Rycerz, Y. Blanter, C.
Beenakker, and A. Morpurgo, Phys. Rev. B \textbf{76}, 235404 (2007).

\bibitem{Wu10} Z. Wu, Z. Zhang, K. Chang, and F.M. Peeters, Nanotechnology
\textbf{21}, 185201 (2010).

\bibitem{Mich11} P. Michetti and P. Recher, Phys. Rev. B \textbf{83}, 125420
(2011).

\bibitem{Arau22} F.R.V. Ara\'{u}jo, D.R. da Costa, A.J.C. Chaves, F.E. B. de
Sousa, and J.M. Pereira Jr., J. Phys.: Condens. Matter \textbf{34}, 125503
(2022).

\bibitem{Saha19FC} A. Saharian, T. Petrosyan, and A. Hovhannisyan, Universe
\textbf{7}, 73 (2021).
\end{thebibliography}
\end{document}